\documentclass[preprint,12pt]{elsarticle}

\usepackage{amssymb}
\usepackage{amsthm}

\usepackage[dvipsnames]{xcolor}

\usepackage{subfigure}

\usepackage{amsmath}

\usepackage{hyperref}

\usepackage{multirow}

\usepackage{bm}

\journal{CMAME}

\begin{document}

\begin{frontmatter}



\title{Overcoming membrane locking in quadratic NURBS-based discretizations of linear Kirchhoff-Love shells: CAS elements}


%

\author[label1]{Hugo Casquero\corref{cor1}}
\ead{casquero@umich.edu}
\author[label1]{Kyle Dakota Mathews}
\address[label1]{Department of Mechanical Engineering, University of Michigan – Dearborn, 4901 Evergreen Road, Dearborn, MI 48128-1491, U.S.A.}

\cortext[cor1]{Corresponding author.}

\begin{abstract}

Quadratic NURBS-based discretizations of the Galerkin method suffer from membrane locking when applied to Kirchhoff-Love shell formulations. Membrane locking causes not only smaller displacements than expected, but also large-amplitude spurious oscillations of the membrane forces. Continuous-assumed-strain (CAS) elements have been recently introduced to remove membrane locking in quadratic NURBS-based discretizations of linear plane curved Kirchhoff rods (Casquero et al., CMAME, 2022). In this work, we generalize CAS elements to vanquish membrane locking in quadratic NURBS-based discretizations of linear Kirchhoff-Love shells. CAS elements bilinearly interpolate the membrane strains at the four corners of each element. Thus, the assumed strains have $C^0$ continuity across element boundaries. To the best of the authors' knowledge, CAS elements are the first assumed-strain treatment to effectively overcome membrane locking in quadratic NURBS-based discretizations of Kirchhoff-Love shells while satisfying the following important characteristics for computational efficiency: (1) No additional degrees of freedom are added, (2) No additional systems of algebraic equations need to be solved, (3) No matrix multiplications or matrix inversions are needed to obtain the stiffness matrix, and (4) The nonzero pattern of the stiffness matrix is preserved. The benchmark problems show that CAS elements, using either $2\times2$ or $3\times3$ Gauss-Legendre quadrature points per element, are an effective locking treatment since this element type results in more accurate displacements for coarse meshes and excises the spurious oscillations of the membrane forces. The benchmark problems also show that CAS elements outperform state-of-the-art element types based on Lagrange polynomials equipped with either assumed-strain or reduced-integration locking treatments.

\end{abstract}




\end{frontmatter}



\section{Introduction}

Isogeometric analysis (IGA) \cite{1003.000, cottrell2009isogeometric} opens the door for a tighter integration between computer-aided design (CAD) and finite element analysis (FEA) of shell structures \cite{wen2023isogeometric, wei2022analysis, casquero2020seam, toshniwal2017smooth, nagy2015numerical, buffa2020minimal, leidinger2019explicit, pasch2021priori}. In addition, the smoothness of splines across element boundaries enables the use of the Galerkin method to discretize the primal formulation of Kirchhoff-Love shells \cite{Kiendl2009, Kiendl2015, casquero2017arbitrary, duong2017new, roohbakhshan2017efficient, leonetti2023mixed}. Kirchhoff-Love shell theory neglects transverse shear strains, which are considered to be negligible as long as $R/t \geq 20$ \cite{bischoff2004models}, where $R$ is the radius of curvature, $t$ is the thickness, and $R/t$ is the slenderness ratio. Thus, Kirchhoff-Love shells do not suffer from transverse shear locking. Nevertheless, Kirchhoff-Love shells still suffer from membrane locking \cite{bieber2018variational, greco2018reconstructed, zou2021galerkin}, which is also the case for Reissner-Mindlin shells \cite{stolarski1983shear, echter2013hierarchic, oesterle2016shear, oesterle2017hierarchic, zou2022efficient} and solid shells \cite{hauptmann1998systematic, liu1998multiple, cardoso2008enhanced, bouclier2013development, bouclier2013efficient, bouclier2015isogeometric}. Membrane locking leads to displacements and bending moments with smaller values than expected and membrane forces with large-amplitude spurious oscillations.

Since our goal is the development of a locking treatment that is adopted in commercial FEA software, we first summarize some of the locking treatments that have been widely adopted in commercial FEA software to overcome membrane locking when using Lagrange polynomials with $C^0$ continuity across element boundaries. Assumed-strain treatments use assumed strains obtained from either projecting or interpolating the compatible strains to vanquish locking \cite{macneal1978simple, hughes1981finite, macneal1982derivation, dvorkin1984continuum}. Reduced-integration treatments vanquish locking by reducing the number of quadrature points, but this introduces spurious energy modes which need to be dealt with using hourglass control \cite{zienkiewicz1971reduced, flanagan1981uniform, belytschko1984hourglass, belytschko1984explicit}. Assumed-strain and reduced-integration treatments are the two locking treatments more widely utilized by the end users of commercial FEA programs \cite{lsdyna, ansysmechanical, abaqus}. These two locking treatments are equivalent to mixed methods under certain conditions \cite{malkus1978mixed, simo1986variational}.

Non-uniform rational B-splines (NURBS) can have up to $C^{p-1}$ continuity across element boundaries, where $p$ is the degree of the basis functions. In contrast, Lagrange polynomials have $C^0$ continuity across element boundaries for any degree $p$. As a result of this different level of continuity across element boundaries, the numerical schemes that are effective in overcoming locking for Lagrange polynomials and NURBS are different. Reduced-integration treatments for Lagrange polynomials use $p$ Gauss-Legendre quadrature points per direction and add hourglass control to handle spurious energy modes \cite{zienkiewicz1971reduced, flanagan1981uniform, belytschko1984hourglass, belytschko1984explicit}. Using $p$ Gauss-Legendre quadrature points per direction in a NURBS-based discretization does not create spurious energy modes, but it is not an effective locking treatment either \cite{adam2014improved, adam2015improved, casquero2022removing, casquero2023trods}. Reduced-integration treatments for NURBS have been developed in which the integration is performed at the patch level instead of at the element level and spurious energy modes may be avoided in linear problems by slightly over-integrating near the patch boundary \cite{adam2015selective, leonetti2018efficient, leonetti2019simplified}. In \cite{hokkanen2020quadrature}, it was pointed out that these reduced patch-wise quadrature rules may still have spurious energy modes in nonlinear problems. Projection-based assumed-strain treatments for Lagrange polynomials obtain the assumed strains by performing $L^2$ projections of the compatible strains onto lower-degree polynomial spaces within each element \cite{hughes1980generalization}. Using a $L^2$ projection for each element results in assumed strains that are discontinuous across element boundaries and it is not an effective locking treatment for NURBS-based discretizations \cite{bouclier2013efficient, greco2018reconstructed, casquero2022removing}. Projection-based assumed-strain treatments for NURBS have been developed in which the $L^2$ projection is performed at the patch level instead of at the element level which enables to obtain assumed strains with the needed continuity patterns across element boundaries \cite{thomas, bouclier2012locking, bouclier2013efficient, zhang2018locking}. However, performing the $L^2$ projection at the patch level requires the inversion of a mass matrix at the patch level to obtain the stiffness matrix at the patch level which results to be a full matrix instead of a sparse matrix. To decrease the computational burden, the reconstructed $\bar{B}$ projections were developed for NURBS-based discretizations \cite{bouclier2013efficient, miao2018bezier, greco2017efficient, greco2018reconstructed, zou2020isogeometric}. The reconstructed $\bar{B}$ projections perform the $L^2$ projections at the element level and then combine the results from different elements to recover the needed continuity patterns of the assumed strains across element boundaries. Nevertheless, reconstructed $\bar{B}$ projections still require matrix multiplications to obtain the stiffness matrix at the patch level and its bandwidth increases. Interpolation-based assumed-strain treatments for Lagrange polynomials define the assumed strains using lower-degree polynomial spaces and interpolate the compatible strains at interior points within each element \cite{bucalem1993higher}. Using interpolation points in the interior of each element results in assumed strains that are discontinuous across element boundaries and it is not an effective locking treatment for NURBS-based discretizations since the spurious oscillations of the stress resultants are not stripped away \cite{greco2017efficient, casquero2022removing, kim2022isogeometric}.


Continuous-assumed-strain (CAS) elements were recently introduced to remove membrane locking in quadratic NURBS-based discretizations of linear plane curved Kirchhoff rods \cite{casquero2022removing}, shear and membrane locking in quadratic NURBS-based discretizations of linear plane curved Timoshenko rods \cite{casquero2023trods}, and volumetric locking in quadratic NURBS-based discretizations of nearly-incompressible linear elasticity \cite{casquero2023le}. CAS elements are an interpolation-based assumed-strain treatment whose interpolation points are located at element boundaries instead of in the interior of each element. In this work, we generalize CAS elements to overcome membrane locking in quadratic NURBS-based discretizations of linear Kirchhoff-Love shells. CAS elements bilinearly interpolate the membrane strains at the four corners of each element. Since the displacement vector given by quadratic NURBS has $C^1$ continuity across element boundaries, the assumed strains obtained have $C^0$ continuity across element boundaries. To the best of the authors' knowledge, CAS elements are the first assumed-strain treatment to effectively overcome membrane locking in quadratic NURBS-based discretizations of Kirchhoff-Love shells while satisfying the following important characteristics for the computational efficiency of the numerical scheme: (1) No additional degrees of freedom are added, (2) No additional systems of algebraic equations need to be solved, (3) No matrix multiplications or matrix inversions are needed to obtain the stiffness matrix, and (4) The nonzero pattern of the stiffness matrix is preserved. Membrane locking causes not only smaller displacements than expected, but also large-amplitude spurious oscillations of the membrane forces. The spurious oscillations of the membrane forces persist for fine meshes that result in accurate displacements \cite{casquero2022removing, casquero2023trods}. Thus, we study the accuracy of both displacements and membrane forces to show that CAS elements verily overcome membrane locking.

The paper is outlined as follows. Section 2 sets forth the mathematical theory of linear Kirchhoff-Love shells. Section 3 summarizes how to solve the problem using compatible-strain (CS) elements. Section 4 develops CAS elements to overcome membrane locking in quadratic NURBS-based discretizations of linear Kirchhoff-Love shells. Section 5 evaluates the performance of CS and CAS elements using either $2\times2$ or $3\times3$ Gauss-Legendre quadrature points per element, including comparisons with exact solutions and state-of-the-art element types based on Lagrange polynomials equipped with either assumed-strain or reduced-integration locking treatments. Sections 5.1, 5.2, 5.3, and 5.4 consider the cylindrical shell strip \cite{zou2021galerkin}, the pinched hemisphere with a hole \cite{macneal1985proposed}, the Scordelis-Lo roof \cite{macneal1985proposed, Belytschko1985}, and the partly clamped hyperbolic paraboloid \cite{chapelle1998fundamental, bathe2000evaluation} as benchmark problems, respectively. Concluding remarks and directions of future work are drawn in Section 6.

\section{Linear Kirchhoff-Love shells}

In this section, we consider Kirchhoff-Love shells with infinitesimal deformations and small strains, that is, we do not consider either geometric nonlinearities or material nonlinearities. The geometry of the shell is defined by its midsurface and its thickness $t$, which we assume to be constant throughout the whole shell. We state the Kirchhoff-Love shell formulation using the Lagrangian description and curvilinear coordinates. For a mathematical derivation of the model, the reader is referred to \cite{lovemathematical, koiter1970mathematical}.

In the following, indices in Greek letters take the values $\{1,2\}$, indices in Latin letters take the values $\{1,2,3\}$, and repeated indices imply summation. Subscript indices indicate covariant quantities while superscript indices indicate contravariant quantities. $(\cdot) \cdot (\cdot)$ and $(\cdot) \times (\cdot)$ denote the dot and cross products of vectors, respectively. $||\cdot||$ denotes the length of a vector. $|\cdot|$ denotes the determinant of a matrix.

\subsection{Kinematics in infinitesimal deformations}

The geometry of the midsurface is defined by the parametric surface $\bm{r} (\bm{\theta}) : [0,1]^2 \mapsto \mathbb{R}^3$, where $\bm{\theta} = (\theta^1, \theta^2)$ are parametric coordinates and $\bm{r}$ is the position vector of a material point on the midsurface. Partial derivatives with respect to the parametric coordinates are indicated by comma, i.e., $(\cdot)_{,\alpha} = \partial (\cdot) / \partial\theta^{\alpha}$. The displacement vector of a material point on the midsurface is defined as $\bm{u} (\bm{\theta}) : [0,1]^2 \mapsto \mathbb{R}^3$. Both $\bm{r}$ and $\bm{u}$ are defined using a global system of Cartesian coordinates. 

Non-unit tangent vectors to the midsurface are obtained by
\begin{equation}
\bm{a}_{\alpha} =  \bm{r}_{,\alpha} \text{.}
\end{equation}
The unit normal vector to the midsurface is obtained by
\begin{equation}
\bm{a}_{3} =  \frac{\bm{a}_1\times\bm{a}_2}{||\bm{a}_1\times\bm{a}_2||} \text{.}
\end{equation}
The local covariant basis of the midsurface is defined as $(\bm{a}_{1},\bm{a}_{2}, \bm{a}_{3})$. The covariant metric coefficients of the midsurface are defined as
\begin{equation}
a_{\alpha\beta}=\bm{a}_{\alpha}\cdot\bm{a}_{\beta} \text{.}
\end{equation}
The contravariant metric coefficients can be computed as the inverse matrix of the covariant coefficients, i.e., $\left[ a^{\alpha\beta} \right] = \left[ a_{\alpha\beta} \right]^{-1}$. The covariant curvature coefficients of the midsurface are defined as
\begin{equation}
b_{\alpha\beta} = \bm{a}_{\alpha, \beta} \cdot \bm{a}_{3} \text{.}
\end{equation}
Mixed curvature coefficients can be obtained using the index raising property of the contravariant metric coefficients, viz.,
\begin{equation}
b^{\alpha}_{\beta} = a^{\alpha\lambda} b_{\lambda\beta} \text{.}
\end{equation}
The covariant coefficients of the membrane strains are defined as
\begin{equation}
\epsilon_{\alpha\beta}= \frac{1}{2} \left( \bm{u}_{,\alpha} \cdot \bm{a}_{\beta} + \bm{u}_{,\beta} \cdot \bm{a}_{\alpha} \right) \text{.}
\end{equation}
The covariant coefficients of the bending pseudo-strains are defined as
\begin{align}
\kappa_{\alpha\beta}=& - \bm{a}_{3} \cdot \bm{u}_{,\alpha\beta} + \frac{1}{||\bm{a}_3|| } [ ( \bm{a}_{\alpha, \beta} \times \bm{a}_{2} ) \cdot \bm{u}_{,1} + (\bm{a}_{1} \times \bm{a}_{\alpha, \beta}) \cdot \bm{u}_{,2} \nonumber \\
& +   \bm{a}_{\alpha, \beta} \cdot \bm{a}_{3} [ ( \bm{a}_{2} \times \bm{a}_{3}) \cdot \bm{u}_{,1} + ( \bm{a}_{3} \times \bm{a}_{1}) \cdot \bm{u}_{,2} ]  ]   \text{.}
\end{align}

\subsection{Linear material}

The membrane forces and the bending moments are the stress resultants of Kirchhoff-Love shells that are obtained from constitutive equations. Here, we consider an isotropic and linear elastic material. The contravariant coefficients of the membrane forces are defined as
\begin{equation}
 n^{\alpha\beta}  = \frac{Et}{1-{\nu}^2} \left[    (1-\nu)  a^{\alpha \lambda} \,  a^{\mu \beta} \,  \epsilon_{\lambda\mu}  + \nu a^{\alpha \beta} \,  a^{\mu \lambda} \,   \epsilon_{\lambda \mu}   \right]   \text{.}
\end{equation}
where $E$ is the Young's modulus and $\nu$ is the Poisson's ratio. The contravariant coefficients of the bending moments are defined as
\begin{equation}
 m^{\alpha\beta}  =  \frac{E{t}^3}{12(1-{\nu}^2)} \left[    (1-\nu)  a^{\alpha \lambda} a^{\mu \beta} \kappa_{\lambda\mu}  + \nu a^{\alpha \beta} a^{\mu \lambda}  \kappa_{\lambda\mu}   \right]   \text{.}
\end{equation}
The contravariant coefficients of the effective membrane forces are defined as
\begin{equation}
 n_{\text{eff}}^{\alpha\beta}  = n^{\alpha\beta} +  m^{\alpha \lambda} b^{\beta}_{\lambda} \text{.}
\end{equation}
The contravariant coefficients $n^{\alpha\beta}$, $m^{\alpha\beta}$, and $n_{\text{eff}}^{\alpha\beta}$ refer to the covariant basis $(\bm{a}_{1},\bm{a}_{2}, \bm{a}_{3})$, which is not necessarily an orthonormal basis. Thus, to obtain these stress resultants in normalized units, these coefficients need to be transformed into a local Cartesian basis using the transformation rules
\begin{align} 
\widehat{n}^{\alpha \beta} & = n^{\gamma\mu} (\mathbf{e}^{\alpha} \cdot \mathbf{a}_{\gamma})  ( \mathbf{a}_{\mu} \cdot \mathbf{e}^{\beta})   \text{,} \\
\widehat{m}^{\alpha \beta} & = m^{\gamma\mu} (\mathbf{e}^{\alpha} \cdot \mathbf{a}_{\gamma})  ( \mathbf{a}_{\mu} \cdot \mathbf{e}^{\beta})              \text{,} \\
\widehat{n}_{\text{eff}}^{\alpha \beta} & = n_{\text{eff}}^{\gamma\mu} (\mathbf{e}^{\alpha} \cdot \mathbf{a}_{\gamma})  ( \mathbf{a}_{\mu} \cdot \mathbf{e}^{\beta})   \text{,}
\end{align}
where $\bm{e}^{\alpha}$ is the $\alpha$th base vector of the local Cartesian basis, $\widehat{n}^{\alpha \beta}$ are the membrane coefficients with respect to the local Cartesian basis, $\widehat{m}^{\alpha \beta}$ are the bending coefficients with respect to the local Cartesian basis, and $\widehat{n}_{\text{eff}}^{\alpha \beta}$ are the effective membrane coefficients with respect to the local Cartesian basis. We choose to compute the local Cartesian basis as follows
\begin{align} 
\bm{e}_{1} &=  \bm{e}^{1} =  \frac{\bm{a}_{1}}{ || \bm{a}_{1} ||} \text{,} \\
\bm{e}_{2} &=  \bm{e}^{2} =  \frac{\bm{a}_{2} - (\bm{a}_{2} \cdot \bm{e}_{1}) \bm{e}_{1}}{ || \bm{a}_{2} - (\bm{a}_{2} \cdot \bm{e}_{1}) \bm{e}_{1} ||}   \text{.}
\end{align}

\subsection{Variational form}

The variational form can be obtained from the principle of virtual work which states that the internal virtual work ($\delta W^{int}$) must be equal to the external virtual work ($\delta W^{ext}$) for any virtual displacement ($\delta \mathbf{u}$), i.e.,
\begin{equation} 
 \delta W^{int} =  \delta W^{ext} \quad \forall \delta \mathbf{u} \text{,}  \label{virtualwork}
\end{equation}
with
\begin{align}
   \delta W^{int}&=  \int_{A}\left(\delta \epsilon_{\alpha \beta} n^{\alpha\beta} + \delta \kappa_{\alpha \beta} m^{\alpha\beta} \right) \, \mathrm dA  \text{,}\label{Wint} \\
   \delta W^{ext}&=  \int_{A} \delta u_i f_i    \, \mathrm dA  \text{,} \label{Wext}
\end{align}
where $d A = \sqrt{|a_{\alpha\beta}|}d\theta^1d\theta^2$ is the differential area, $A$ is the area of the midsurface, $\delta \epsilon_{\alpha \beta}$ are the virtual covariant membrane strain coefficients, $\delta \kappa_{\alpha \beta}$ are the virtual covariant bending pseudo-strain coefficients, and $f_i$ is the $i$th component of a vector defined in a global system of Cartesian coordinates that represents a load per unit area acting on the midsurface.

The total strain energy of the shell is defined as
\begin{equation} 
 E_T = E_m + E_b \text{,}  \label{Et}
\end{equation}
with
\begin{align}
   E_m &=  \frac{1}{2} \int_{A}  \epsilon_{\alpha \beta} n^{\alpha\beta}  \, \mathrm dA  \text{,}\label{Em} \\
   E_b &= \frac{1}{2} \int_{A} \kappa_{\alpha \beta} m^{\alpha\beta}   \, \mathrm dA  \text{,} \label{Eb}
\end{align}
where $E_m$ is the membrane strain energy of the shell and $E_b$ is the bending strain energy of the shell.

\section{Compatible-strain (CS) elements}

The geometry of the midsurface is represented as a linear combination of NURBS basis functions, viz.,
\begin{equation}
\bm{r} (\bm{\theta})= \sum_{A=1}^{n_{cp}} N_A (\bm{\theta}) \bm{Q}_A  \text{,}
\end{equation}
where $\bm{Q}_A$ is the $A$-th control point and $n_{cp}$ is the total number of control points. In this work, we use open knot vectors with no repeated interior knots and quadratic basis functions. For the details of how to define a geometry using NURBS basis functions and how to perform $h$-refinement using the knot insertion algorithm, the reader is referred to \cite{cottrell2009isogeometric}. Using the isoparametric concept, the displacement vector is discretized as follows
\begin{equation}
\bm{u}^h (\bm{\theta}) = \sum_{A=1}^{n_{cp}} N_A (\bm{\theta}) \bm{U}_A  \text{,}
\end{equation}
where $\bm{U}_A$ is the $A$-th control variable of the displacement vector. Using the Bubnov-Galerkin method, the virtual displacements are discretized as $\delta \bm{u}^h (\bm{\theta}) \in \text{span} \{ N_{A}(\bm{\theta})\}_{A=1}^{n_{cp}}$.

The discretization choices made above completely define the element stiffness matrix of CS elements, viz.,
\begin{equation}
\mathbf{k} = \mathbf{k}_\epsilon  + \mathbf{k}_\kappa \text{,}
\end{equation}
\begin{equation} 
 \mathbf{k}_{\epsilon} =  \left[ k_{\epsilon,ibjc} \right], \quad \mathbf{k}_{\kappa} =  \left[ k_{\kappa,ibjc} \right] \text{,} 
\end{equation}
where $\mathbf{k}$ is the element stiffness matrix, $\mathbf{k}_{\epsilon}$ is the element membrane stiffness matrix, and $\mathbf{k}_{\kappa}$ is the element bending stiffness matrix. The explicit expression of $k_{\epsilon,ibjc}$ is given by
\begin{align} 
  k_{\epsilon,ibjc} & = \frac{Et}{4(1+\nu)}  \int_{A^e}  N_{b,\alpha}   \bm{e}_i \cdot \bm{a}_{\beta} a^{\alpha \lambda}   a^{\mu \beta}  N_{c,\lambda}   \bm{e}_j \cdot \bm{a}_{\mu}   \, \mathrm dA  \nonumber \\
  & + \frac{Et}{4(1+\nu)}  \int_{A^e}  N_{b,\alpha}   \bm{e}_i \cdot \bm{a}_{\beta} a^{\alpha \lambda}   a^{\mu \beta}  N_{c,\mu}   \bm{e}_j \cdot \bm{a}_{\lambda}   \, \mathrm dA \nonumber \\
  & + \frac{Et}{4(1+\nu)}  \int_{A^e}  N_{b,\beta}   \bm{e}_i \cdot \bm{a}_{\alpha} a^{\alpha \lambda}   a^{\mu \beta}  N_{c,\lambda}   \bm{e}_j \cdot \bm{a}_{\mu}   \, \mathrm dA  \nonumber \\
  & + \frac{Et}{4(1+\nu)}  \int_{A^e}  N_{b,\beta}   \bm{e}_i \cdot \bm{a}_{\alpha} a^{\alpha \lambda}   a^{\mu \beta}  N_{c,\mu}   \bm{e}_j \cdot \bm{a}_{\lambda}   \, \mathrm dA \nonumber \\ \displaybreak[4]
  & + \frac{Et\nu}{4(1-\nu^2)}  \int_{A^e}  N_{b,\alpha}   \bm{e}_i \cdot \bm{a}_{\beta} a^{\alpha \beta}   a^{\mu \lambda}  N_{c,\lambda}   \bm{e}_j \cdot \bm{a}_{\mu}   \, \mathrm dA  \nonumber \\
  & + \frac{Et\nu}{4(1-\nu^2)}  \int_{A^e}  N_{b,\alpha}   \bm{e}_i \cdot \bm{a}_{\beta} a^{\alpha \beta}   a^{\mu \lambda}  N_{c,\mu}   \bm{e}_j \cdot \bm{a}_{\lambda}   \, \mathrm dA \nonumber \\
  & + \frac{Et\nu}{4(1-\nu^2)}  \int_{A^e}  N_{b,\beta}   \bm{e}_i \cdot \bm{a}_{\alpha} a^{\alpha \beta}   a^{\mu \lambda} N_{c,\lambda}   \bm{e}_j \cdot \bm{a}_{\mu}   \, \mathrm dA  \nonumber \\
  & + \frac{Et\nu}{4(1-\nu^2)}  \int_{A^e}  N_{b,\beta}   \bm{e}_i \cdot \bm{a}_{\alpha} a^{\alpha \beta}   a^{\mu \lambda}  N_{c,\mu}   \bm{e}_j \cdot \bm{a}_{\lambda}   \, \mathrm dA \text{,} 
\end{align}
where $A^e$ is the midsurface area of element $e$. The explicit expression of $k_{\kappa,ibjc}$ can be obtained analogously and it is omitted here for brevity. Following standard FEA paraphernalia, the integrals above are computed performing change of variables from the parametric coordinates $(\theta^1, \theta^2)$ to the parent element with coordinates $(\widehat{\theta^1}, \widehat{\theta^2})\in [-1,1]^2$. The assembly of the  $n_{el}$ element stiffness matrices into the global stiffness matrix is performed using conventional connectivity arrays as explained in \cite{Hughes2012, cottrell2009isogeometric}, where $n_{el}$ is the total number of elements in the mesh.

\section{Continuous-assumed-strain (CAS) elements}

The covariant membrane strain coefficients of a CS element have the following expression

\begin{equation} 
\epsilon^h_{\alpha\beta} (\bm{\theta}) = \frac{1}{2} \left( \bm{u}^h_{,\alpha} ((\bm{\theta}) \cdot \bm{a}_{\beta} (\bm{\theta}) + \bm{u}^h_{,\beta} (\bm{\theta}) \cdot \bm{a}_{\alpha} (\bm{\theta}) \right) \text{.} 
\end{equation}

Leveraging the $C^1$ continuity across element boundaries of the geometry and the displacement vector given by quadratic NURBS, the linear interpolation of the covariant membrane strain coefficients at the knots in each direction results in assumed strains with $C^0$ continuity across element boundaries. Thus, the assumed covariant membrane strain coefficients of a CAS element are defined as follows
\begin{equation} 
\epsilon^{\text{CAS}}_{\alpha\beta} (\bm{\theta}) = \sum_{l=1}^{4} L_{l} (\bm{\theta}) \epsilon^h_{\alpha\beta} (\bm{\theta}^{e}_{l})   \label{epsilonCAS} \text{,} 
\end{equation}
with
\begin{align}
& \bm{\theta}^{e}_{1} = (\theta^{1e}_{1}, \theta^{2e}_{1}), \; \; \; \bm{\theta}^{e}_{2} = (\theta^{1e}_{2}, \theta^{2e}_{1}), \; \; \; \bm{\theta}^{e}_{3} = (\theta^{1e}_{1}, \theta^{2e}_{2}), \; \; \; \bm{\theta}^{e}_{4} = (\theta^{1e}_{2}, \theta^{2e}_{2}), & \\
& L_{1} (\bm{\theta})  =  \frac{\theta^{1e}_{2} - \theta^1}{\theta^{1e}_{2} - \theta^{1e}_{1}} \frac{\theta^{2e}_{2} - \theta^2}{\theta^{2e}_{2} - \theta^{2e}_{1}} \text{,} \; \; \;
L_{2} (\bm{\theta})  =  \frac{\theta^1 - \theta^{1e}_{1}}{\theta^{1e}_{2} - \theta^{1e}_{1}} \frac{\theta^{2e}_{2} - \theta^2}{\theta^{2e}_{2} - \theta^{2e}_{1}} \text{,} & \nonumber \\
& L_{3} (\bm{\theta}) =  \frac{\theta^{1e}_{2} - \theta^1}{\theta^{1e}_{2} - \theta^{1e}_{1}} \frac{\theta^2 - \theta^{2e}_{1}}{\theta^{2e}_{2} - \theta^{2e}_{1}} \text{,} \; \; \;
L_{4} (\bm{\theta}) =  \frac{\theta^1 - \theta^{1e}_{1}}{\theta^{1e}_{2} - \theta^{1e}_{1}} \frac{\theta^2 - \theta^{2e}_{1}}{\theta^{2e}_{2} - \theta^{2e}_{1}} \text{,}  & \label{epsilonCAS5}
\end{align}
where $\theta^{1e}_{1}$ and $\theta^{1e}_{2}$ are the parametric coordinates of the two knots that define element $e$ in the parametric direction given by $\theta^{1}$, $\theta^{2e}_{1}$ and $\theta^{2e}_{2}$ are the parametric coordinates of the two knots that define element $e$ in the parametric direction given by $\theta^{2}$, and $L_{l}$ is a bilinear Lagrange polynomial.

Using the assumed strains proposed in Eq. \eqref{epsilonCAS}, the element stiffness matrix of CAS elements is obtained as follows
\begin{equation}
\mathbf{k}^{\text{CAS}}= \mathbf{k}^{\text{CAS}}_\epsilon  + \mathbf{k}_\kappa \text{,}
\end{equation}
\begin{equation} 
 \mathbf{k}^{\text{CAS}}_{\epsilon} =  \left[ k^{\text{CAS}}_{\epsilon,ibjc} \right], \quad \mathbf{k}_{\kappa} =  \left[ k_{\kappa,ibjc} \right] \text{,} 
\end{equation}
where $\mathbf{k}^{\text{CAS}}$ is the element stiffness matrix of CAS elements, $\mathbf{k}^{\text{CAS}}_{\epsilon}$ is the element membrane stiffness matrix of CAS elements, and $\mathbf{k}_{\kappa}$ is the same element bending stiffness matrix as that of CS elements. The explicit expression of $k^{\text{CAS}}_{\epsilon,ibjc}$ is given by
\begin{align} 
 & k^{\text{CAS}}_{\epsilon,ibjc} = \nonumber \\ 
 & + \frac{Et}{4(1+\nu)} \sum_{l=1}^{4} \sum_{m=1}^{4}  \int_{A^e} L_{l} (\bm{\theta}) N_{b,\alpha} (\bm{\theta}^{e}_{l})   \bm{e}_i \cdot \bm{a}_{\beta} (\bm{\theta}^{e}_{l})   a^{\alpha \lambda} (\bm{\theta})   a^{\mu \beta} (\bm{\theta}) L_{m} (\bm{\theta}) N_{c,\lambda}  (\bm{\theta}^{e}_{m})   \bm{e}_j \cdot \bm{a}_{\mu}  (\bm{\theta}^{e}_{m})  \, \mathrm dA  \nonumber \\ 
 &  + \frac{Et}{4(1+\nu)} \sum_{l=1}^{4} \sum_{m=1}^{4}  \int_{A^e} L_{l} (\bm{\theta}) N_{b,\alpha} (\bm{\theta}^{e}_{l})   \bm{e}_i \cdot \bm{a}_{\beta} (\bm{\theta}^{e}_{l})   a^{\alpha \lambda} (\bm{\theta})   a^{\mu \beta} (\bm{\theta}) L_{m} (\bm{\theta}) N_{c,\mu}  (\bm{\theta}^{e}_{m})   \bm{e}_j \cdot \bm{a}_{\lambda}  (\bm{\theta}^{e}_{m})  \, \mathrm dA  \nonumber \\ 
  & + \frac{Et}{4(1+\nu)} \sum_{l=1}^{4} \sum_{m=1}^{4}  \int_{A^e} L_{l} (\bm{\theta}) N_{b,\beta} (\bm{\theta}^{e}_{l})   \bm{e}_i \cdot \bm{a}_{\alpha} (\bm{\theta}^{e}_{l})   a^{\alpha \lambda} (\bm{\theta})   a^{\mu \beta} (\bm{\theta}) L_{m} (\bm{\theta}) N_{c,\lambda}  (\bm{\theta}^{e}_{m})   \bm{e}_j \cdot \bm{a}_{\mu}  (\bm{\theta}^{e}_{m})  \, \mathrm dA  \nonumber \\ 
 &  + \frac{Et}{4(1+\nu)} \sum_{l=1}^{4} \sum_{m=1}^{4}  \int_{A^e} L_{l} (\bm{\theta}) N_{b,\beta} (\bm{\theta}^{e}_{l})   \bm{e}_i \cdot \bm{a}_{\alpha} (\bm{\theta}^{e}_{l})   a^{\alpha \lambda} (\bm{\theta})   a^{\mu \beta} (\bm{\theta}) L_{m} (\bm{\theta}) N_{c,\mu}  (\bm{\theta}^{e}_{m})   \bm{e}_j \cdot \bm{a}_{\lambda}  (\bm{\theta}^{e}_{m})  \, \mathrm dA  \nonumber \\ 
 & + \frac{Et \nu}{4(1-\nu^2)} \sum_{l=1}^{4} \sum_{m=1}^{4}  \int_{A^e} L_{l} (\bm{\theta}) N_{b,\alpha} (\bm{\theta}^{e}_{l})   \bm{e}_i \cdot \bm{a}_{\beta} (\bm{\theta}^{e}_{l})   a^{\alpha \beta} (\bm{\theta})   a^{\mu \lambda} (\bm{\theta}) L_{m} (\bm{\theta}) N_{c,\lambda}  (\bm{\theta}^{e}_{m})   \bm{e}_j \cdot \bm{a}_{\mu}  (\bm{\theta}^{e}_{m})  \, \mathrm dA  \nonumber \\ 
 &  + \frac{Et \nu}{4(1-\nu^2)} \sum_{l=1}^{4} \sum_{m=1}^{4}  \int_{A^e} L_{l} (\bm{\theta}) N_{b,\alpha} (\bm{\theta}^{e}_{l})   \bm{e}_i \cdot \bm{a}_{\beta} (\bm{\theta}^{e}_{l})   a^{\alpha \beta} (\bm{\theta})   a^{\mu \lambda} (\bm{\theta}) L_{m} (\bm{\theta}) N_{c,\mu}  (\bm{\theta}^{e}_{m})   \bm{e}_j \cdot \bm{a}_{\lambda}  (\bm{\theta}^{e}_{m})  \, \mathrm dA  \nonumber \\ 
  & + \frac{Et \nu}{4(1-\nu^2)} \sum_{l=1}^{4} \sum_{m=1}^{4}  \int_{A^e} L_{l} (\bm{\theta}) N_{b,\beta} (\bm{\theta}^{e}_{l})   \bm{e}_i \cdot \bm{a}_{\alpha} (\bm{\theta}^{e}_{l})   a^{\alpha \beta} (\bm{\theta})   a^{\mu \lambda} (\bm{\theta}) L_{m} (\bm{\theta}) N_{c,\lambda}  (\bm{\theta}^{e}_{m})   \bm{e}_j \cdot \bm{a}_{\mu}  (\bm{\theta}^{e}_{m})  \, \mathrm dA  \nonumber \\ 
 &  + \frac{Et \nu}{4(1-\nu^2)} \sum_{l=1}^{4} \sum_{m=1}^{4}  \int_{A^e} L_{l} (\bm{\theta}) N_{b,\beta} (\bm{\theta}^{e}_{l})   \bm{e}_i \cdot \bm{a}_{\alpha} (\bm{\theta}^{e}_{l})   a^{\alpha \beta} (\bm{\theta})   a^{\mu \lambda} (\bm{\theta}) L_{m} (\bm{\theta}) N_{c,\mu}  (\bm{\theta}^{e}_{m})   \bm{e}_j \cdot \bm{a}_{\lambda}  (\bm{\theta}^{e}_{m})  \, \mathrm dA  \text{.} 
\end{align}
For CAS elements, the computation of the integrals to obtain each element stiffness matrix and the assembly of each element stiffness matrix into the global stiffness matrix follows the same steps as those summarized for CS elements in the last paragraph of the preceding section. Note that CAS elements do not require any additional global or element matrix operations such as matrix inversions or matrix multiplications. The locking treatment is applied at the element level and the nonzero pattern of the global stiffness matrix is preserved.

\section{Numerical experiments}

In this section, we perform numerical investigations using the discretizations
described in Sections 3 and 4. Unless mentioned otherwise, $3 \times 3$ Gauss-Legendre quadrature points per element are used to compute all the integrals. The code used to perform these simulations has been
developed on top of the PetIGA framework \cite{dalcin2016petiga}, which adds NURBS discretization
capabilities and integration of forms to the scientific library PETSc \cite{petsc-web-page}.

\subsection{Cylindrical shell strip}

\begin{figure} [t!] 
 \centering
 \includegraphics[scale=0.3]{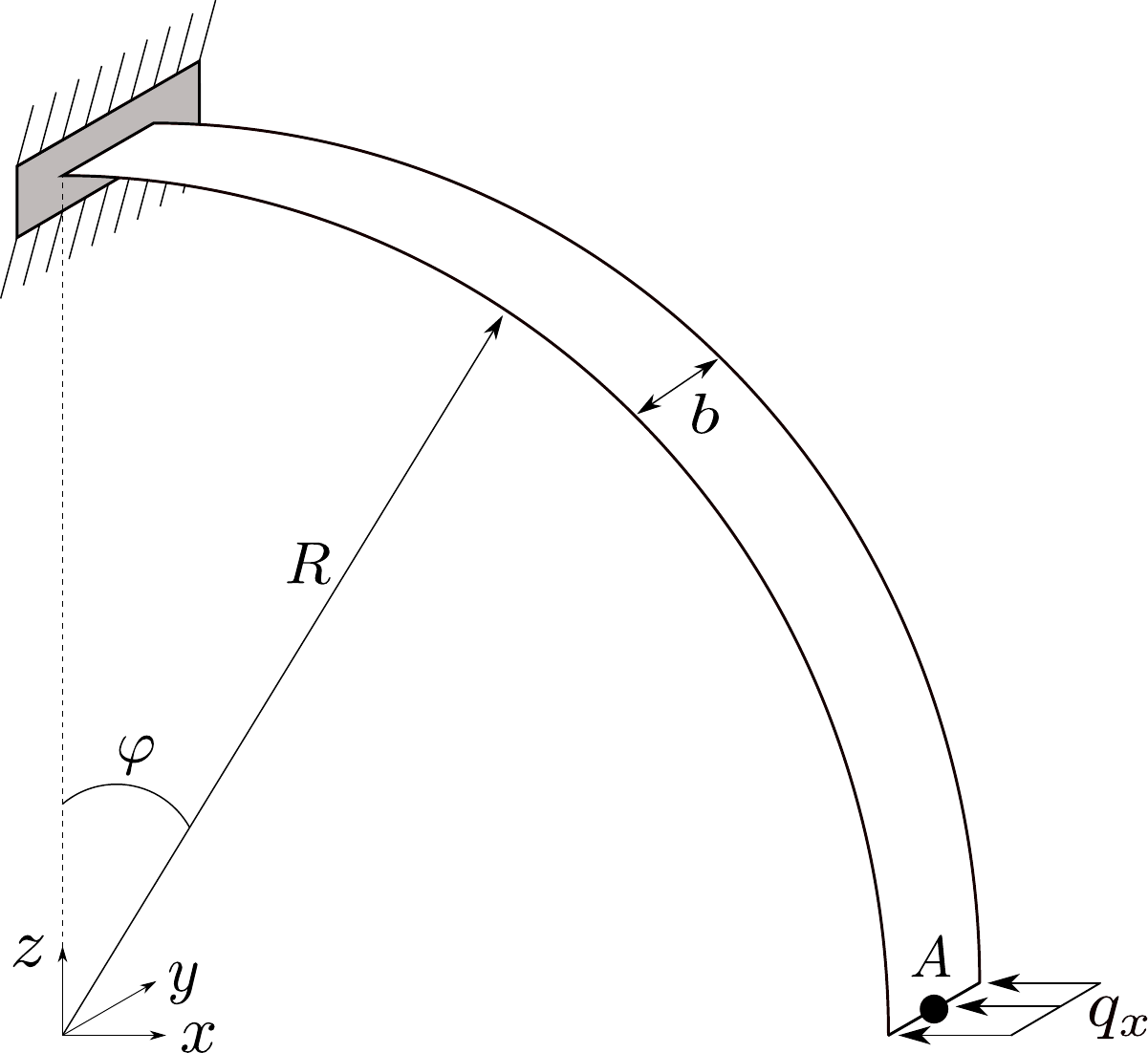}
\caption{Geometry, boundary conditions, and applied load for the cylindrical shell strip.} 
\label{beam2rotgeo}
\end{figure}

\begin{figure} [t!] 
 \centering
 \subfigure[Deflection]{\includegraphics[scale=0.55]{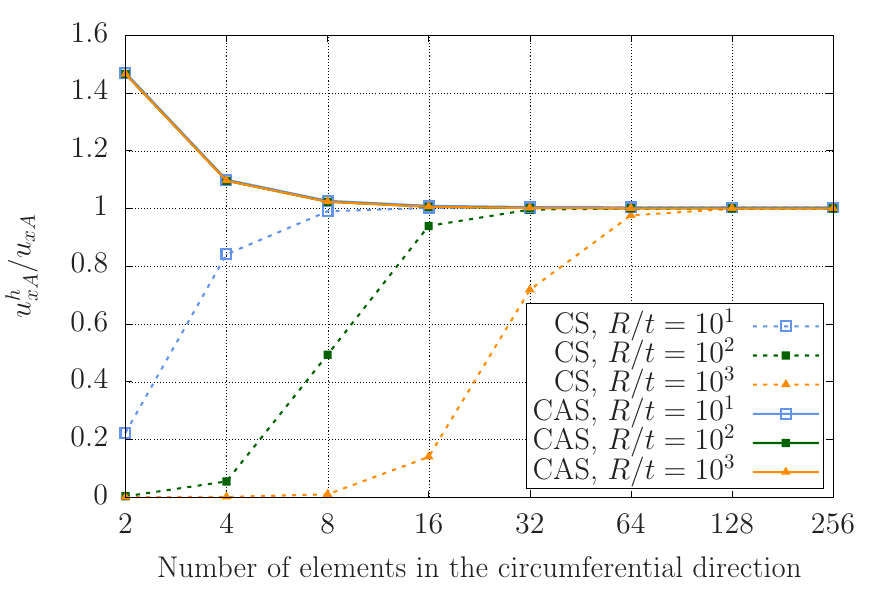}}
 \subfigure[Membrane force in the circumferential direction]{\includegraphics[scale=0.55]{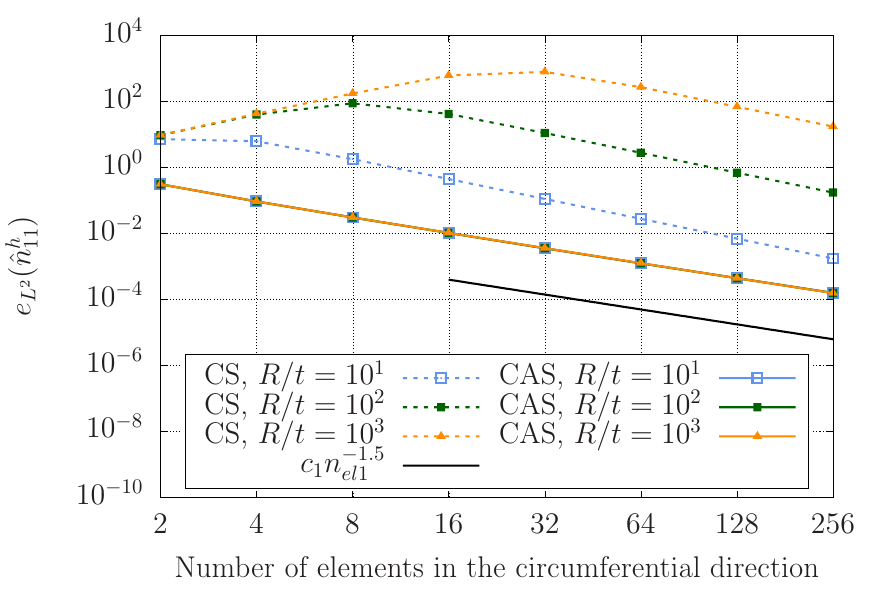}} \\
 \subfigure[Bending moment in the circumferential direction]{\includegraphics[scale=0.55]{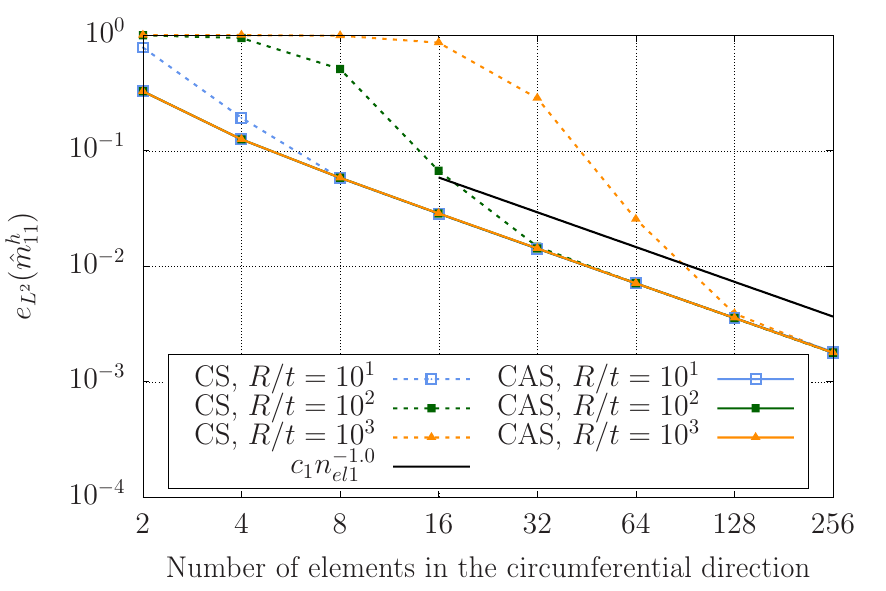}} 
  \subfigure[Membrane strain energy]{\includegraphics[scale=0.55]{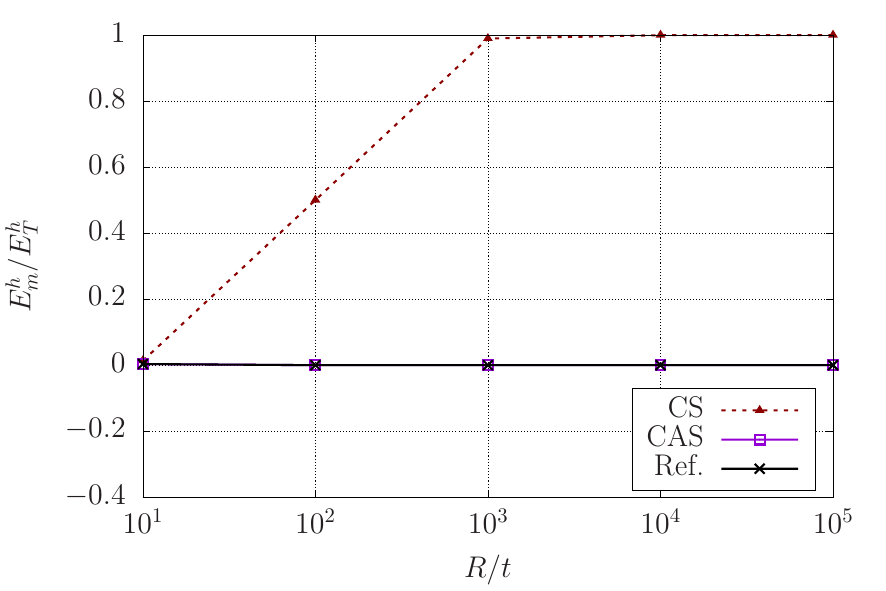}} \\
\caption{(Color online) Cylindrical shell strip. (a) Convergence of the deflection for different slenderness ratios using CS and CAS elements.  (b) Convergence in $L^2$ norm of the membrane force in the circumferential direction for different slenderness ratios using CS and CAS elements. (c) Convergence in $L^2$ norm of the bending moment in the circumferential direction for different slenderness ratios using CS and CAS elements. (d) Ratio of the membrane strain energy to the total strain energy for different slenderness ratios using 8 CS elements and 8 CAS elements in the circumferential direction.}
\label{beam2rotplots}
\end{figure}

The first numerical investigation considers a cylindrical shell strip clamped at one end and free at the other end. A line load, i.e., a load per unit length is applied at the free end in the radial direction. The geometry, the boundary conditions, and the applied load are shown in Fig. \ref{beam2rotgeo}. The next values are used in this example:
\begin{equation} 
 q_x =  -0.1 t^3,  \quad R = 10.0,  \quad  b = 1.0,  \quad E = 1.0 \times 10^{3},  \quad \nu = 0.0 \text{.} 
\end{equation}
In order to consider different values of the slenderness ratio, three values are used for the thickness in this example, namely, $t=1.0$, $t=0.1$, and $t=0.01$. For $t=1.0$, $R/t = 10$ which is considered to be too thick to use Kirchhoff-Love theory according to  \cite{bischoff2004models}. We consider this slenderness ratio in this example to show that even a quite thick shell can suffer from membrane locking. Despite its apparent simplicity, this benchmark problem is challenging to solve numerically since it is a bending-dominated problem. For this problem, we can obtain the exact solution of the effective membrane force in the circumferential direction and the bending moment in the circumferential direction by enforcing static equilibrium on the shell. Specifically, we obtain $\hat{n}^{11}_{\text{eff}} = q_x \cos \varphi $ and $\hat{m}^{11} =  -q_x R \cos \varphi $, where the angle $\varphi$ is shown in Fig. \ref{beam2rotgeo}. The exact solution of $\hat{n}^{11}$ can be obtained combining the preceding two exact solutions. Thus, we can use these exact solutions to study the convergence in $L^2$ norm of the membrane force in the circumferential direction and the bending moment in the circumferential direction. In order to do so, we define the relative errors in $L^2$ norm of the membrane force in the circumferential direction and the bending moment in the circumferential direction as
\begin{align}
   e_{L^2}(\hat{n}^h_{11}) &=  \frac{ \sqrt{\int_{A} \left( \hat{n}^h_{11} - \hat{n}_{11} \right)^2 \, \mathrm dA }}{ \sqrt{ \int_{A}  \hat{n}_{11}^2 \, \mathrm dA} }   \text{,} \\
   e_{L^2}(\hat{m}^h_{11}) &=  \frac{ \sqrt{\int_{A} \left( \hat{m}^h_{11} - \hat{m}_{11} \right)^2 \, \mathrm dA }}{ \sqrt{ \int_{A}  \hat{m}_{11}^2 \, \mathrm dA} } \text{,}
\end{align}
respectively. Since we are solving fourth-order partial differential equations with quadratic basis functions, the optimal asymptotic convergence rates of $e_{L^2}(\hat{n}^h_{11})$ and $e_{L^2}(\hat{m}^h_{11})$ are 2 and 1, respectively \cite{Hughes2012}.

\begin{figure} [t!] 
 \centering
 \subfigure[CS elements]{\includegraphics[scale=0.55]{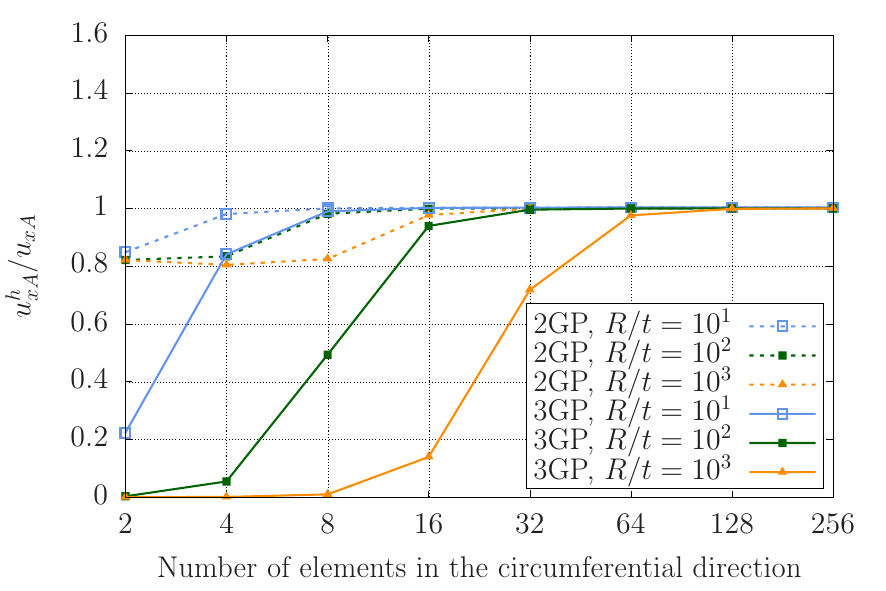}}
 \subfigure[CAS elements]{\includegraphics[scale=0.55]{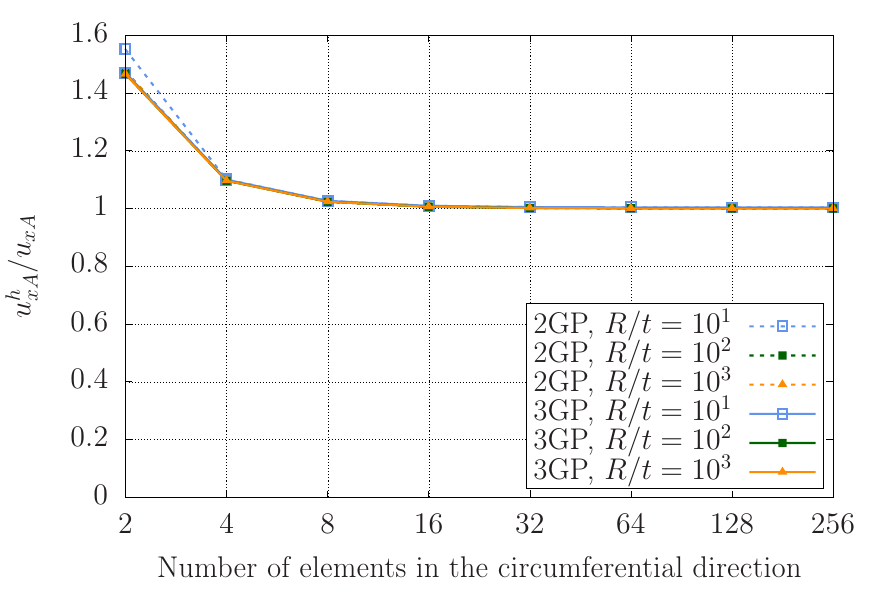}} \\
 \subfigure[CS elements]{\includegraphics[scale=0.55]{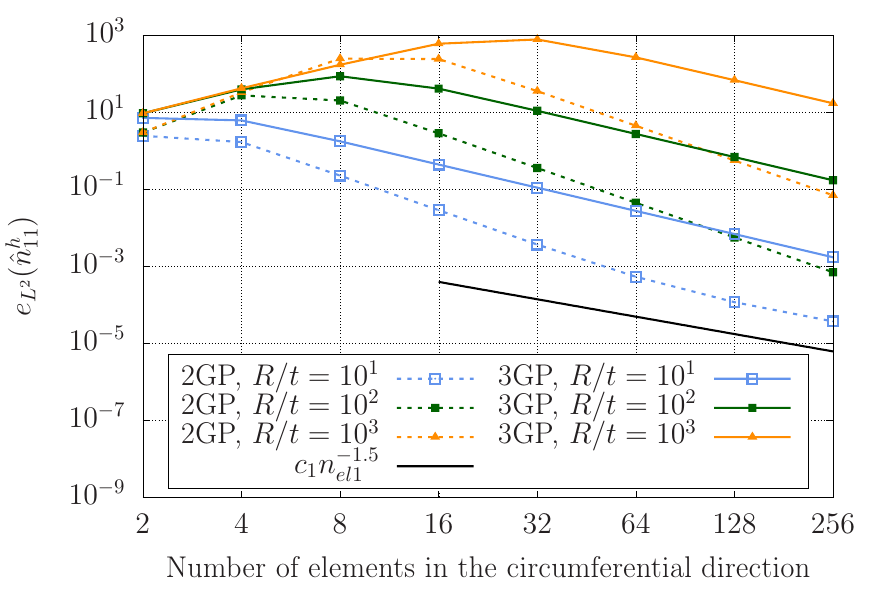}}
 \subfigure[CAS elements]{\includegraphics[scale=0.55]{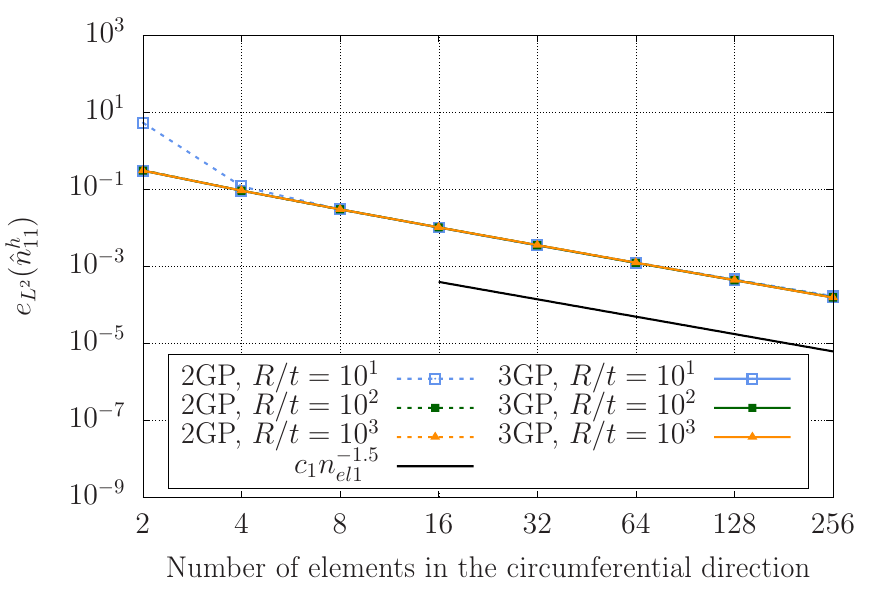}} \\
 \subfigure[CS elements]{\includegraphics[scale=0.55]{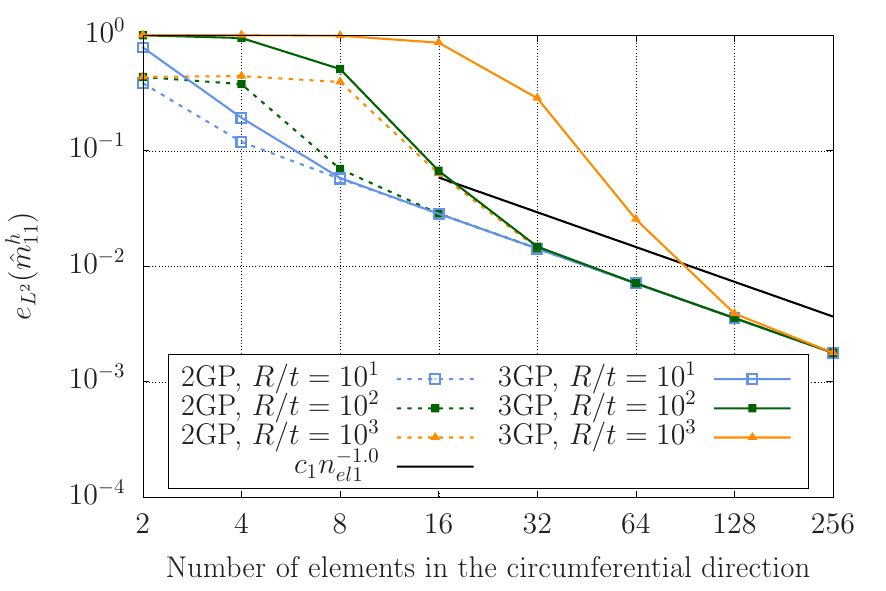}}
 \subfigure[CAS elements]{\includegraphics[scale=0.55]{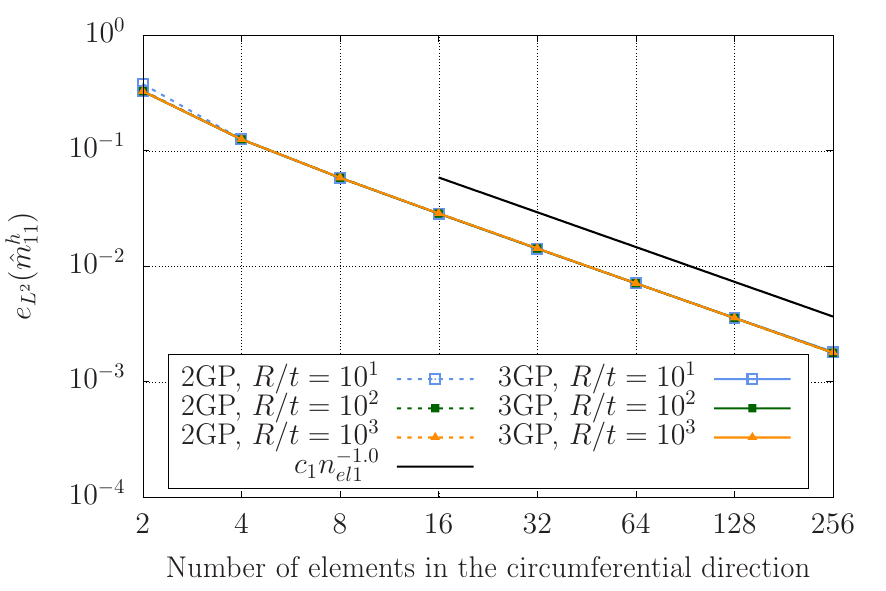}} \\
\caption{(Color online) Cylindrical shell strip. (a)-(b) Convergence of the deflection for different slenderness ratios using CS and CAS elements with $2\times2$ and $3\times3$ Gauss-Legendre quadrature points. (c)-(d) Convergence in $L^2$ norm of the membrane force in the circumferential direction for different slenderness ratios using CS and CAS elements with $2\times2$ and $3\times3$ Gauss-Legendre quadrature points. (e)-(f) Convergence in $L^2$ norm of the bending moment in the circumferential direction for different slenderness ratios using CS and CAS elements with $2\times2$ and $3\times3$ Gauss-Legendre quadrature points.}
\label{beam2rot2GP3GPplots}
\end{figure}


We initiate our convergence study with a uniform mesh composed of two quadratic elements in the circumferential direction and one element in the other direction. The midsurface of the shell is represented exactly since we are using quadratic NURBS. After that, we perform uniform $h$-refinement seven times in the circumferential direction (i.e., we keep one element in the other direction throughout all the refinement levels). Using CS elements and CAS elements, Fig. \ref{beam2rotplots} a) plots the convergence of the radial displacement of the point $A$ indicated in Fig. \ref{beam2rotgeo}. The displacement values are normalized by the solution obtained using 256 CS elements of degree 9 in the circumferential direction, which is used as a reference solution. The reference values of the radial displacement at point A are $-9.4561 \times 10^{-1}$, $-9.4250 \times 10^{-1}$, and $-9.4247 \times 10^{-1}$ for $R/t =10$ , $10^2$, and $10^3$, respectively. Fig. \ref{beam2rotplots} b) and c) plot the convergence in $L^2$ norm of the membrane force in the circumferential direction and the bending moment in the circumferential direction, respectively. $n_{el1}$ represents the number of elements in the circumferential direction. As shown in Fig. \ref{beam2rotplots} a), b), and c), the convergence of CAS elements is independent of the slenderness ratio for the broad range of $R/t$ values considered while the convergence of CS elements heavily deteriorates as the slenderness ratio increases. Fig. \ref{beam2rotplots} b) reveals an anomalous behavior in the convergence of the membrane force in the circumferential direction using CS elements, namely, the relative error in $L^2$ norm of the membrane force in the circumferential direction increases as uniform $h$-refinement is performed multiple times (note that for many mesh resolutions and slenderness ratios the relative error of the membrane force is greater than $100\%$). This anomalous behavior, caused by membrane locking, has also been reported in \cite{greco2013b, zou2021galerkin, casquero2022removing, casquero2023trods}. For both coarse and fine meshes, the relative error in $L^2$ norm of the membrane force in the circumferential direction obtained with CAS elements is several orders of magnitude smaller than the relative error in $L^2$ norm of the membrane force in the circumferential direction obtained with CS elements. In \cite{greco2017efficient, casquero2022removing}, the global $\bar{B}$ method, the reconstructed $\bar{B}$ method, the reconstructed ANS method, and CAS elements lead to an asymptotic convergence rate for the $L^2$ norm of the membrane force equal to 1.5 when applied to $C^1$-continuous quadratic NURBS discretizations of linear plane Kirchhoff rods. Thus, CAS elements leading to an asymptotic convergence rate for the $L^2$ norm of the membrane force equal to 1.5 when applied to $C^1$-continuous quadratic NURBS discretizations of linear Kirchhoff-Love shells was the expected outcome. Fig. \ref{beam2rotplots} d) plots the ratio of the membrane strain energy to the total strain energy as the slenderness ratio increases using 8 CS elements and 8 CAS elements in the circumferential direction. The numerical solution using 8 CAS elements in the circumferential direction overlaps with the reference solution, which uses 256 CS elements of degree 9 in the circumferential direction. However, as the slenderness ratio increases, membrane locking causes the introduction of spurious membrane energy in the numerical solution obtained using 8 CS elements in the circumferential direction.

\begin{figure} [t!] 
 \centering
 \subfigure[Reference]{\includegraphics[scale=0.08]{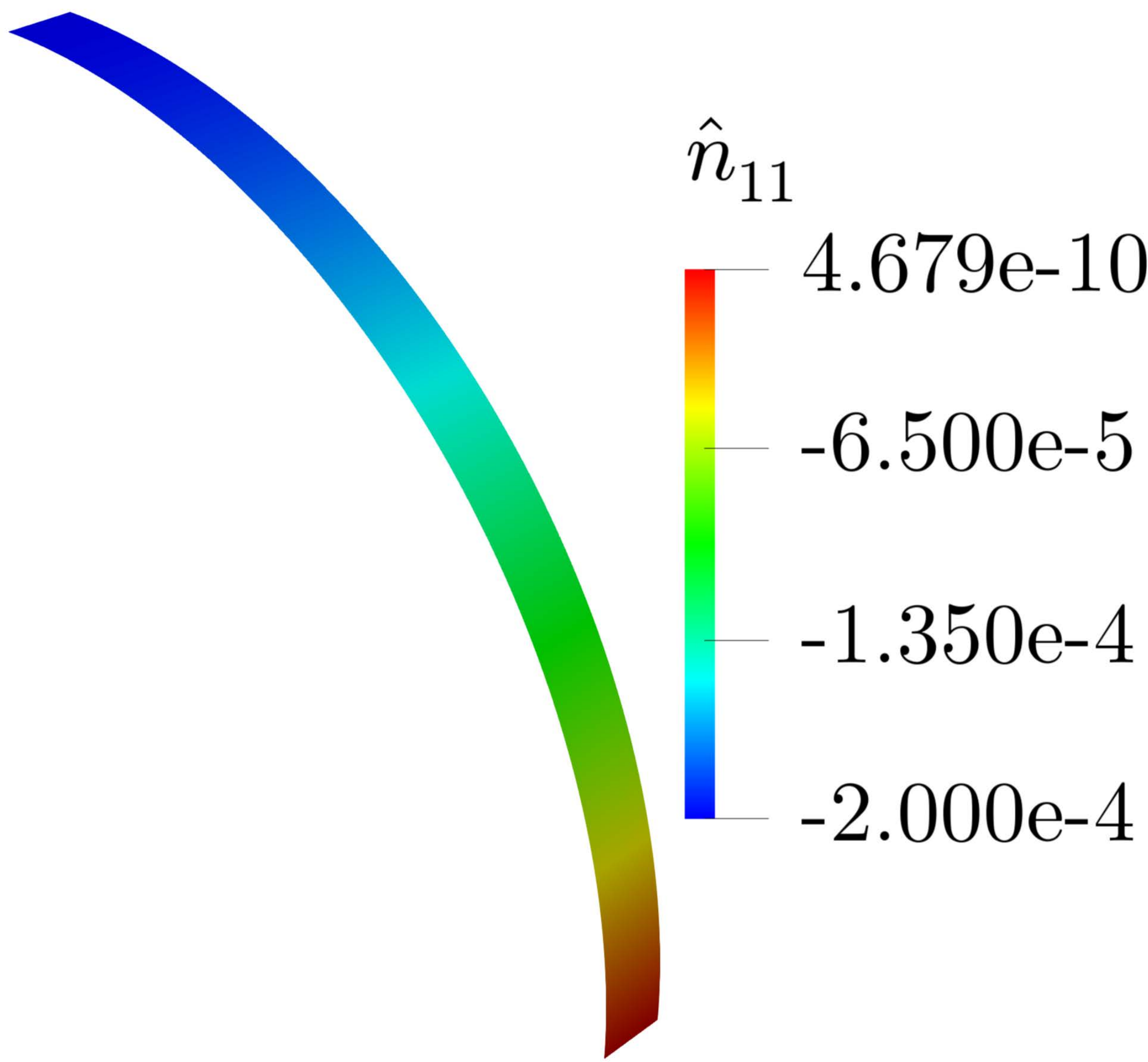}} \hspace*{+5mm}
 \subfigure[8 CAS elements, 3GP]{\includegraphics[scale=0.08]{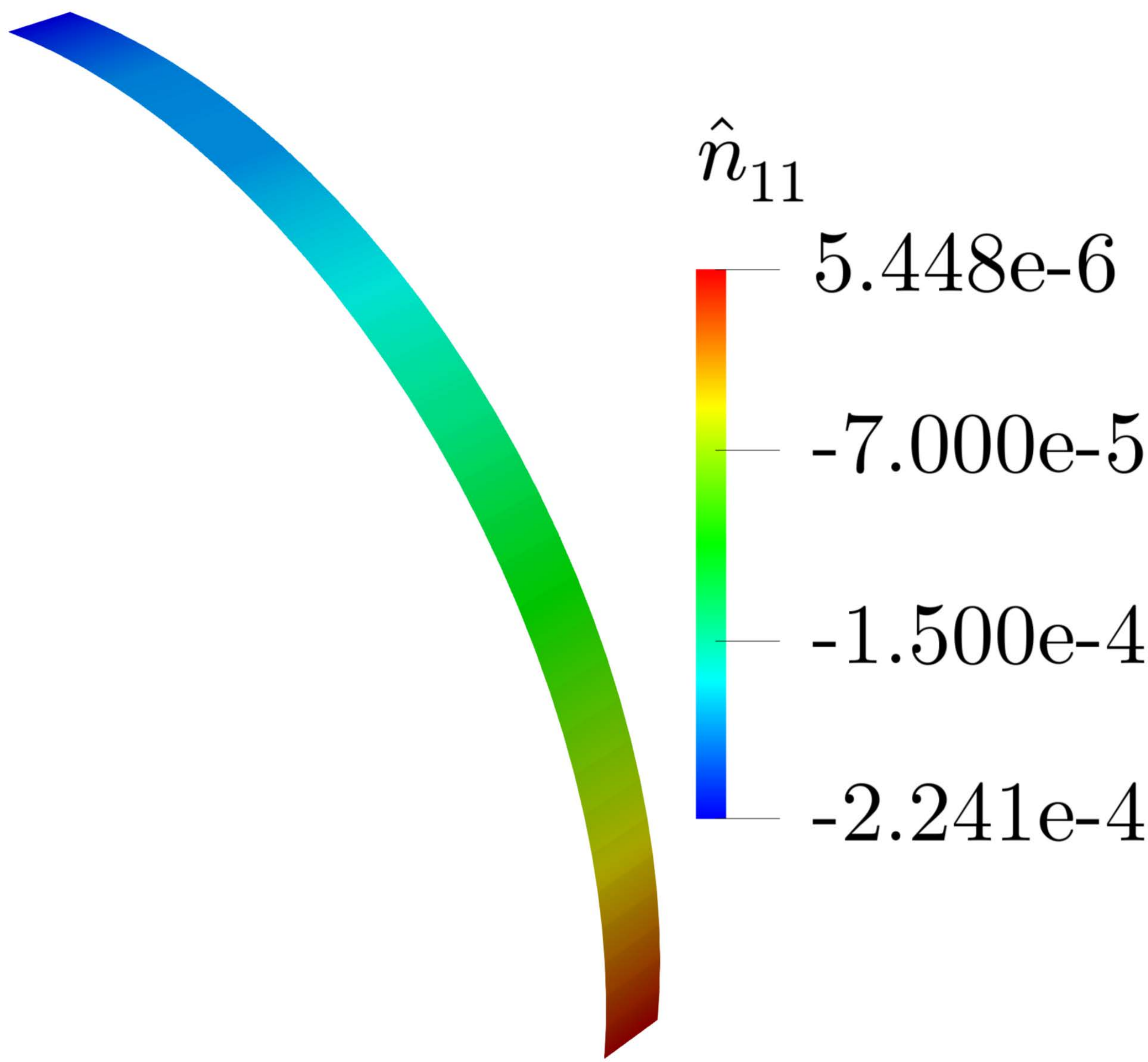}} \hspace*{+5mm}
 \subfigure[8 CAS elements, 2GP]{\includegraphics[scale=0.08]{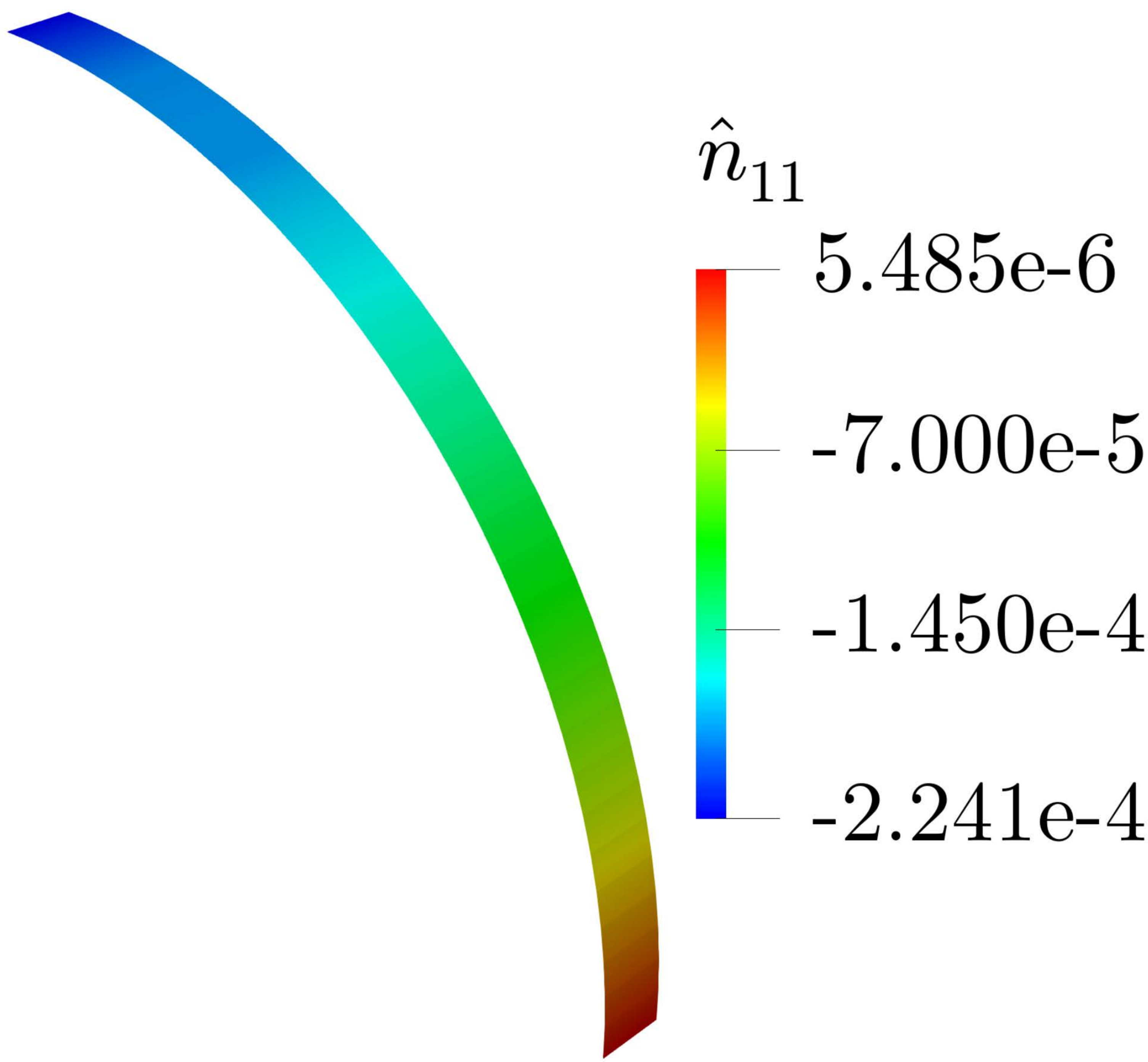}} \\
 \subfigure[8 CS elements, 3GP]{\includegraphics[scale=0.08]{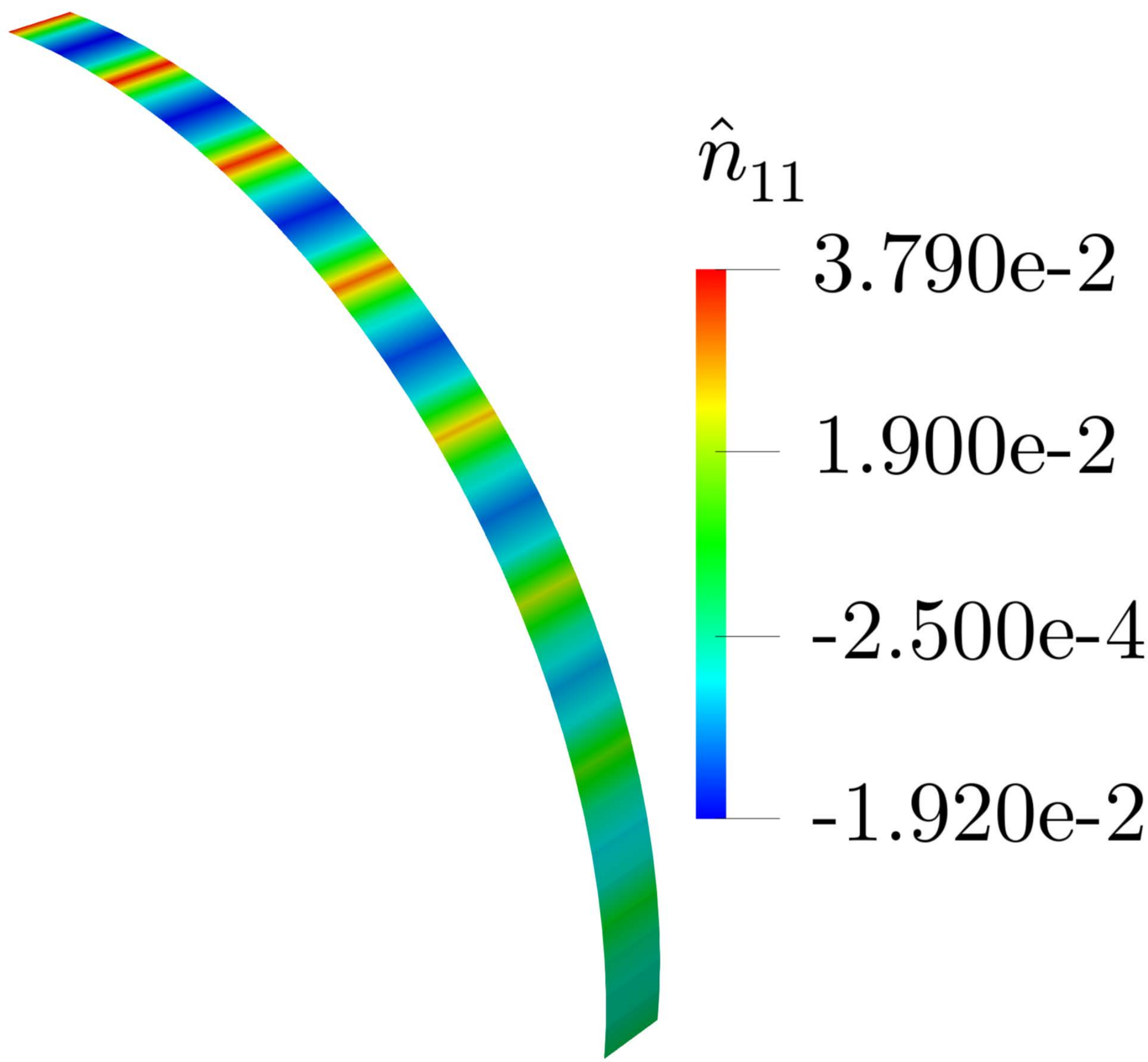}} \hspace*{+5mm}
 \subfigure[16 CS elements, 3GP]{\includegraphics[scale=0.08]{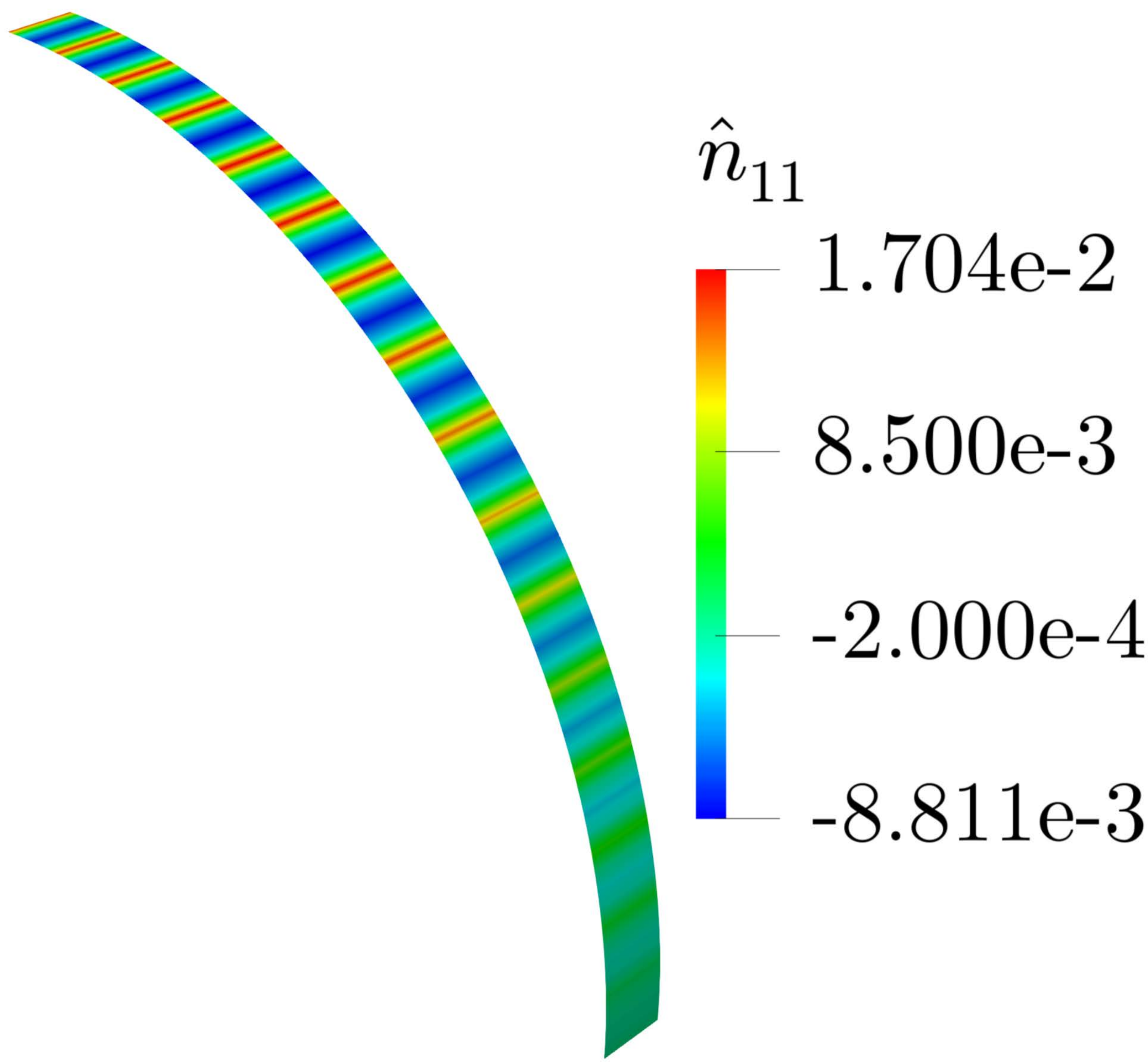}} \hspace*{+5mm}
 \subfigure[32 CS elements, 3GP]{\includegraphics[scale=0.08]{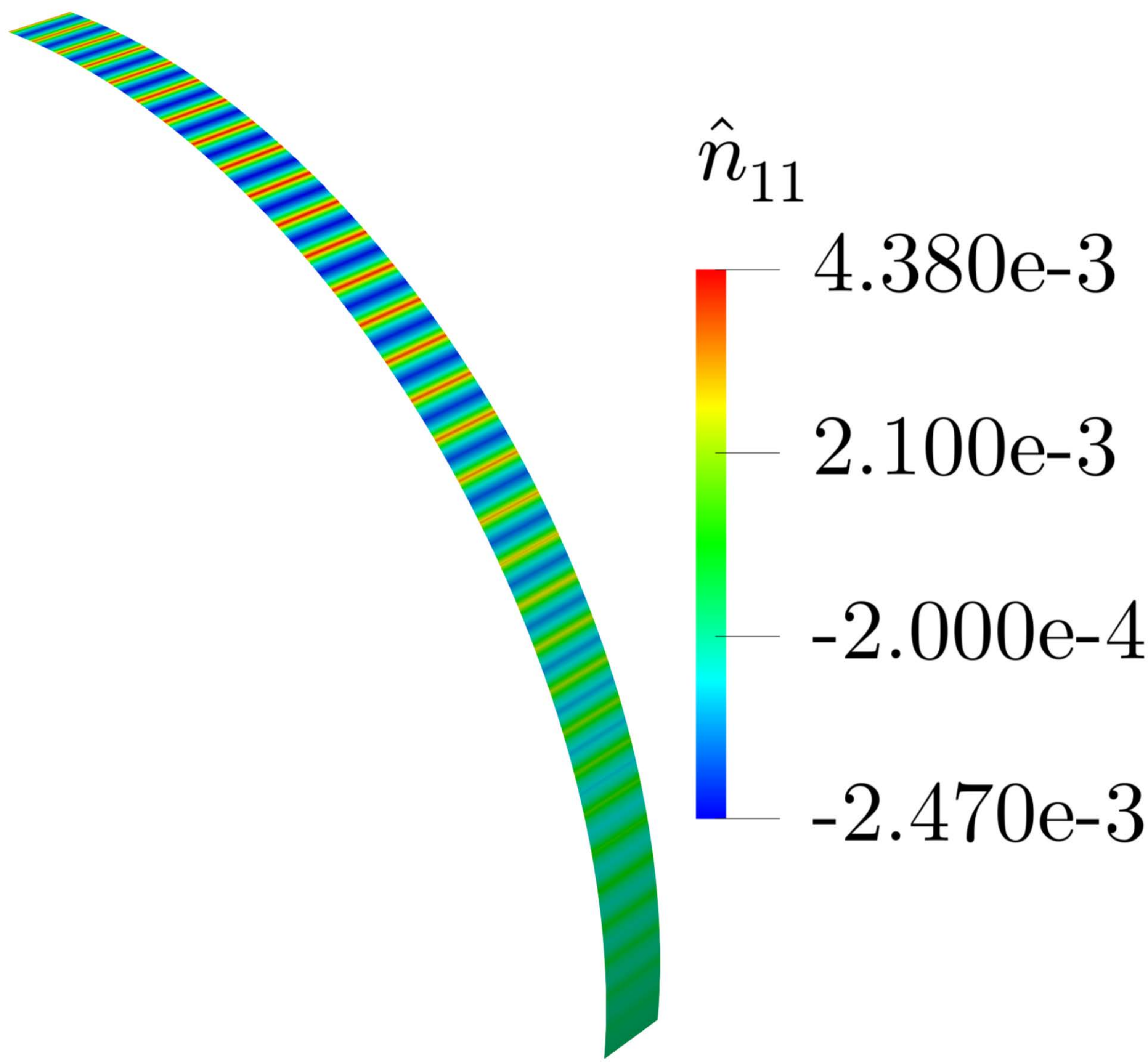}}  \\
 \subfigure[8 CS elements, 2GP]{\includegraphics[scale=0.08]{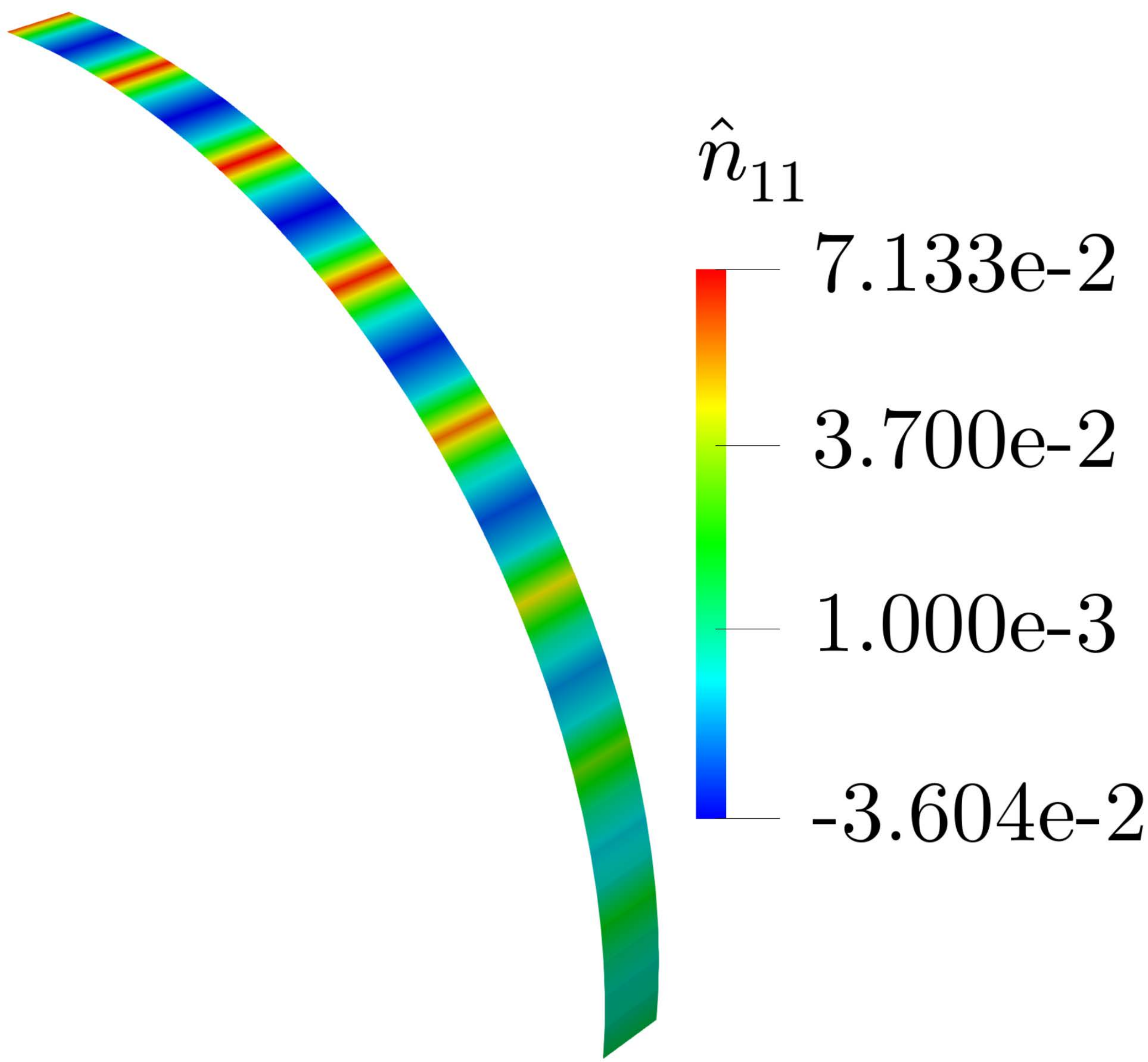}} \hspace*{+5mm}
 \subfigure[16 CS elements, 2GP]{\includegraphics[scale=0.08]{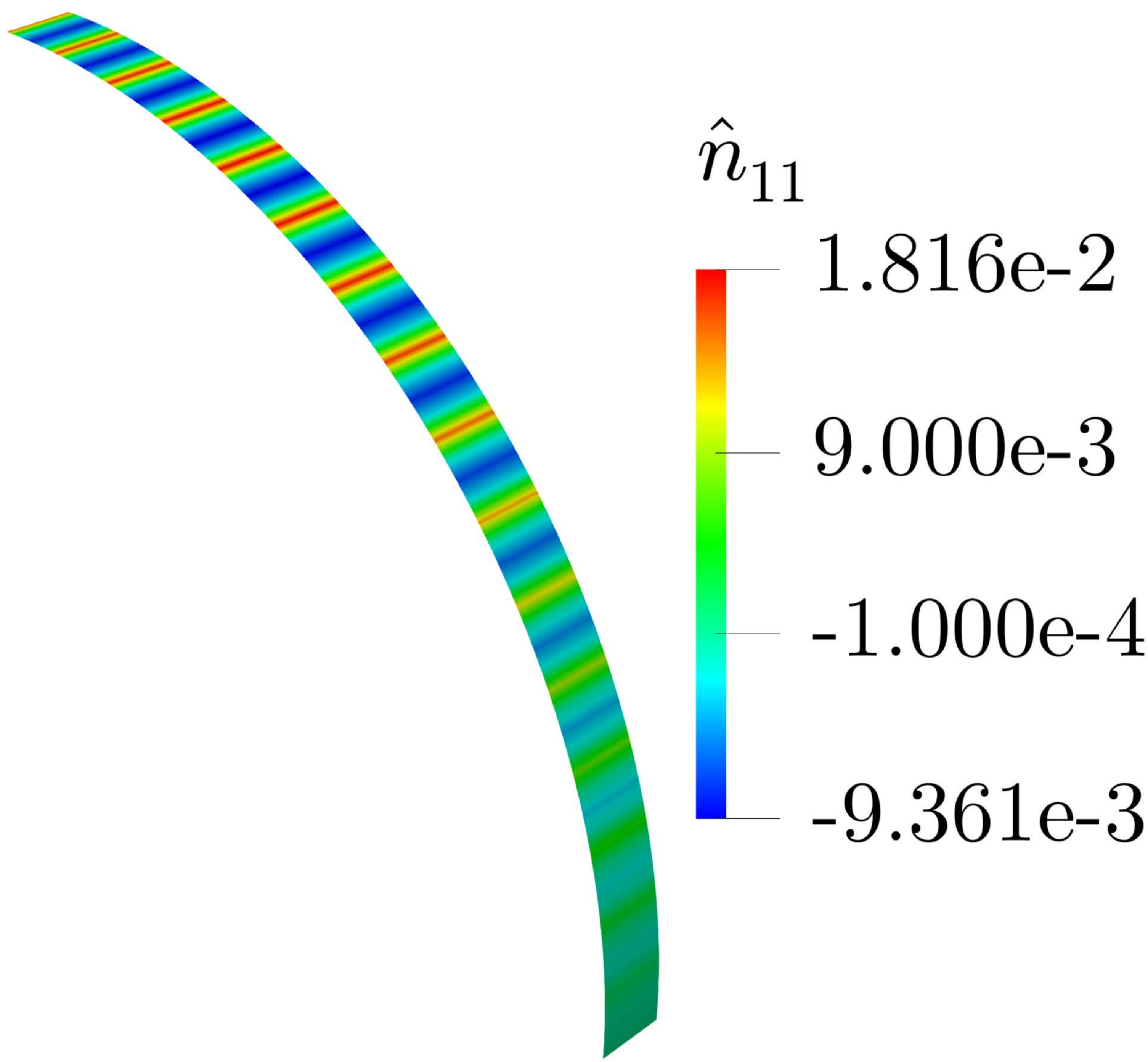}} \hspace*{+5mm}
 \subfigure[32 CS elements, 2GP]{\includegraphics[scale=0.08]{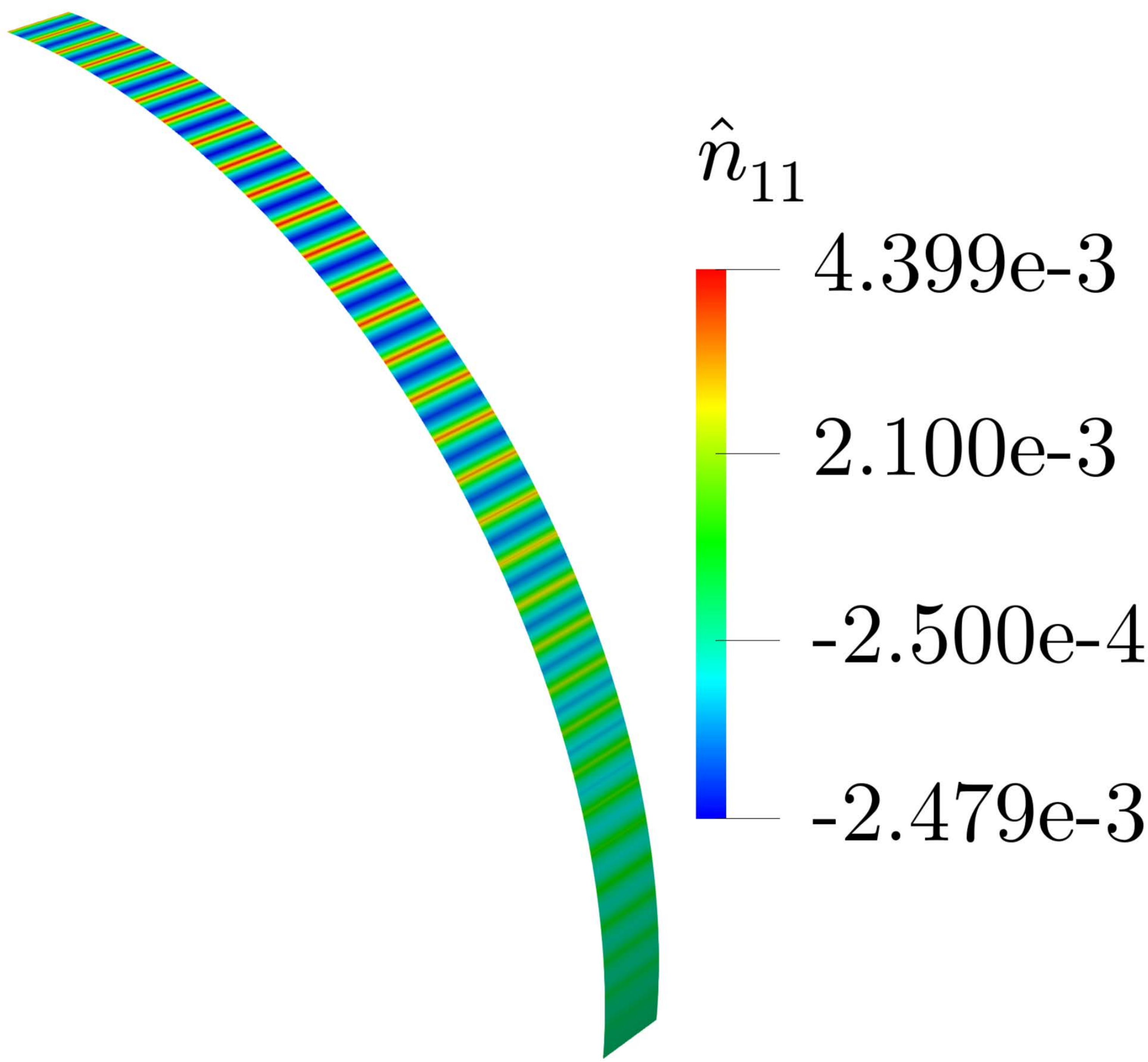}} \\
\caption{(Color online) Membrane force in the circumferential direction for the cylindrical shell strip and $R/t=10^{2}$. (a) 256 CS elements of degree 9 in the circumferential direction are used as reference solution. (b)-(c) 8 CAS elements in the circumferential direction using either $3\times3$ or $2\times2$  Gauss-Legendre quadrature points are free of spurious oscillations. (d)-(i) 8, 16, and 32 CS elements in the circumferential direction using either $3\times3$ or $2\times2$  Gauss-Legendre quadrature points have spurious oscillations. Note the different scales used in each plot.}
\label{beam2rotmf}
\end{figure}

As shown in \cite{leonetti2019simplified}, when using reduced patch-wise integration rules, the continuity of the chosen integration space must not be greater than the continuity of the strains so as to reproduce constant stress states. The bending strains are discontinuous across element boundaries when discretizing Kirchhoff-Love shells with $C^1$-continuous quadratic NURBS. Therefore, the continuity of the integration space must be discontinuous across element boundaries to exactly reproduce constant stress states, which is needed to obtain accurate results. In this case, the integration is no longer patch-wise, but element-wise instead and coincides with the Gauss-Legendre quadrature rules. Thus, using both CS and CAS elements, we now solve this problem using $2\times2$ Gauss-Legendre quadrature points per element (2GP) as opposed to using $3\times3$ Gauss-Legendre quadrature points per element (3GP) to compute all the integrals. As shown in Fig. \ref{beam2rot2GP3GPplots}, CAS elements result in essentially the same accuracy regardless of whether 2 quadrature points per direction or 3 quadrature points per direction are used. Therefore, 2 quadrature points per direction can be used to decrease the computational time. However, CS elements with 2 quadrature points per direction are still a locking-prone discretization.

Fig. \ref{beam2rotmf} plots the distribution of the membrane force in the circumferential direction for the slenderness ratio $R/t = 10^2$. Fig. \ref{beam2rotmf} a) plots the reference solution which is obtained using 256 CS elements of degree 9 in the circumferential direction. Fig. \ref{beam2rotmf} b) and c) plot the numerical solution obtained using 8 CAS elements in the circumferential direction with 3 and 2 quadrature points per direction, respectively. These two numerical solutions are free from spurious oscillations. Fig. \ref{beam2rotmf} d), e), and f) plot the numerical solution obtained using 8, 16, and 32 CS elements in the circumferential direction with 3 quadrature points per direction, respectively. Fig. \ref{beam2rotmf} g), h), and i) plot the numerical solution obtained using 8, 16, and 32 CS elements in the circumferential direction with 2 quadrature points per direction, respectively. These six numerical solutions suffer from large-amplitude spurious oscillations caused by membrane locking. As shown in Fig. \ref{beam2rot2GP3GPplots}, 32 CS elements in the circumferential direction result in accurate displacement values for $R/t = 10^2$, but the membrane force in the circumferential direction undergoes large-amplitude spurious oscillations nonetheless. Thus, whenever investigating whether or not a certain discretization suffers from membrane locking, it is not enough to only study the accuracy of the displacements, the accuracy of the membrane forces must be studied as well. Note that the amplitude of the oscillations obtained with 2 quadrature points per direction is greater than the amplitude of the oscillations obtained with 3 quadrature points despite that the relative error in $L^2$ norm of the membrane force in the circumferential direction obtained with 2 quadrature points is smaller than the relative error in $L^2$ norm of the membrane force in the circumferential direction obtained with 3 quadrature points as shown in Fig. \ref{beam2rot2GP3GPplots} c). 

\begin{figure} [t!] 
 \centering
 \subfigure[Before applying symmetry]{\includegraphics[scale=0.3]{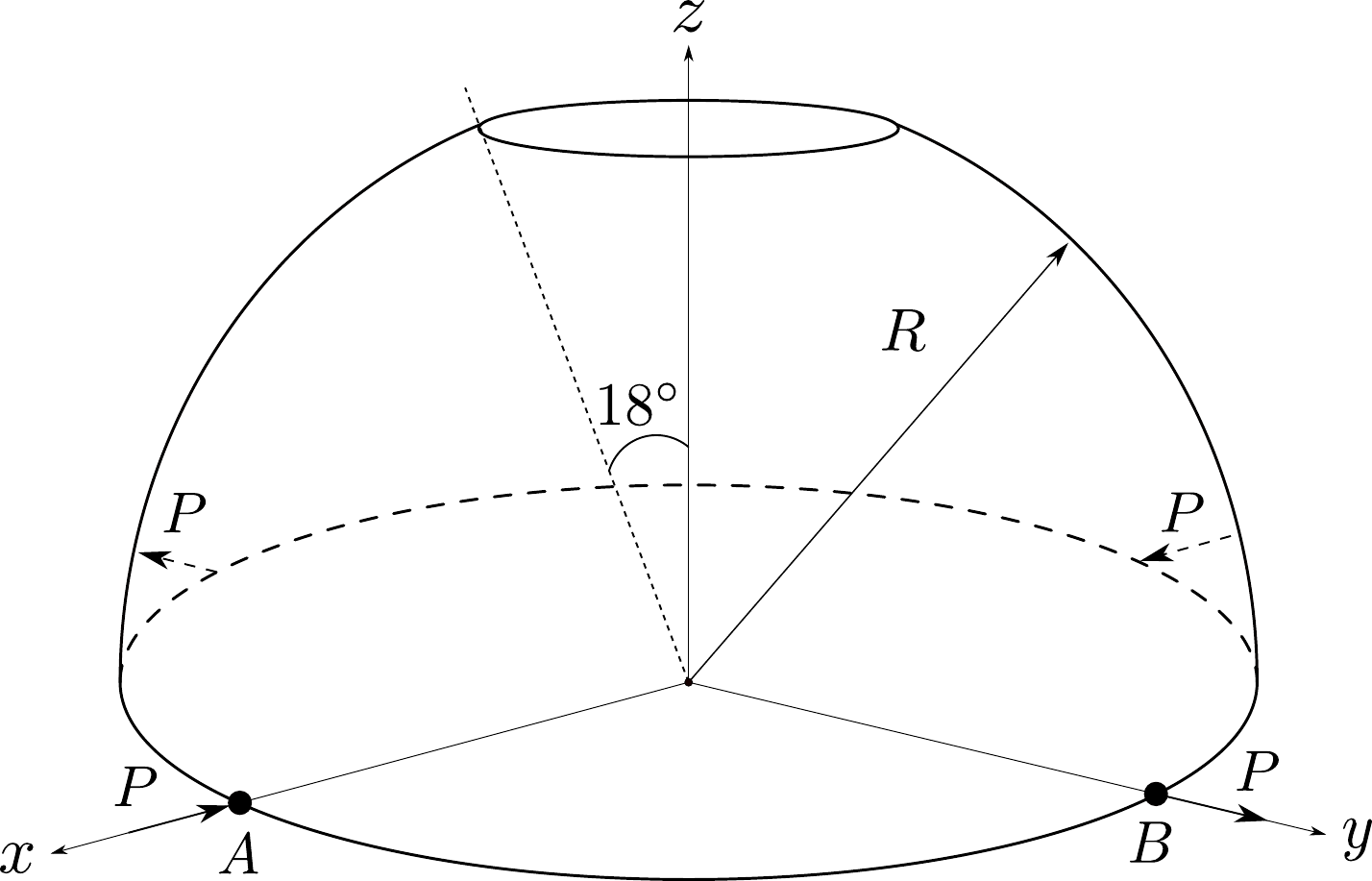}} \hspace*{+10mm}
 \subfigure[After applying symmetry]{\includegraphics[scale=0.3]{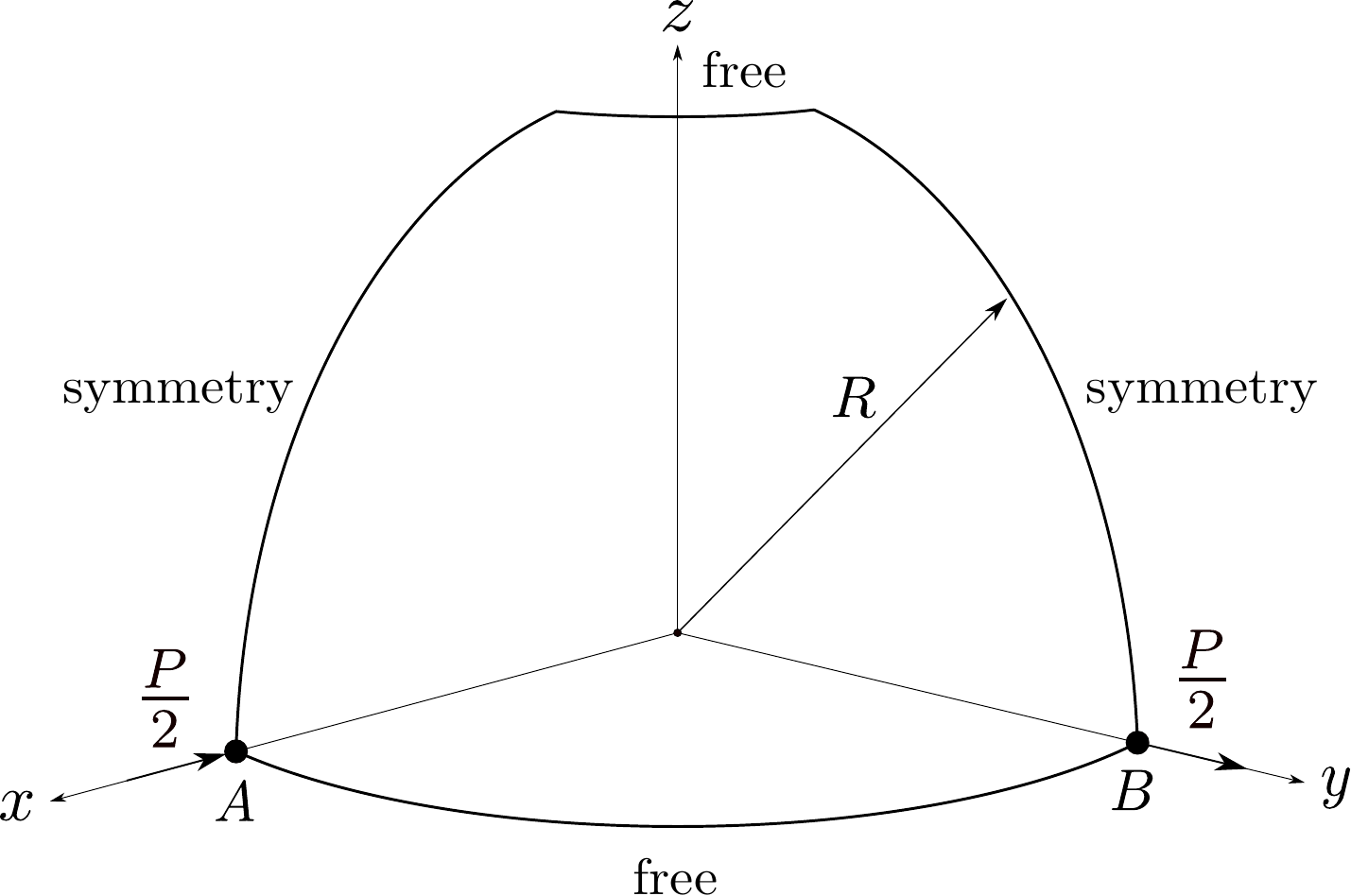}}
\caption{Geometry, boundary conditions, and applied load for the pinched hemisphere with a hole.} 
\label{pinchedhemispheregeo}
\end{figure}

\subsection{Pinched hemisphere with a hole}

\begin{figure} [t!] 
 \centering
 \subfigure[Deflection]{\includegraphics[scale=0.55]{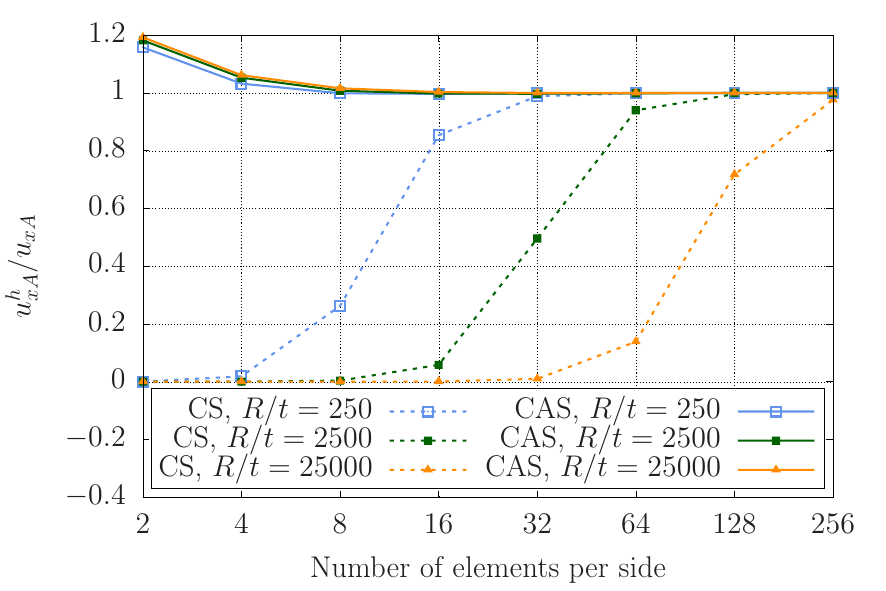}}
 \subfigure[Membrane strain energy]{\includegraphics[scale=0.55]{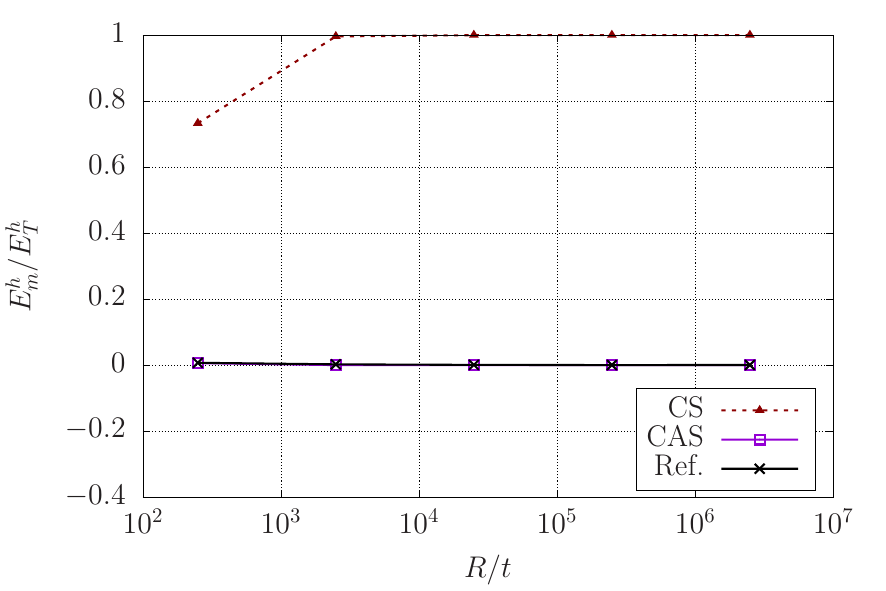}} \\
\caption{(Color online) Pinched hemisphere with a hole. (a) Convergence of the deflection for different slenderness ratios using CS and CAS elements. (b) Ratio of the membrane strain energy to the total strain energy for different slenderness ratios using $8^2$ CS elements and $8^2$ CAS elements. The numerical solution using $8^2$ CAS elements overlaps with the reference solution, which uses $256^2$ CS elements of degree 9.}
\label{pinchedhemisphereplots}
\end{figure}

The second numerical investigation considers a pinched hemisphere with a hole under four point loads perpendicular to the midsurface as shown in Fig. \ref{pinchedhemispheregeo} a). Given the symmetry of this problem, we solve for one quarter of the geometry with the appropriate boundary conditions and point loads as shown in Fig. \ref{pinchedhemispheregeo} b). The next values are used in this example
\begin{equation} 
 P =  31250 t^3,  \quad R = 10.0,  \quad E = 6.825\times10^{7},  \quad \nu = 0.3 \text{.} 
\end{equation}
In order to consider different values of the slenderness ratio, the thickness values $t=4.0\times10^{-2}$, $t=4.0\times10^{-3}$, and $t=4.0\times10^{-4}$ are used. This benchmark problem is a doubly-curved bending-dominated problem. Thus, this problem is particularly prone to membrane locking.

\begin{figure} [t!] 
 \centering
 \subfigure[Reference]{\includegraphics[scale=0.085]{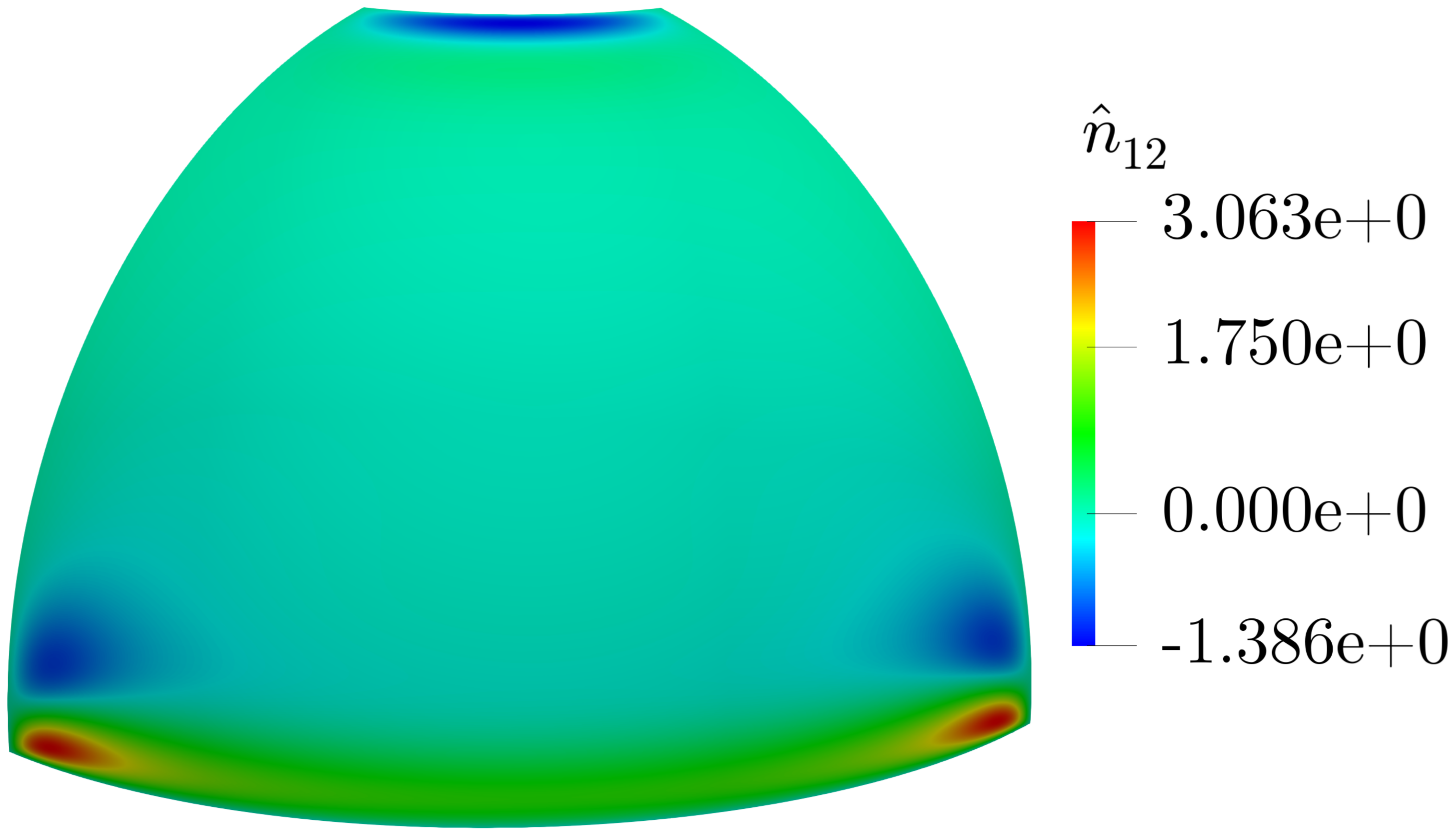}} \hspace*{+8mm}
 \subfigure[$16^2$ CAS elements]{\includegraphics[scale=0.085]{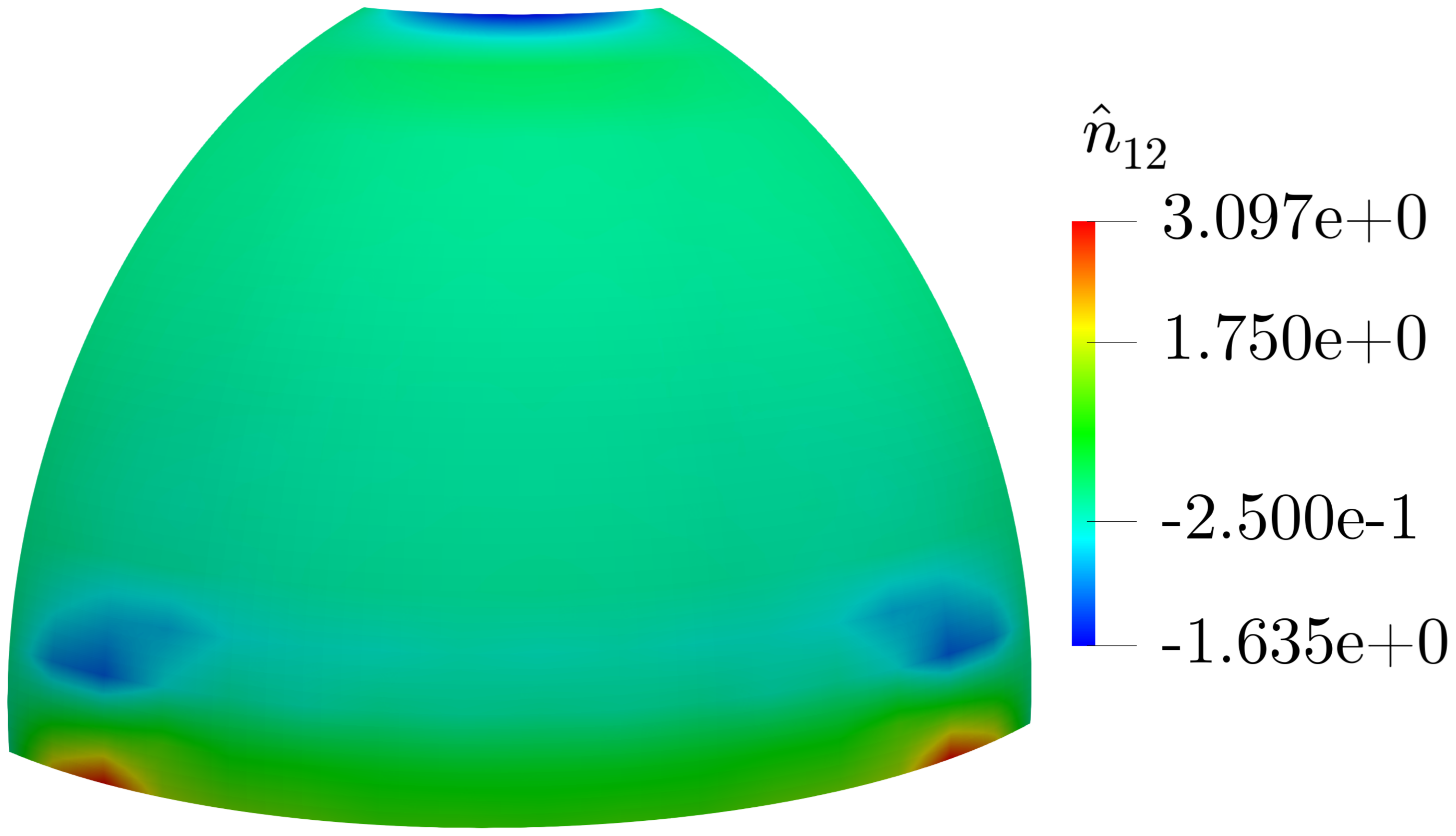}} \\
 \subfigure[$16^2$ CS elements]{\includegraphics[scale=0.085]{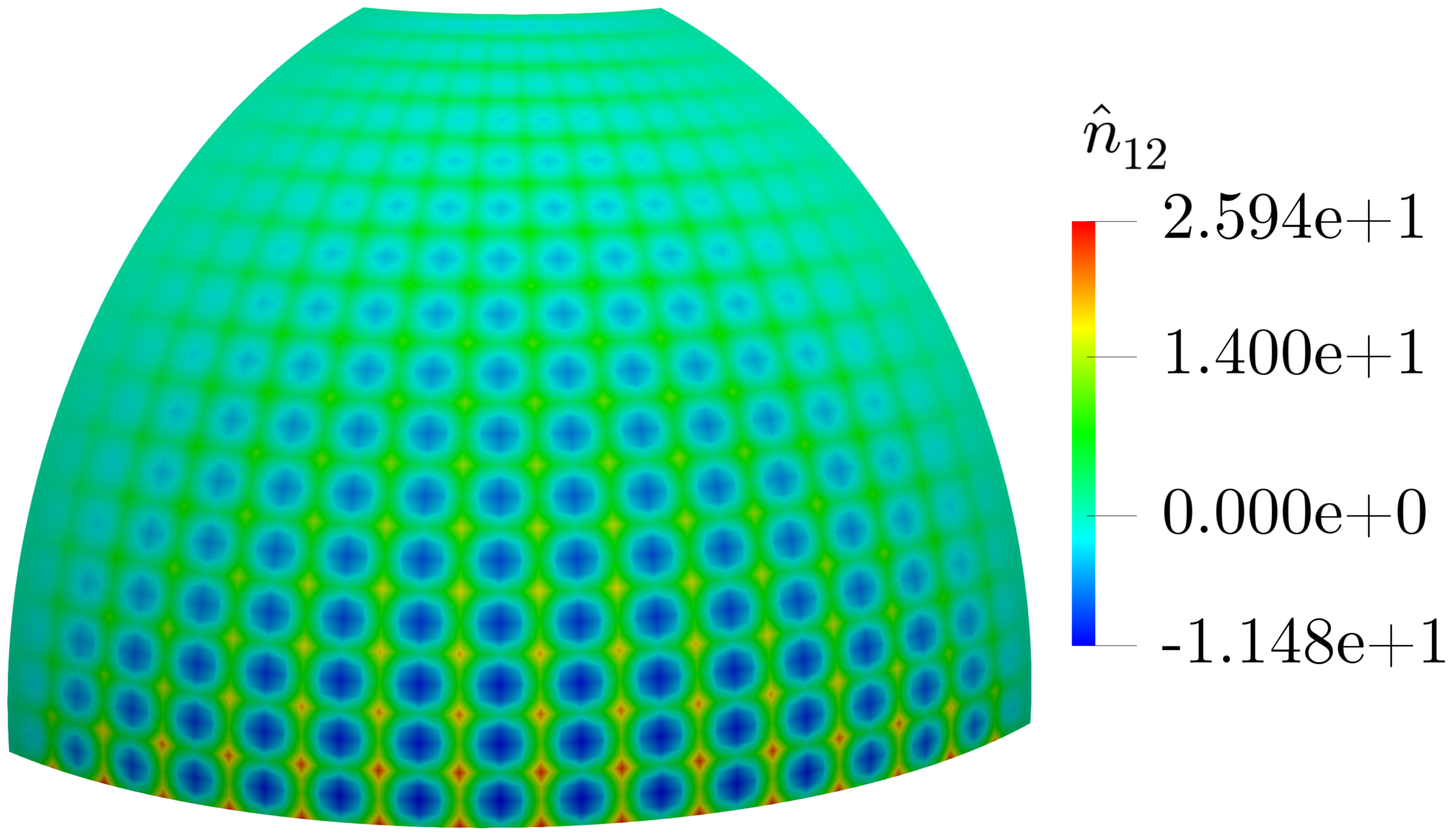}} \hspace*{+8mm}
 \subfigure[$32^2$ CS elements]{\includegraphics[scale=0.085]{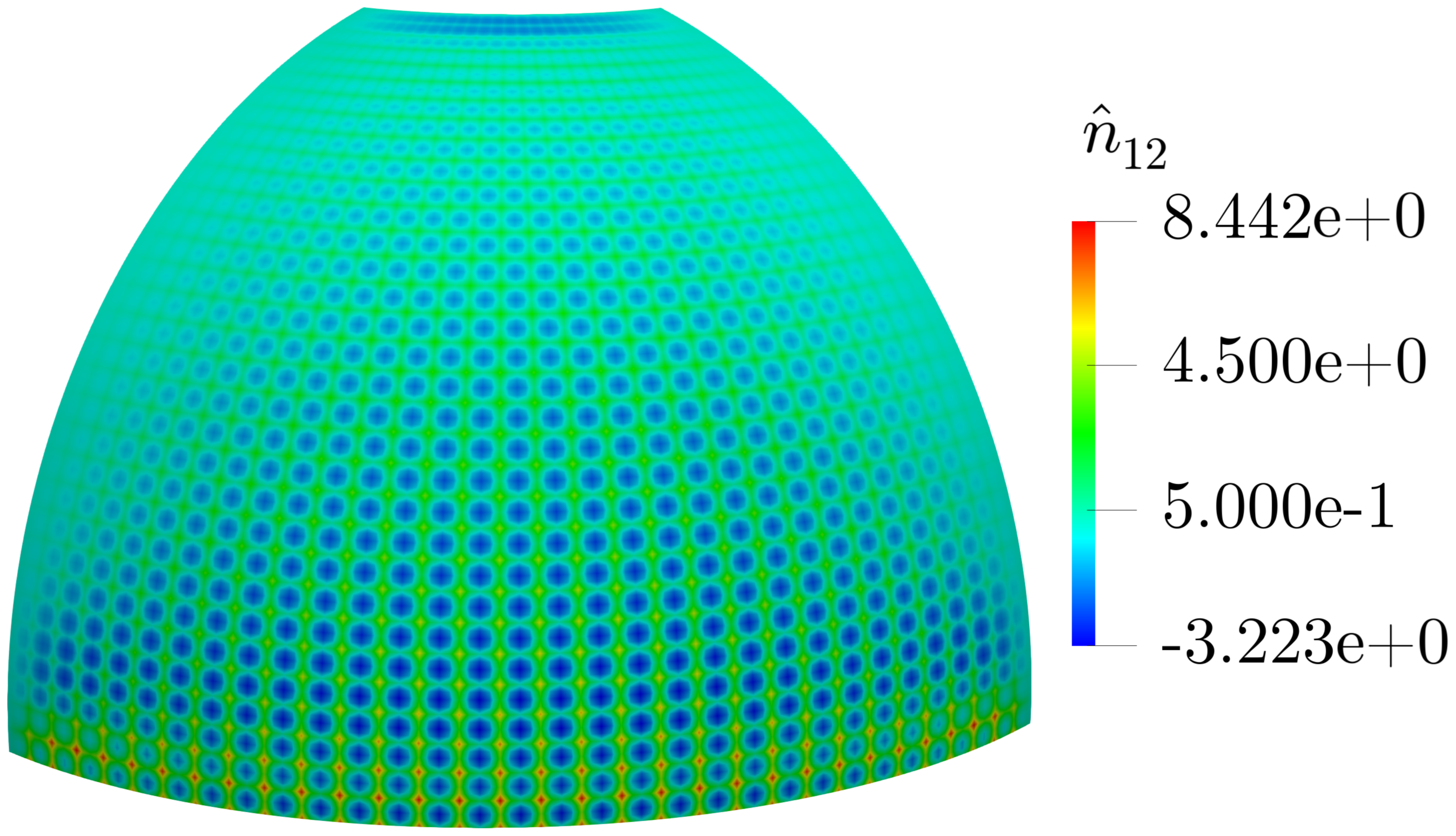}} \\
 \subfigure[$64^2$ CS elements]{\includegraphics[scale=0.085]{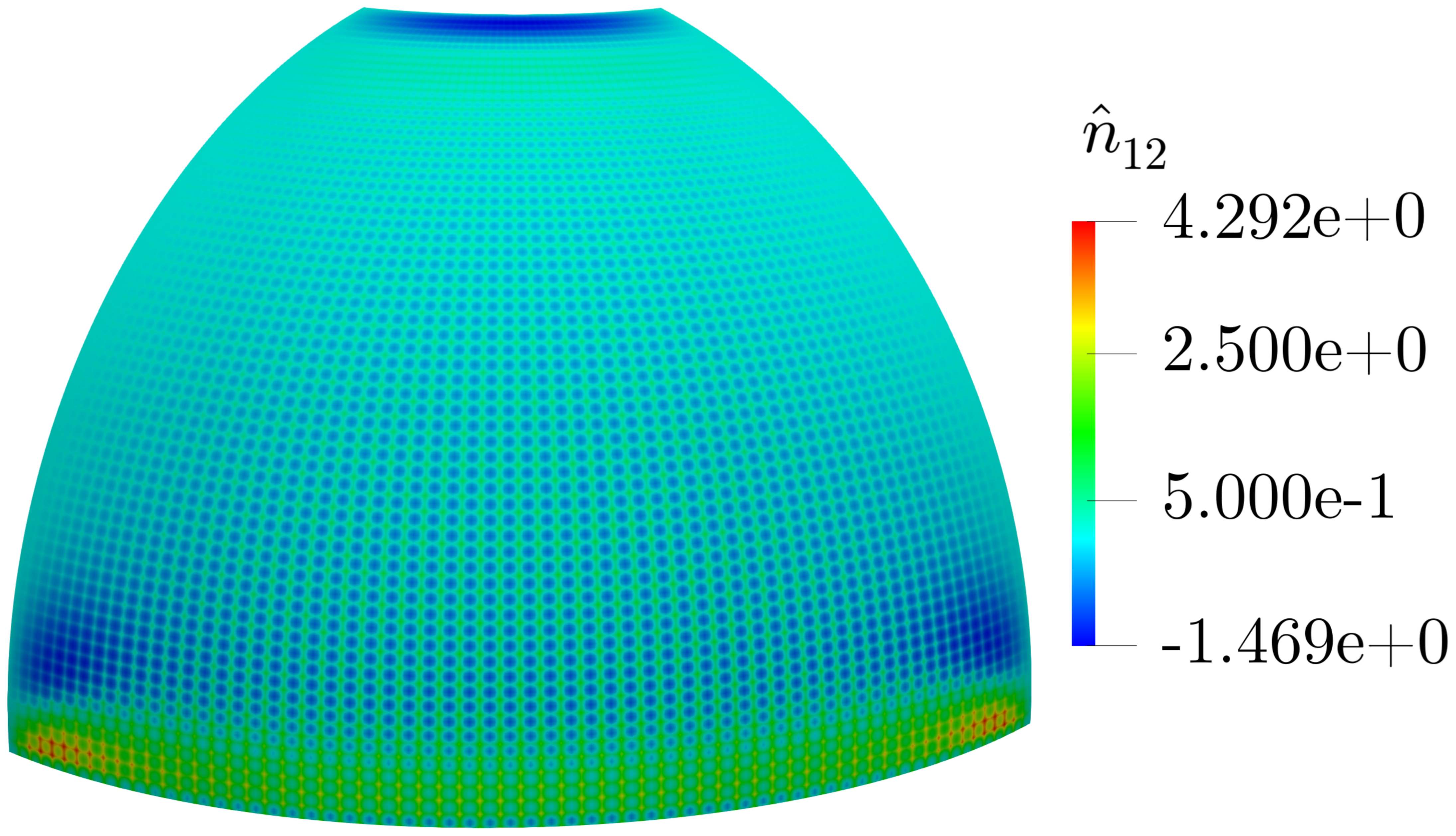}} \hspace*{+8mm}
 \subfigure[$128^2$ CS elements]{\includegraphics[scale=0.085]{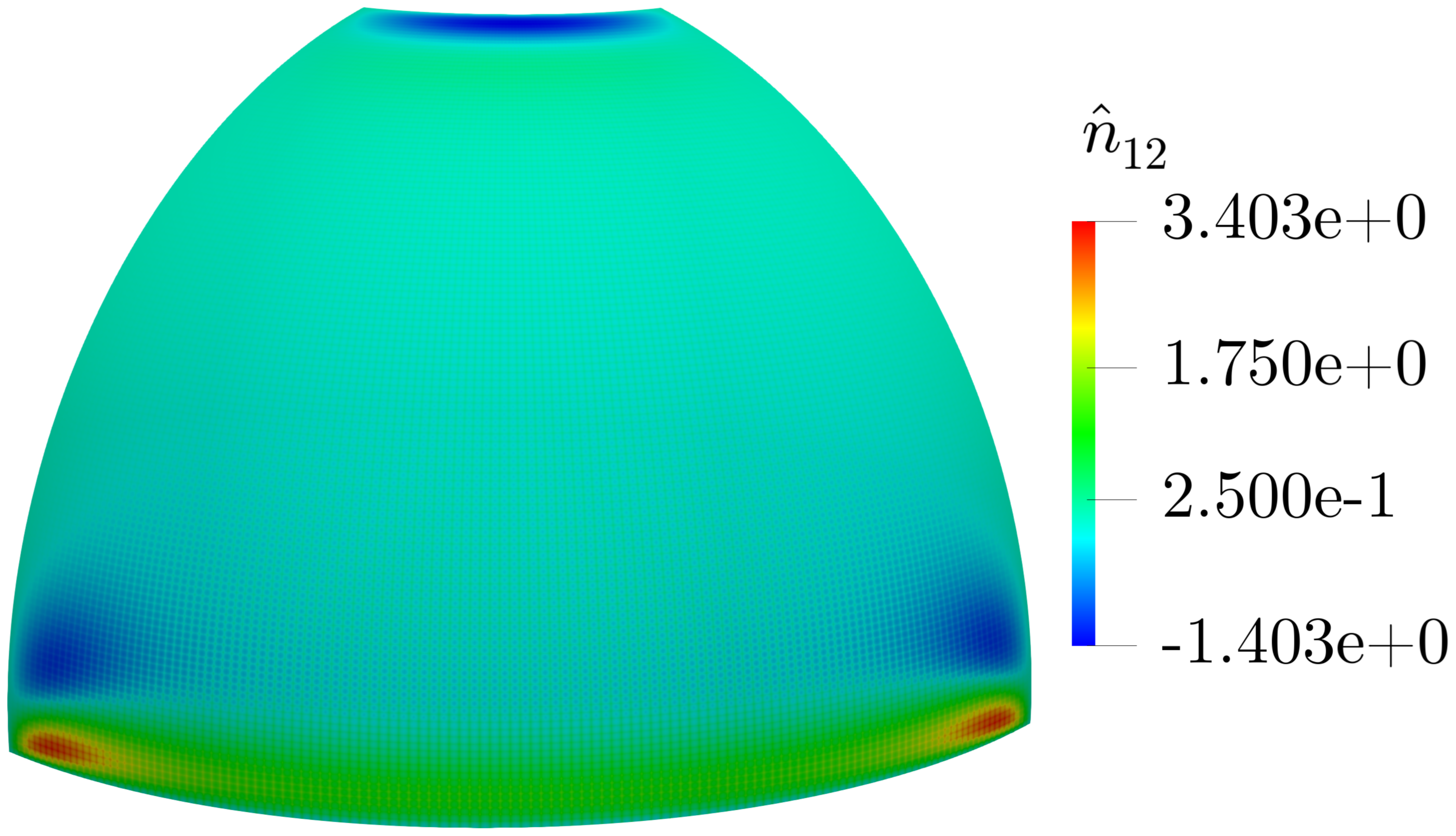}} \\
\caption{(Color online) In-plane shear force of the pinched hemisphere with a hole for $R/t=250$. (a) $256^2$ CS elements of degree 9 are used as reference solution. (b) $16^2$ CAS elements are free of spurious oscillations. (c)-(f) $16^2$, $32^2$, $64^2$, and $128^2$ CS elements have spurious oscillations. Note the different scales used in each plot.}
\label{pinchedhemispheremf}
\end{figure}

We initiate our convergence study with a uniform mesh composed of $2^2$ quadratic elements. The midsurface of the shell is represented exactly since we are using quadratic NURBS. After that, we perform uniform $h$-refinement seven times. Using CS elements and CAS elements, Fig. \ref{pinchedhemisphereplots} a) plots the convergence of the radial displacement of the point $A$ indicated in Fig. \ref{pinchedhemispheregeo}. The displacement values are normalized by the reference solution, which is obtained using $256^2$ CS elements of degree 9. The reference values of the radial displacement at point A are $-9.3521 \times 10^{-2}$, $-9.1594 \times 10^{-2}$, and $-9.0817 \times 10^{-2}$ for $R/t = 2.5 \times 10^2$, $2.5 \times 10^3$, and $2.5 \times 10^4$, respectively. As shown in Fig. \ref{pinchedhemisphereplots} a), the convergence of CAS elements is independent of the slenderness ratio for the broad range of $R/t$ values considered while the convergence of CS elements heavily deteriorates as the slenderness ratio increases. Fig. \ref{pinchedhemisphereplots} b) plots the ratio of the membrane strain energy to the total strain energy as the slenderness ratio increases using $8^2$ CS elements and $8^2$ CAS elements. The numerical solution using $8^2$ CAS elements overlaps with the reference solution, which uses $256^2$ CS elements of degree 9. However, membrane locking causes the introduction of spurious membrane energy in the numerical solution obtained using $8^2$ CS elements for all the slenderness ratios considered.

Fig. \ref{pinchedhemispheremf} plots the distribution of the in-plane shear force for the slenderness ratio $R/t = 250$. Fig. \ref{pinchedhemispheremf} a) plots the reference solution which is obtained using $256^2$ CS elements of degree 9. Fig. \ref{pinchedhemispheremf} b) plots the numerical solution obtained using $16^2$ CAS elements. This numerical solution is free from spurious oscillations which shows the effectiveness of CAS elements in vanquishing membrane locking. Fig. \ref{pinchedhemispheremf}  c), d), e), and f) plot the numerical solution obtained using $16^2$, $32^2$, $64^2$, and $128^2$ CS elements, respectively, which suffer from spurious oscillations caused by membrane locking. Note that $32^2$ and $64^2$ CS elements for the slenderness ratio $R/t = 250$ result in accurate displacements (see Fig. \ref{pinchedhemisphereplots}), but they still suffer from large-amplitude spurious oscillations of the in-plane shear
force (see Fig. \ref{pinchedhemispheremf}).

\begin{figure} [t!] 
 \centering
 \subfigure[$R/t = 2.5 \times 10^2$]{\includegraphics[scale=0.55]{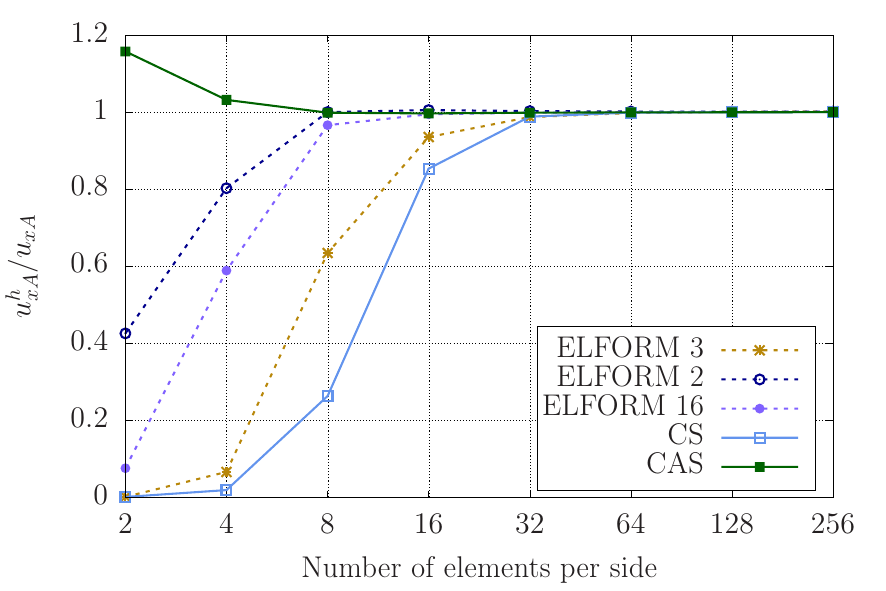}}
 \subfigure[$R/t = 2.5 \times 10^3$]{\includegraphics[scale=0.55]{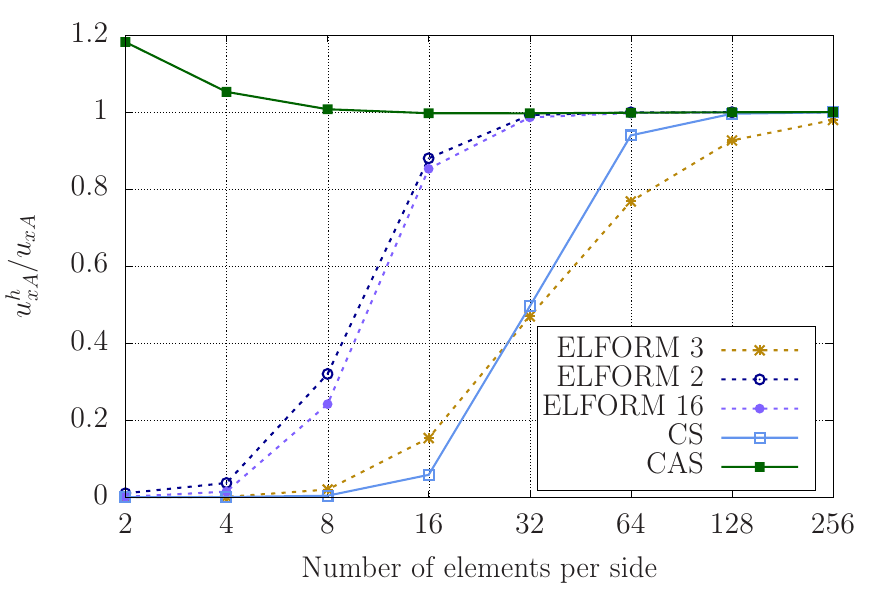}} \\
\caption{(Color online) Pinched hemisphere with a hole. Convergence of the deflection for different slenderness ratios using CS elements, CAS elements, and element types based on Lagrange polynomials from the commercial program LS-DYNA.}
\label{pinchedhemisphereELFORMs}
\end{figure}

Fig. \ref{pinchedhemisphereELFORMs} compares the performance of CAS elements with that of state-of-the-art element types based on Lagrange polynomials. ELFORM 2 is a Reissner-Mindlin shell element based on bilinear Lagrange polynomials that uses reduced integration with hourglass control to treat locking. ELFORM 2 is available in the commercial software LS-DYNA within the keyword SECTION\_SHELL and it is based on References \cite{belytschko1981explicit, belytschko1984explicit}. ELFORM 16 is a Reissner-Mindlin shell element based on bilinear Lagrange polynomials that uses assumed strains to treat locking. ELFORM 16 is available in the commercial software LS-DYNA within the keyword SECTION\_SHELL and it is based on References 
\cite{engelmann1989simple, pian1984rational, simo1986variational, dvorkin1984continuum}. ELFORM 3 is a solid shell element based on trilinear Lagrange polynomials that uses assumed strains to treat locking. ELFORM 3 is available in the commercial software LS-DYNA within the keyword SECTION\_TSHELL and it is based on References 
\cite{kam1985use, liu1994multiple, liu1998multiple}. As shown in Fig. \ref{pinchedhemisphereELFORMs}, CAS elements outperform all the considered element types based on Lagrange polynomials.

\begin{figure} [t!] 
 \centering
 \includegraphics[scale=0.3]{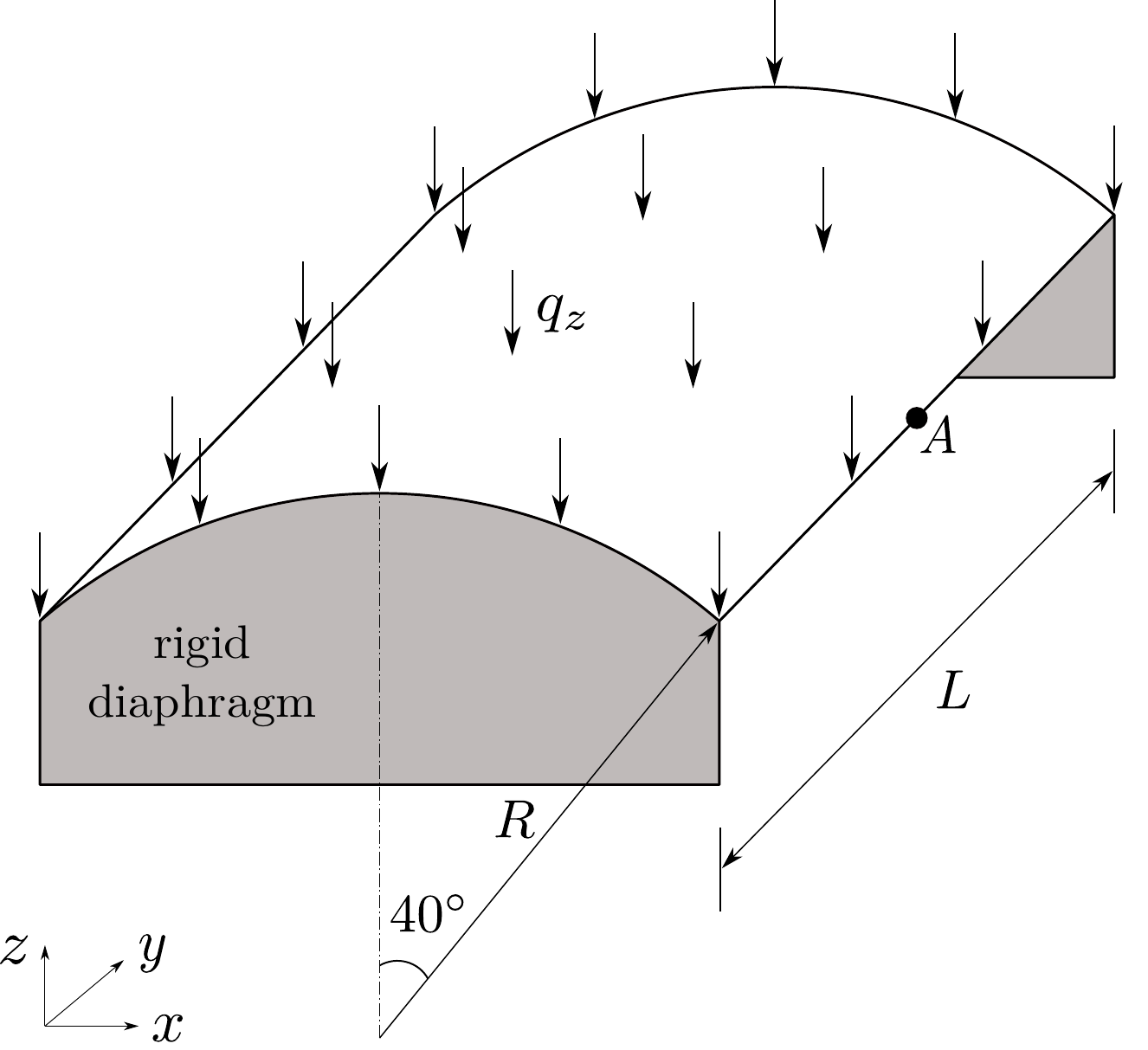}
\caption{Geometry, boundary conditions, and applied load for the Scordelis-Lo roof.} 
\label{scordelisloroofgeo}
\end{figure}

\subsection{Scordelis-Lo roof}

The third numerical investigation considers a cylinder under a distributed load in the vertical direction as shown in Fig. \ref{scordelisloroofgeo}. The next values are used in this example
\begin{equation} 
 q_{z} =  90.0,  \quad R = 25.0,  \quad L = 50.0,  \quad E = 4.32\times10^{8},  \quad \nu = 0 \text{.} 
\end{equation}
In order to consider different values of the slenderness ratio, the thickness values $t=2.5\times10^{-1}$ and $t=2.5\times10^{-2}$ are used. This benchmark problem is not a bending-dominated problem. In \cite{chapelle2010finite}, an asymptotic analysis was performed that showed that $E_b/E_T$ and $E_m/E_T$ tend to $3/8$ and $5/8$ as $R/t$ tends to infinity. It was also shown that the rate with which $E_T$ increases as $R/t$ tends to infinity is $7/4$.

\begin{figure} [t!] 
 \centering
 \subfigure[Deflection]{\includegraphics[scale=0.55]{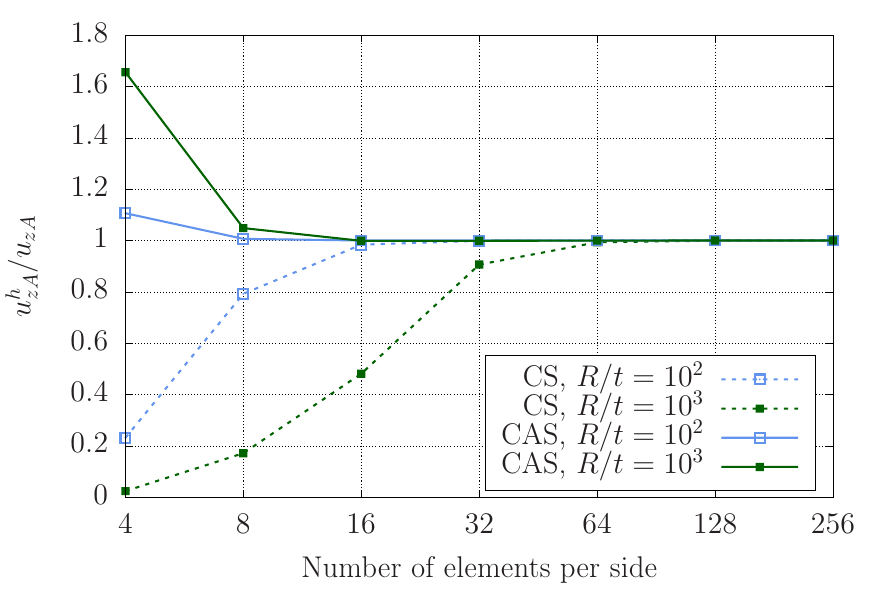}}
 \subfigure[Total strain energy]{\includegraphics[scale=0.55]{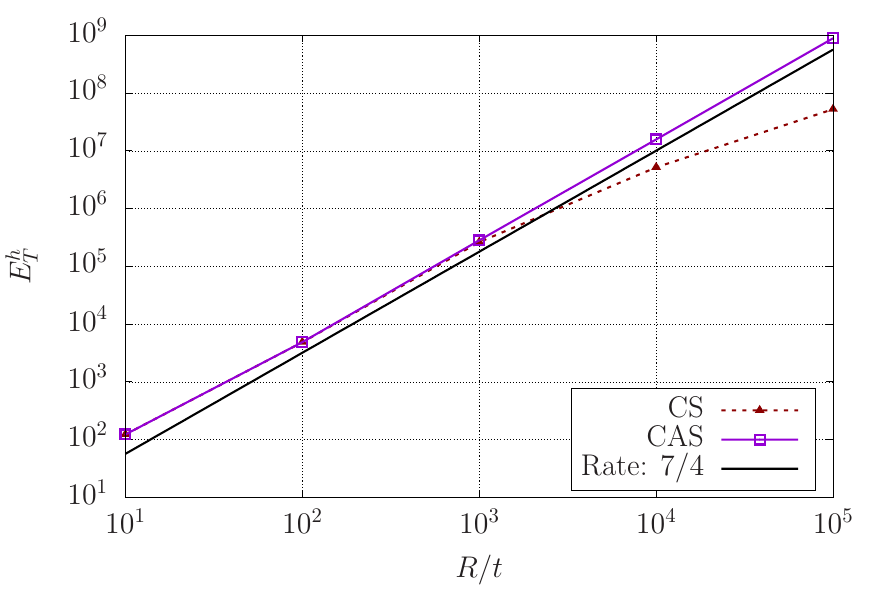}} \\
 \subfigure[Membrane strain energy]{\includegraphics[scale=0.55]{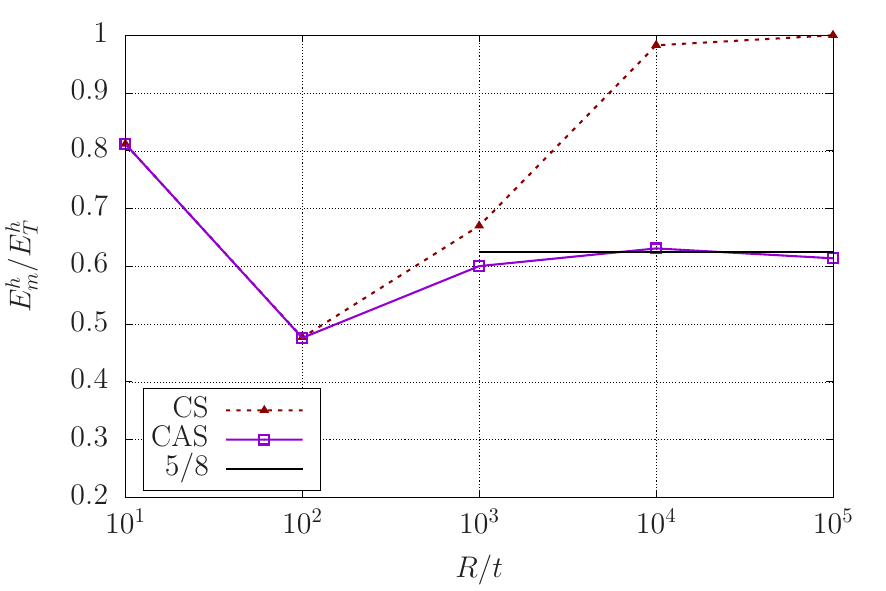}}
 \subfigure[Bending strain energy]{\includegraphics[scale=0.55]{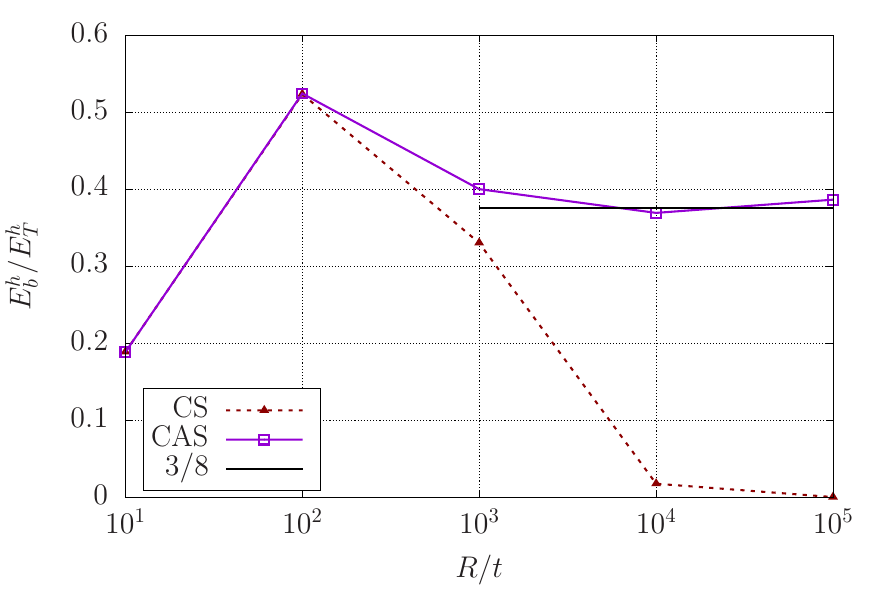}} \\
\caption{(Color online) Scordelis-Lo roof. (a) Convergence of the deflection for different slenderness ratios using CS and CAS elements. (b) Total strain energy for different slenderness ratios using $32^2$ CS elements and $32^2$ CAS elements. (c) Ratio of the membrane strain energy to the total strain energy for different slenderness ratios using $32^2$ CS elements and $32^2$ CAS elements. (d) Ratio of the bending strain energy to the total strain energy for different slenderness ratios using $32^2$ CS elements and $32^2$ CAS elements.}
\label{scordelisloroofplots}
\end{figure}

\begin{figure} [t!] 
 \centering
 \subfigure[Reference, $R/t=10^{2}$]{\includegraphics[scale=0.085]{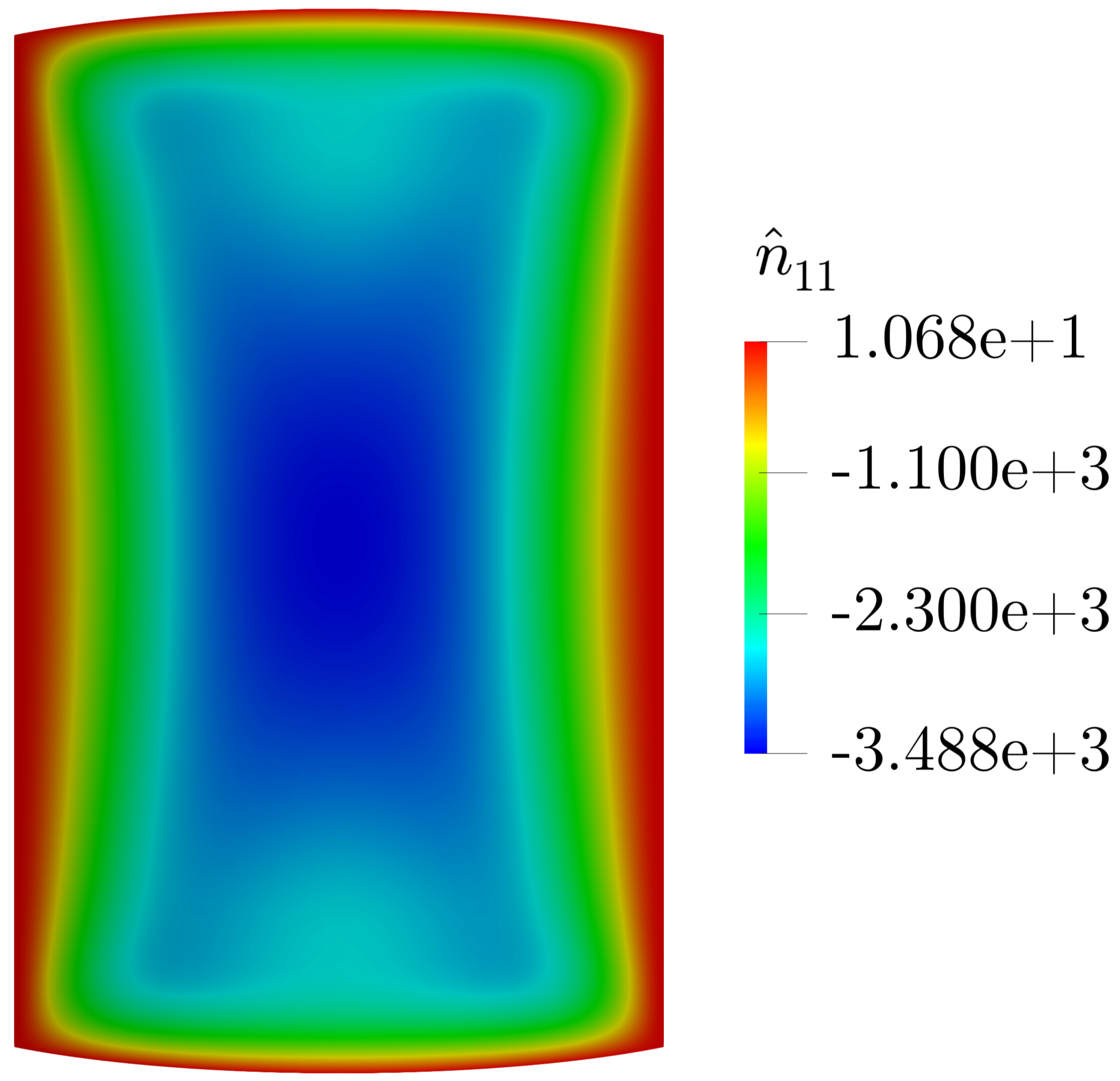}} \hspace*{+15mm}
 \subfigure[Reference, $R/t=10^{3}$]{\includegraphics[scale=0.085]{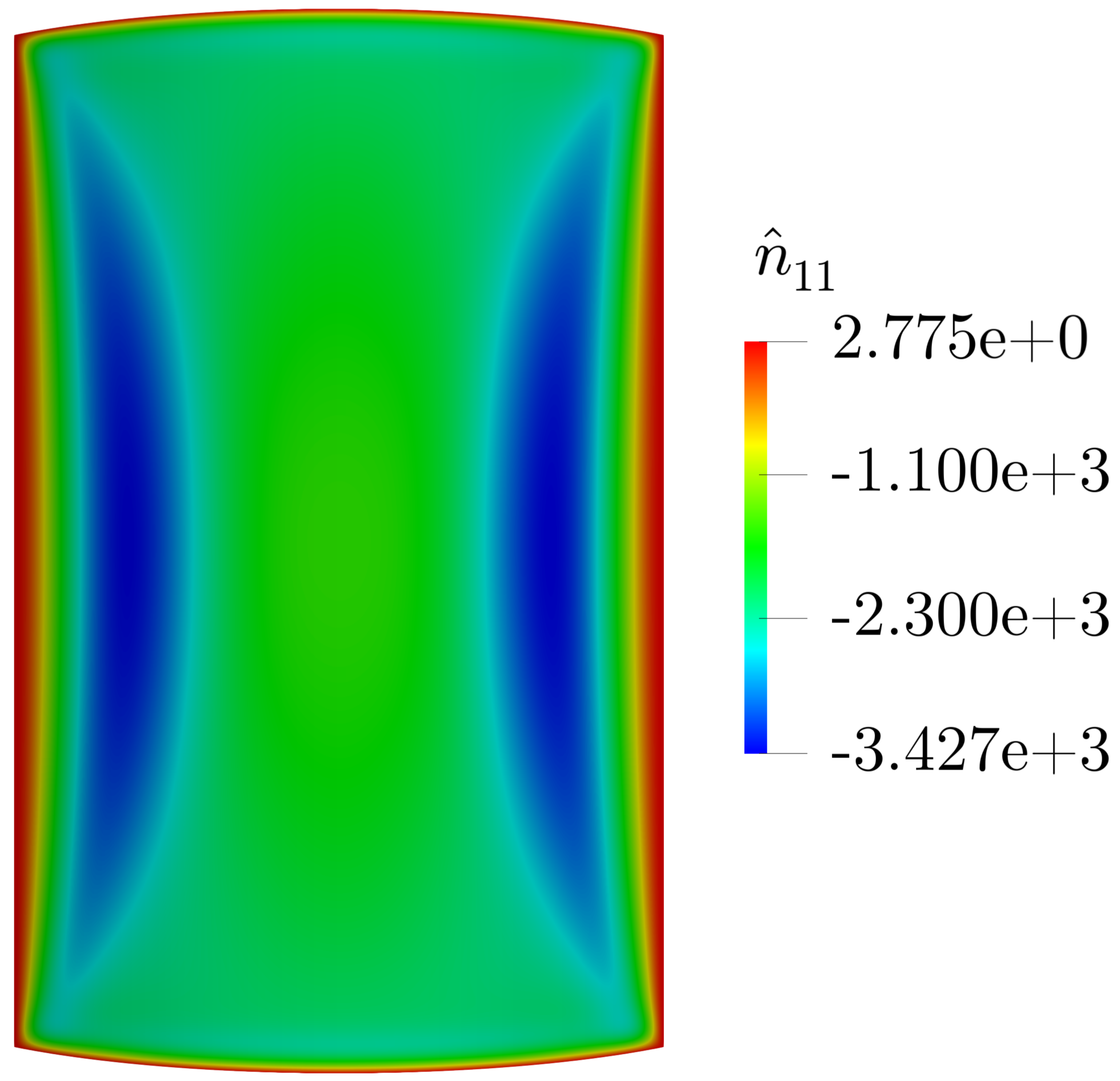}} \\
 \subfigure[$16^2$ CS elements, $R/t=10^{2}$]{\includegraphics[scale=0.085]{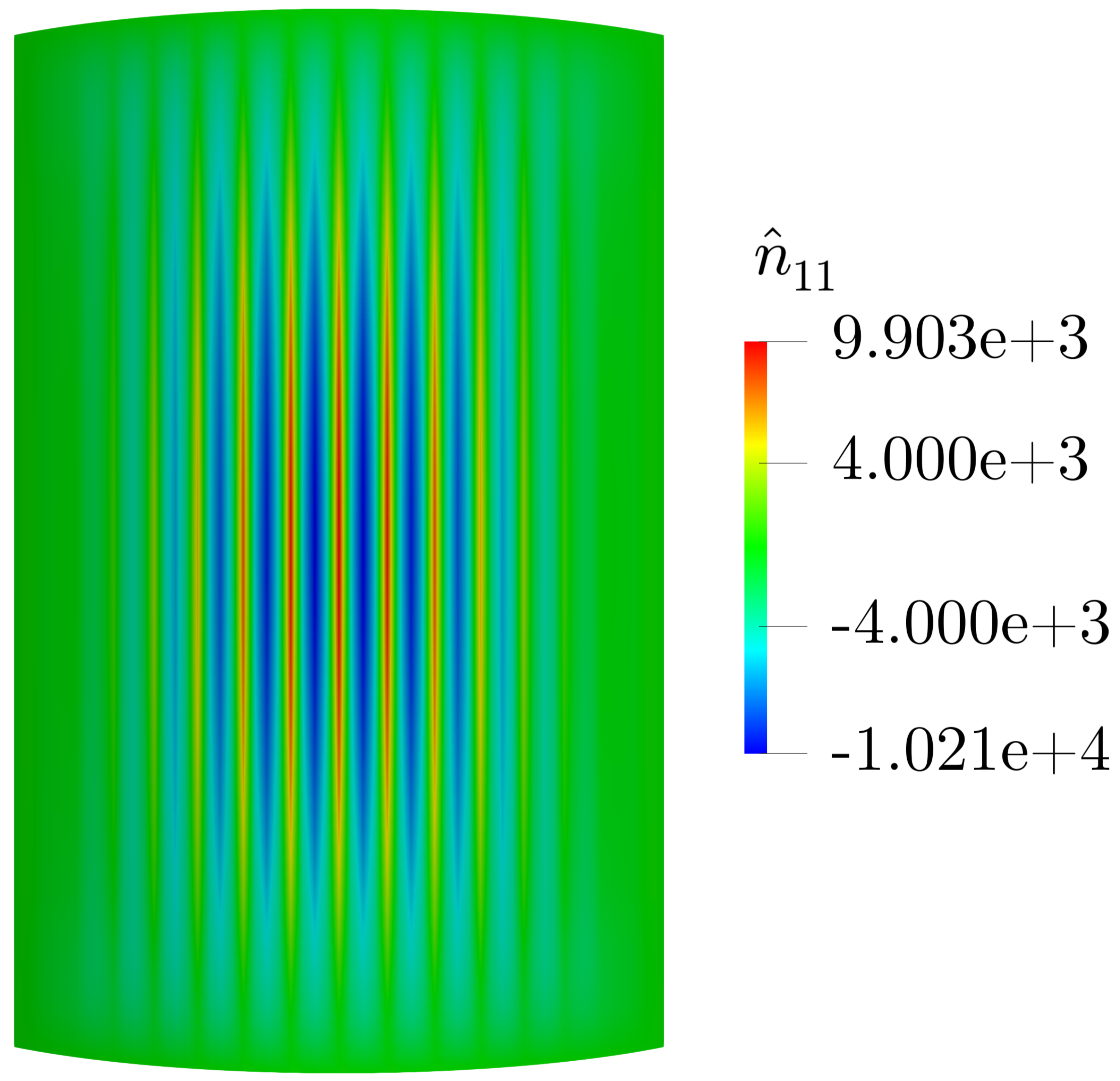}} \hspace*{+15mm}
 \subfigure[$32^2$ CS elements, $R/t=10^{3}$]{\includegraphics[scale=0.085]{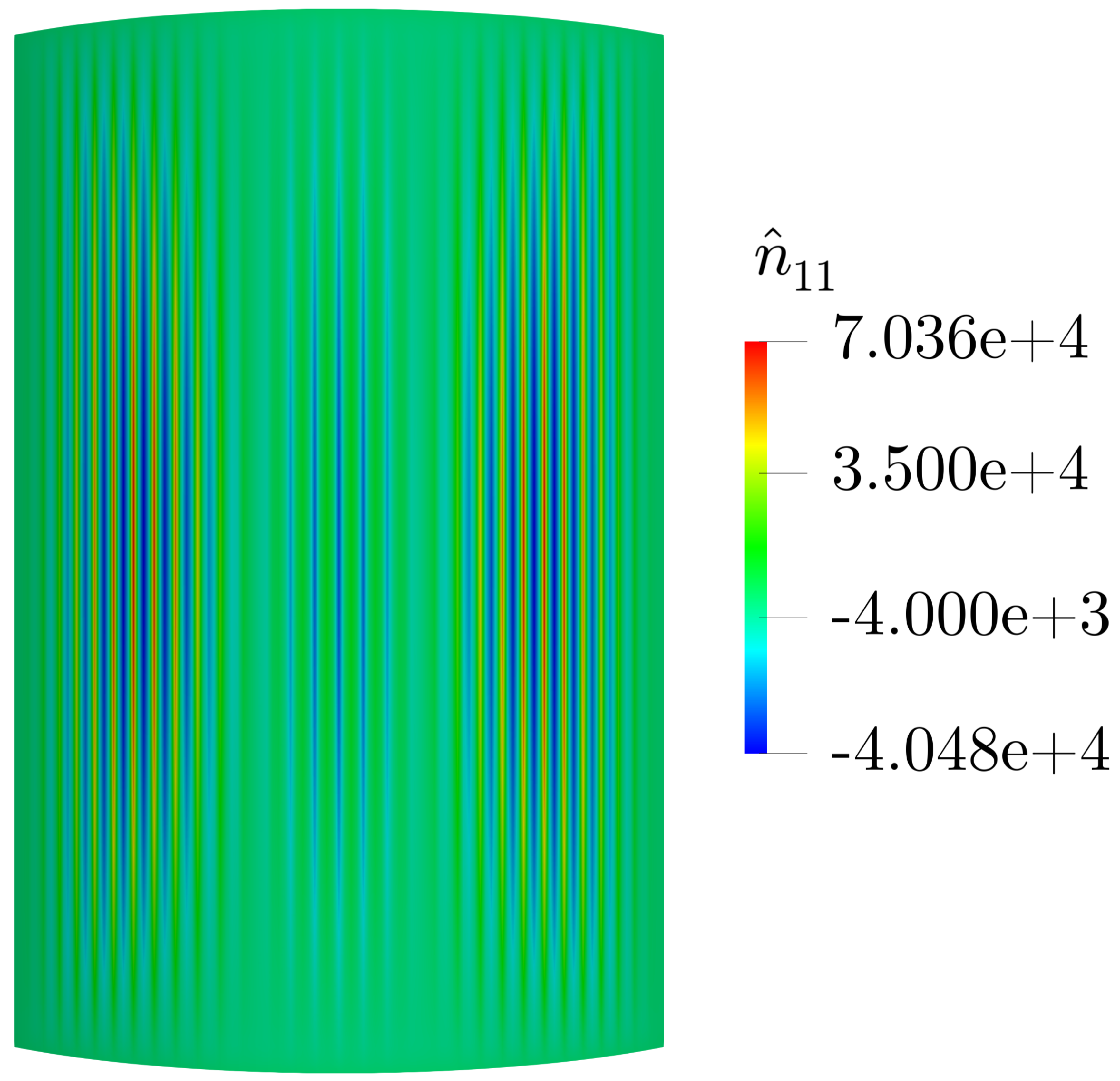}} \\
 \subfigure[$16^2$ CAS elements, $R/t=10^{2}$]{\includegraphics[scale=0.085]{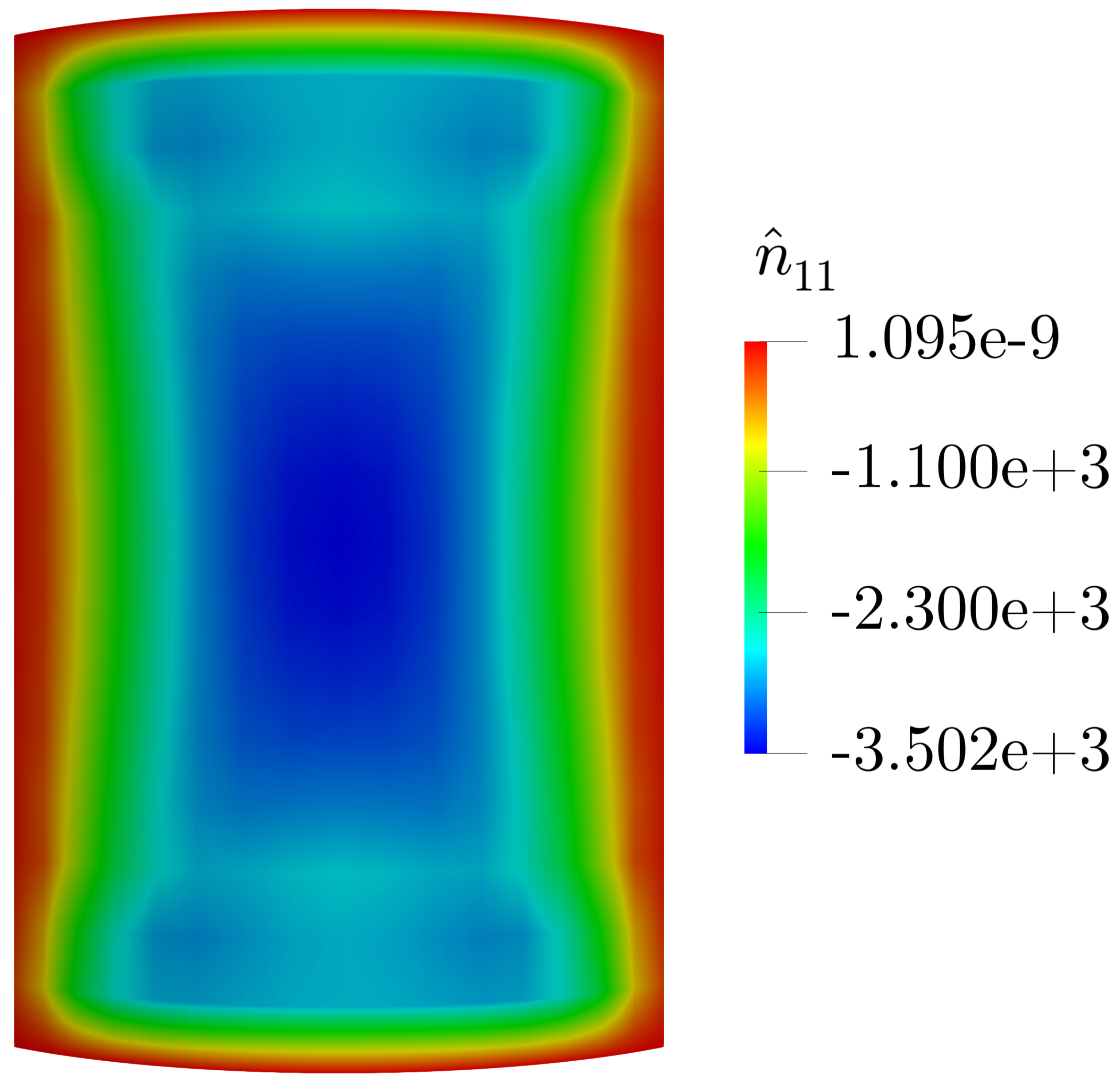}}  \hspace*{+15mm}
 \subfigure[$32^2$ CAS elements, $R/t=10^{3}$]{\includegraphics[scale=0.085]{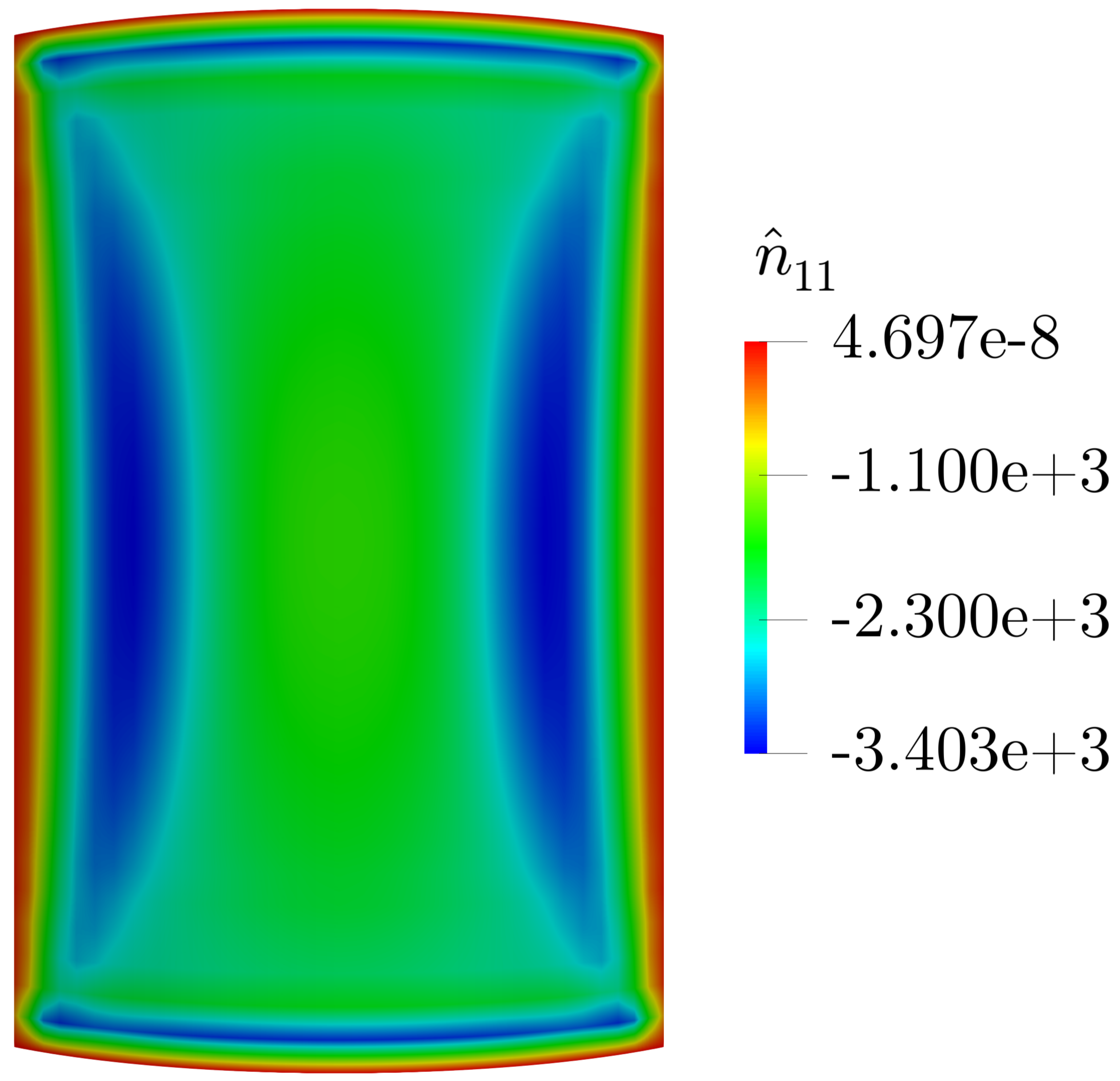}} \\
\caption{(Color online) Membrane force in the circumferential direction for the Scordelis-Lo roof. (a)-(b) $256^2$ CS elements of degree 9 are used as reference solution for $R/t=10^{2}$ and $R/t=10^{3}$. (c)-(d) $16^2$ and $32^2$ CS elements undergo spurious oscillations for $R/t=10^{2}$ and $R/t=10^{3}$, respectively. (e)-(f) $16^2$ and $32^2$ CAS elements are free of spurious oscillations for $R/t=10^{2}$ and $R/t=10^{3}$, respectively. Note the different scales used in each plot.}
\label{scordelisloroofmf}
\end{figure}

Even though the geometry has two planes of symmetry, we run the simulations on the whole geometry without applying symmetry. We initiate our convergence study with a uniform mesh composed of $4^2$ quadratic elements for the whole geometry. The midsurface of the shell is represented exactly since we are using quadratic NURBS. After that, we perform uniform $h$-refinement six times. Using CS elements and CAS elements, Fig. \ref{scordelisloroofplots} a) plots the convergence of the vertical displacement of the point $A$ indicated in Fig. \ref{scordelisloroofgeo}. The displacement values are normalized by the solution obtained using $256^2$ CS elements of degree 9, which is used as a reference solution. The reference values of the vertical displacement at point A are $-3.0059 \times 10^{-1}$ and $-3.2010 \times 10^{+1}$ for $R/t = 10^2$ and $10^3$, respectively. As reported in \cite{greco2018reconstructed}, the membrane and bending strain energies tend to localize near the free edge as the slenderness ratio tends to infinity, which leads to the formation of boundary layers. Thus, even in a locking-free discretization, it is expected that more elements are needed to obtain a certain level of accuracy as the slenderness ratio increases. To the best of the authors' knowledge, no discretization method has shown to result in levels of accuracy independent of the slenderness ratio for this benchmark problem. As shown in Fig. \ref{scordelisloroofplots} a), CAS elements result in significantly more accurate displacement values for coarse meshes than CS elements. Fig. \ref{scordelisloroofplots} b) plots the total strain energy as the slenderness ratio increases using $32^2$ CS elements and $32^2$ CAS elements. Fig. \ref{scordelisloroofplots} c) plots the ratio of the membrane strain energy to the total strain energy as the slenderness ratio increases using $32^2$ CS elements and $32^2$ CAS elements. Fig. \ref{scordelisloroofplots} d) plots the ratio of the bending strain energy to the total strain energy as the slenderness ratio increases using $32^2$ CS elements and $32^2$ CAS elements. As shown in Fig. \ref{scordelisloroofplots} b)-d), the numerical solution using $32^2$ CAS elements is in agreement with the asymptotic analysis performed in \cite{chapelle2010finite} while the numerical solution obtained using $32^2$ CS elements is not in agreement with the asymptotic analysis due to membrane locking.

Fig. \ref{scordelisloroofmf} plots the distribution of the membrane force in the circumferential direction for the slenderness ratios $R/t = 10^2$ and $10^3$. Fig. \ref{scordelisloroofmf} a) and b) plot the reference solutions for the slenderness ratios $R/t = 10^2$ and $10^3$, respectively, which are obtained using $256^2$ CS elements of degree 9. Fig. \ref{scordelisloroofmf} c) and d) plot the numerical solutions obtained for the slenderness ratio $R/t = 10^2$ using $16^2$ CS elements and for the slenderness ratio $R/t = 10^3$ using $32^2$ CS elements, respectively. Fig. \ref{scordelisloroofmf} e) and f) plot the numerical solutions obtained for the slenderness ratio $R/t = 10^2$ using $16^2$ CAS elements and for the slenderness ratio $R/t = 10^3$ using $32^2$ CAS elements, respectively. As shown in Fig. \ref{scordelisloroofmf}, the numerical solutions obtained using CS elements suffer from large-amplitude spurious oscillations caused by membrane locking while the numerical solutions obtained using CAS elements are free from spurious oscillations and closely resemble the reference solutions. 

\begin{table}[t!]
\caption{Vertical displacement at point A for the Scordelis-Lo roof.}
\label{tablescordelisloroof}
\bigskip
 \centering
 \renewcommand{\arraystretch}{1.2}
  \begin{tabular}{c@{\hspace{5.0mm}}  c@{\hspace{5.0mm}}  c@{\hspace{5.0mm}}  c@{\hspace{5.0mm}}  c@{\hspace{5.0mm}}  c@{\hspace{5.0mm}}}
  \hline
  Number of elements per side &  & 5 & 10 & 15 & 20 \\
  \hline
  \multirow{5}{*}{$R/t=10^{2}$} & CS & -0.11513 & -0.27152 & -0.29432 & -0.29852 \\
  & CAS & -0.31102 & -0.30133 & -0.30070 & -0.30059 \\
  & $Rl-\bar{B}$ & -0.29406 & -0.29948 & -0.30012 & -0.30033 \\
  & ref. & & & & -0.30059 \\
  \hline
  \multirow{5}{*}{$R/t=10^{3}$} & CS & -1.46212 & -8.23648 & -14.13189 & -20.44103 \\
  & CAS & -41.60341 & -32.50624 & -32.00124 & -31.94970 \\
  & $Rl-\bar{B}$ & -23.49781 & -31.40780 & -31.69664 & -31.82118 \\
  & ref. & & & & -32.01045 \\
  \hline  
 \end{tabular}     
\end{table}

Table \ref{tablescordelisloroof} compares the performance of CAS elements with that of the reconstructed $\bar{B}$ projections developed for NURBS-based discretizations of linear Kirchhoff-Love shells in \cite{greco2018reconstructed}. In both \cite{greco2018reconstructed} and in this work, the whole geometry of the roof is considered in the simulations. The reconstructed $\bar{B}$ projections perform the $L^2$ projections at the element level and then combine the results from different elements to recover the continuity patterns of the assumed strains across element boundaries. As shown in Table \ref{tablescordelisloroof}, CAS elements and reconstructed $\bar{B}$ projections result in very similar levels of accuracy for the different mesh resolutions considered. However, for a given mesh, reconstructed $\bar{B}$ projections are significantly more computationally expensive than CAS elements since they require solving systems of algebraic equations at the element level, matrix multiplications to obtain the stiffness matrix at the patch level, and the bandwidth of the stiffness matrix at the patch level is increased.

\begin{figure} [t!] 
 \centering
  \subfigure[$R/t = 10^2$]{\includegraphics[scale=0.55]{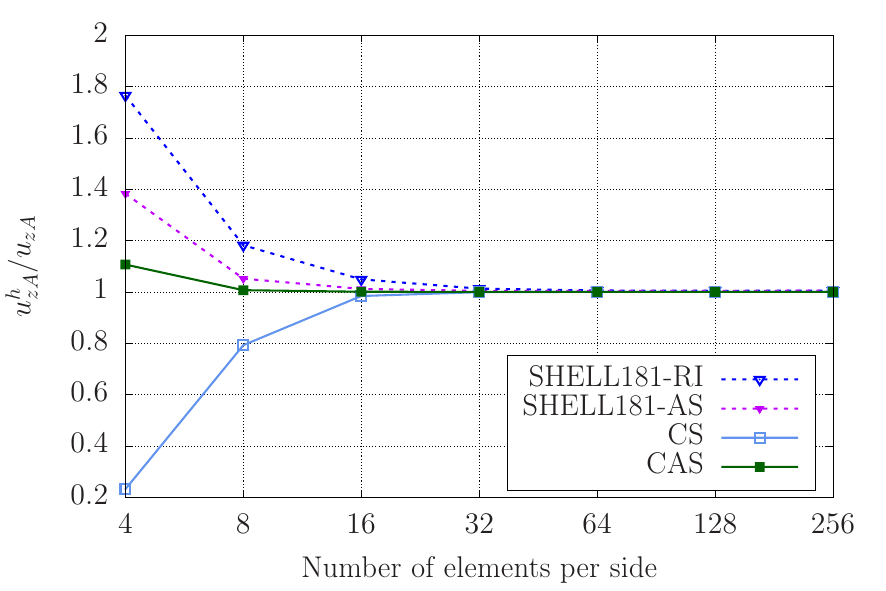}}
 \subfigure[$R/t = 10^3$]{\includegraphics[scale=0.55]{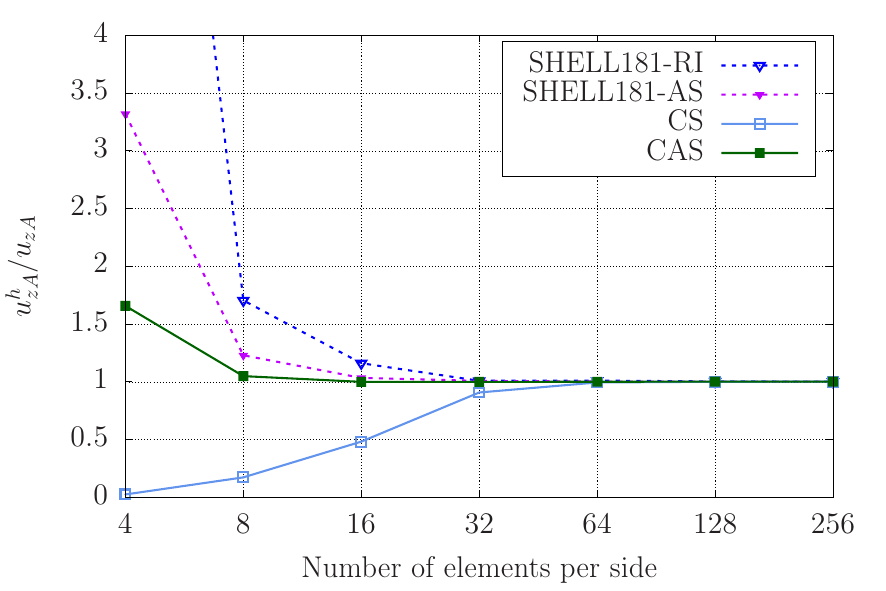}} \\
\caption{(Color online) Scordelis-Lo roof. Convergence of the deflection for different slenderness ratios using CS elements, CAS elements, and element types based on Lagrange polynomials from the commercial program Ansys Mechanical.}
\label{scordelisloroofELFORMs}
\end{figure}

Fig. \ref{scordelisloroofELFORMs} compares the performance of CAS elements with that of state-of-the-art element types based on Lagrange polynomials. SHELL181-RI is a Reissner-Mindlin shell element based on bilinear Lagrange polynomials that uses reduced integration with hourglass control to treat locking. SHELL181-RI is available in the commercial software Ansys Mechanical. SHELL181-AS is a Reissner-Mindlin shell element based on bilinear Lagrange polynomials that uses assumed strains to treat locking. SHELL181-AS is available in the commercial software Ansys Mechanical and it is based on References 
\cite{dvorkin1984continuum, bathe1986formulation}. As shown in Fig. \ref{scordelisloroofELFORMs}, CAS elements perform better than these element types based on Lagrange polynomials.

\subsection{Partly clamped hyperbolic paraboloid}

\begin{figure} [t!] 
 \centering
 \subfigure[Before applying symmetry]{\includegraphics[scale=0.3]{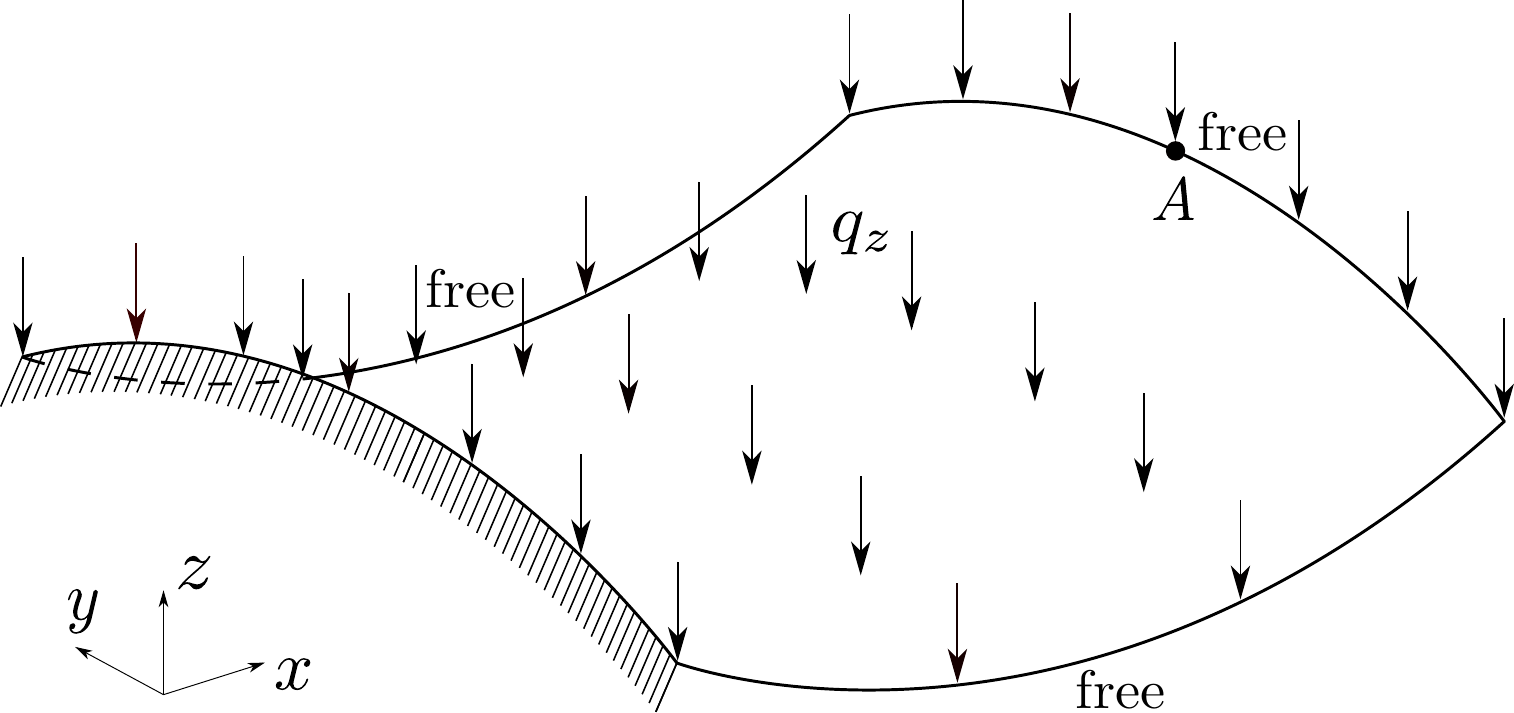}} \hspace*{+10mm}
 \subfigure[After applying symmetry]{\includegraphics[scale=0.3]{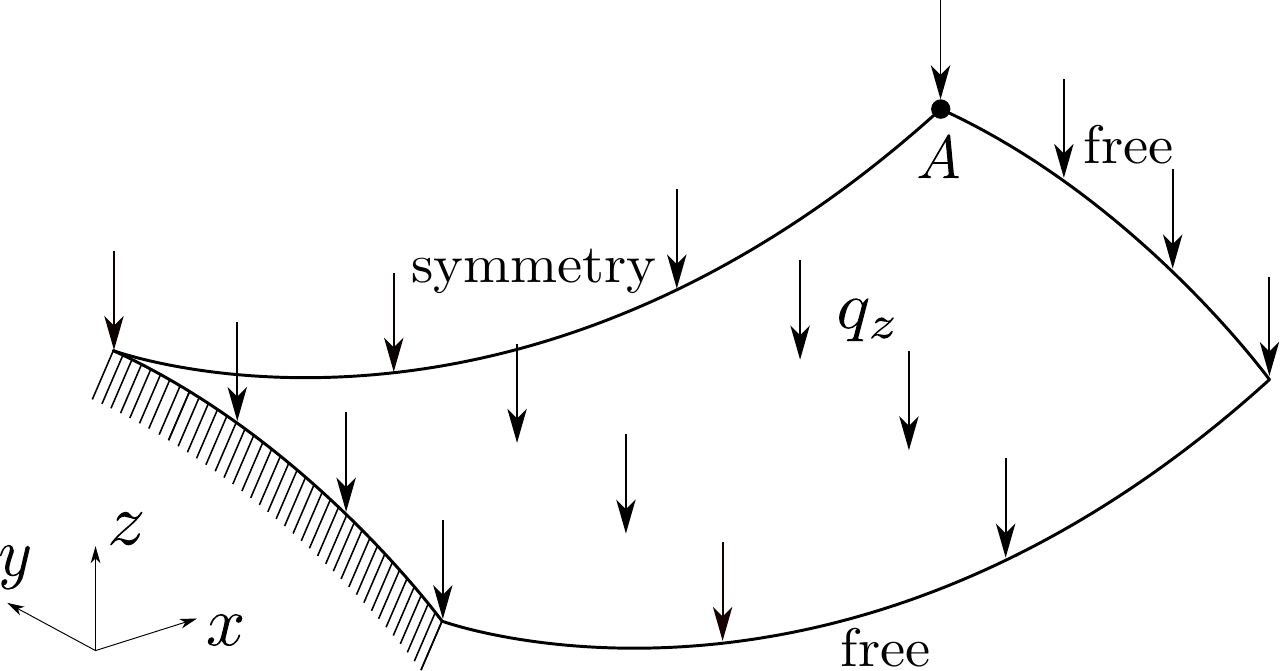}} \\
\caption{Geometry, boundary conditions, and applied load for the partly clamped hyperbolic paraboloid.} 
\label{hyperbolicparaboloidgeo}
\end{figure}

The fourth numerical investigation considers a hyperbolic paraboloid clamped on one side under a distributed load in the vertical direction as shown in Fig. \ref{hyperbolicparaboloidgeo} a). The geometry of the midsurface is defined by 
\begin{equation} 
 z =  x^2 - y^2,  \quad -L/2 \leq x \leq L/2,  \quad -L/2 \leq y \leq L/2 \text{.} 
\end{equation}
Given the symmetry of this problem, we solve for one half of the geometry with the appropriate boundary conditions as shown in Fig. \ref{hyperbolicparaboloidgeo} b). The next values are used in this example
\begin{equation} 
 q_{z} =  8000.0t,  \quad L = 1.0,  \quad E = 2.0\times10^{11},  \quad \nu = 0.3 \text{.} 
\end{equation}
In order to consider different slenderness, the thickness values $t=10^{-2}$, $t=10^{-3}$, and $t=10^{-4}$ are used. As in \cite{bathe2000evaluation, bouclier2013efficient, greco2018reconstructed}, the $L/t$ ratio is used as a measure of slenderness. This benchmark problem is a bending-dominated problem with non-constant curvatures.

\begin{figure} [t!] 
 \centering
 \subfigure[Deflection]{\includegraphics[scale=0.55]{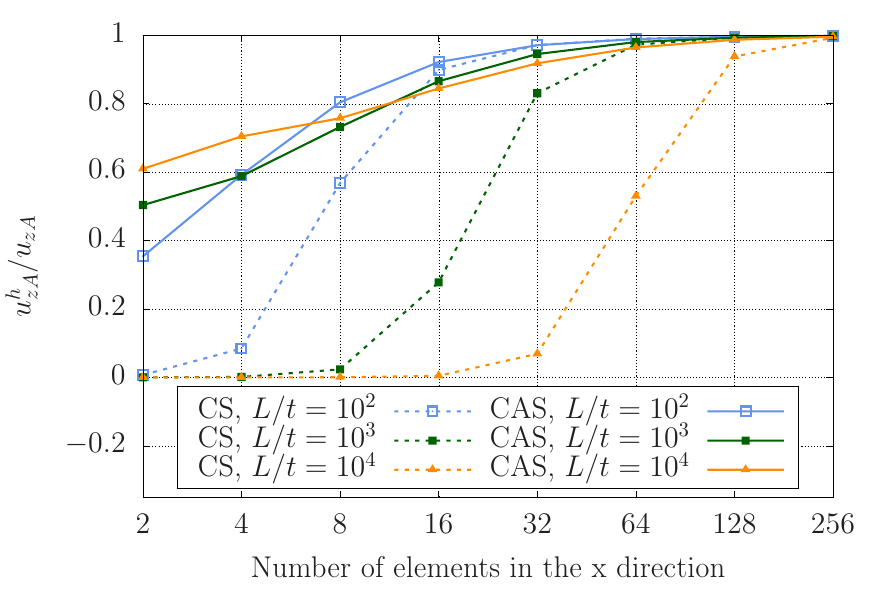}}
 \subfigure[Membrane strain energy]{\includegraphics[scale=0.55]{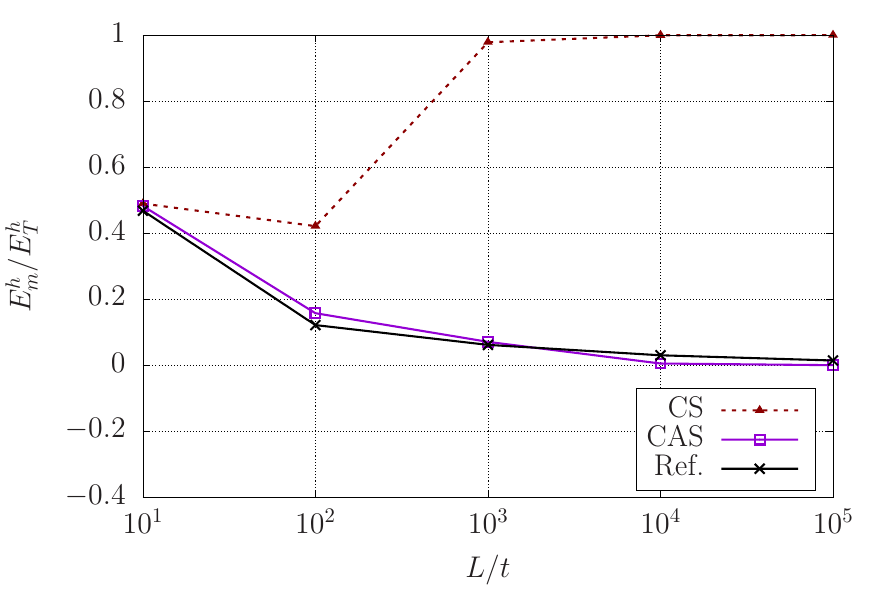}} \\
\caption{(Color online) Partly clamped hyperbolic paraboloid. (a) Convergence of the deflection for different $L/t$ ratios using CS and CAS elements. (b) Ratio of the membrane strain energy to the total strain energy for different $L/t$ ratios using $8 \times 4$ CS elements and $8 \times 4$ CAS elements. The numerical solution using $8 \times 4$ CAS elements closely follows the reference solution, which uses $256 \times 128$ CS elements of degree 9.}
\label{hyperbolicparaboloidplots}
\end{figure}

We initiate our convergence study with a uniform mesh composed of $2 \times 1$ quadratic elements (2 elements in the $x$ direction and 1 element in the $y$ direction). The midsurface of the shell is represented exactly since we are using quadratic NURBS. After that, we perform uniform $h$-refinement seven times. Using CS elements and CAS elements, Fig. \ref{hyperbolicparaboloidplots} a) plots the convergence of the vertical displacement of the point $A$ indicated in Fig. \ref{hyperbolicparaboloidgeo}. The displacement values are normalized by the solution obtained using $256 \times 128$ CS elements of degree 9, which is used as a reference solution. The reference values of the vertical displacement at point A are $-9.3128\times 10^{-5}$, $-6.3957 \times 10^{-3}$, and $-5.3059 \times 10^{-1}$ for $L/t = 10^2$, $10^3$, and $10^4$, respectively. As shown in Fig. \ref{hyperbolicparaboloidplots} a), the convergence of CS elements heavily deteriorates as the $L/t$ ratio increases while the convergence of CAS elements is similar for the broad range of $L/t$ values considered. Fig. \ref{hyperbolicparaboloidplots} b) plots the ratio of the membrane strain energy to the total strain energy as the $L/t$ ratio increases using $8 \times 4$ CS elements and $8 \times 4$ CAS elements. The numerical solution using $8 \times 4$ CAS elements closely follows  the reference solution, which uses $256 \times 128$ CS elements of degree 9. However, membrane locking causes the introduction of spurious membrane energy in the numerical solution obtained using $8 \times 4$ CS elements as the $L/t$ ratio increases.

\begin{figure} [t!] 
 \centering
 \subfigure[CS elements]{\includegraphics[scale=0.55]{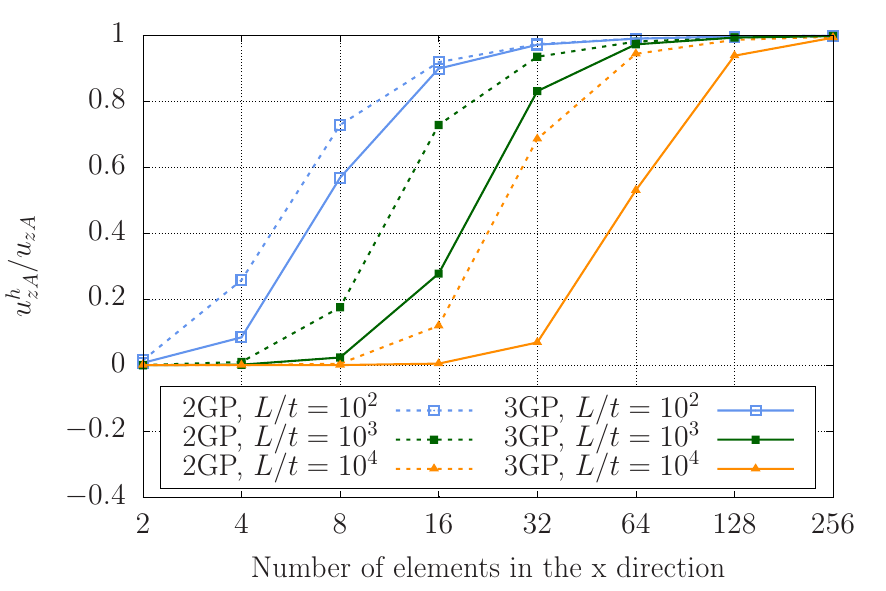}}
 \subfigure[CAS elements]{\includegraphics[scale=0.55]{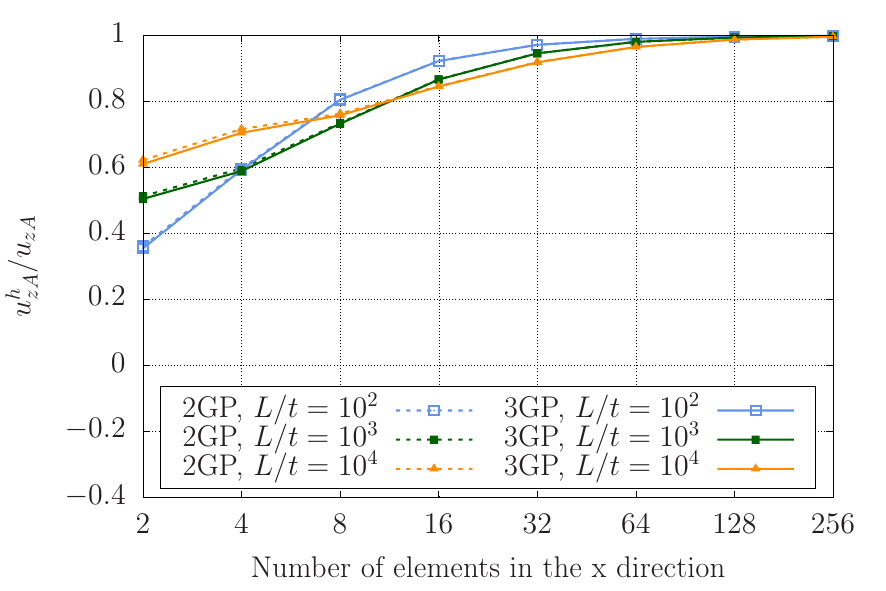}} \\
 \subfigure[CS elements]{\includegraphics[scale=0.55]{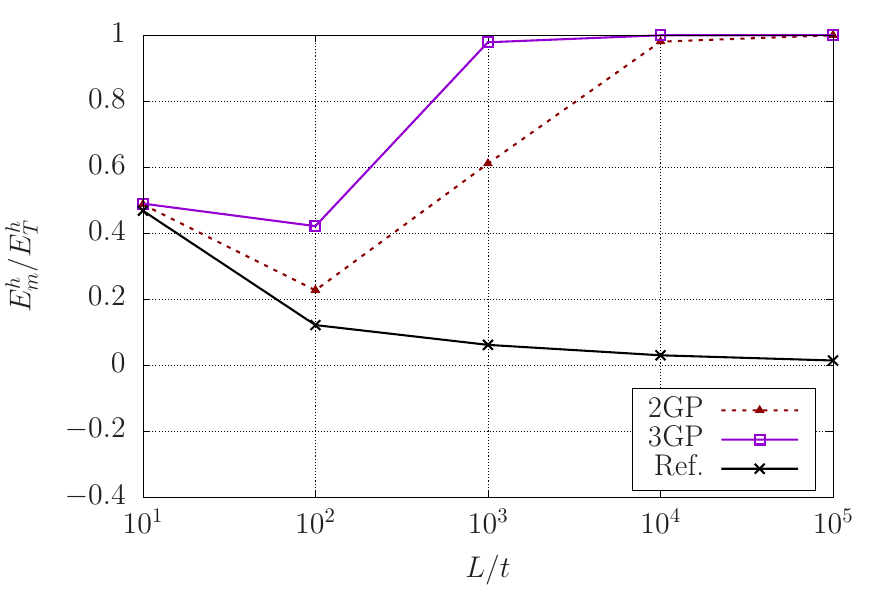}}
 \subfigure[CAS elements]{\includegraphics[scale=0.55]{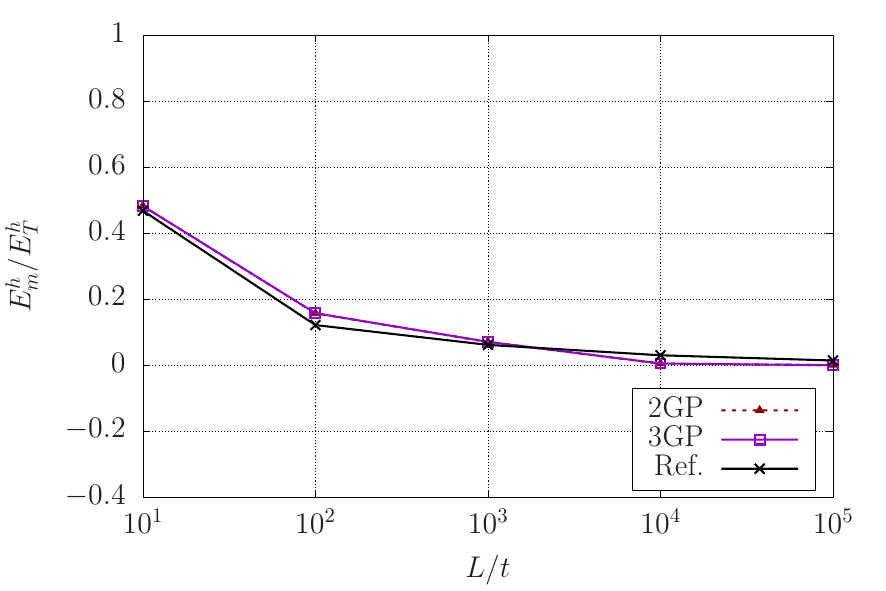}} \\
\caption{(Color online) Partly clamped hyperbolic paraboloid. (a) Convergence of the deflection for different $L/t$ ratios using CS elements with $2 \times 2$ and $3 \times 3$ Gauss-Legendre quadrature points. (b) Convergence of the deflection for different $L/t$ ratios using CAS elements with $2 \times 2$ and $3 \times 3$ Gauss-Legendre quadrature points. (c) Ratio of the membrane strain energy to the total strain energy for different $L/t$ ratios using $8 \times 4$ CS elements with $2 \times 2 $ and $3 \times 3$ Gauss-Legendre quadrature points. (d) Ratio of the membrane strain energy to the total strain energy for different $L/t$ ratios using $8 \times 4$ CAS elements with $2 \times 2$ and $3 \times 3$ Gauss-Legendre quadrature points.}
\label{hyperbolicparaboloidGPplots}
\end{figure}

We also solve this problem using $2\times2$ Gauss-Legendre quadrature points per element to compute all the integrals as opposed to using $3\times3$ Gauss-Legendre quadrature points per element. As shown in Fig. \ref{hyperbolicparaboloidGPplots}, CAS elements result in essentially the same accuracy regardless of whether 2 quadrature points per direction or 3 quadrature points per direction are used. Therefore, 2 quadrature points per direction can be used to decrease the computational time. However, CS elements with 2 quadrature points per direction are still a locking-prone discretization.

\section{Conclusions and future work}

In this work, linear Kirchhoff-Love shells are used to investigate how to effectively overcome membrane locking in quadratic NURBS-based discretizations of shells. We developed an interpolation-based assumed-strain treatment, named continuous-assumed-strain (CAS) elements, that vanquishes membrane locking by bilinearly interpolating the membrane strains at the four corners of each element. This locking treatment results in assumed strains with $C^0$ continuity across element boundaries. The effects of membrane locking are not only smaller displacements than expected, but also large-amplitude spurious oscillations of membrane forces. The spurious oscillations of membrane forces can take place for fine meshes for which the displacement values are already accurate. Therefore, evaluating the efficacy of a locking treatment by only studying the accuracy of the displacements is not enough, the accuracy of the membrane forces needs to be studied as well. CAS elements, using either $2\times2$ or $3\times3$ Gauss-Legendre quadrature points per element, are an effective locking treatment since this element type results in more accurate displacements for coarse meshes and gets rid of the spurious oscillations of the membrane forces. CAS elements are computationally efficient since the computational cost scarcely increases with respect to the locking-prone CS elements. In addition, CAS elements outperform state-of-the-art element types based on Lagrange polynomials equipped with either assumed-strain or reduced-integration locking treatments. Future research directions include extending the proposed locking treatment to nonlinear Kirchhoff-Love shells as well as other types of shells such as Reissner-Mindlin shells and solid shells.

\section*{Acknowledgements}

H. Casquero and K. D. Mathews were partially supported by the NSF grant CMMI-2138187, Honda Motor Co., and Ansys Inc.


\bibliographystyle{elsarticle-num} 
\bibliography{./Bibliography}

\begin{thebibliography}{10}
\expandafter\ifx\csname url\endcsname\relax
  \def\url#1{\texttt{#1}}\fi
\expandafter\ifx\csname urlprefix\endcsname\relax\def\urlprefix{URL }\fi
\expandafter\ifx\csname href\endcsname\relax
  \def\href#1#2{#2} \def\path#1{#1}\fi

\bibitem{1003.000}
T.~J.~R. Hughes, J.~A. Cottrell, Y.~Bazilevs, Isogeometric analysis: {CAD},
  finite elements, {NURBS}, exact geometry and mesh refinement, Computer
  Methods in Applied Mechanics and Engineering 194 (2005) 4135--4195.

\bibitem{cottrell2009isogeometric}
J.~A. Cottrell, T.~J.~R. Hughes, Y.~Bazilevs, Isogeometric analysis: {T}oward
  integration of {CAD} and {FEA}, John Wiley \& Sons, 2009.

\bibitem{wen2023isogeometric}
Z.~Wen, M.~S. Faruque, X.~Li, X.~Wei, H.~Casquero, Isogeometric analysis using
  {G}-spline surfaces with arbitrary unstructured quadrilateral layout,
  Computer Methods in Applied Mechanics and Engineering 408 (2023) 115965.

\bibitem{wei2022analysis}
X.~Wei, X.~Li, K.~Qian, T.~J.~R. Hughes, Y.~J. Zhang, H.~Casquero,
  Analysis-suitable unstructured {T}-splines: {M}ultiple extraordinary points
  per face, Computer Methods in Applied Mechanics and Engineering 391 (2022)
  114494.

\bibitem{casquero2020seam}
H.~Casquero, X.~Wei, D.~Toshniwal, A.~Li, T.~J.~R. Hughes, J.~Kiendl, Y.~J.
  Zhang, Seamless integration of design and {K}irchhoff--{L}ove shell analysis
  using analysis-suitable unstructured {T}-splines, Computer Methods in Applied
  Mechanics and Engineering 360 (2020) 112765.

\bibitem{toshniwal2017smooth}
D.~Toshniwal, H.~Speleers, T.~J.~R. Hughes, Smooth cubic spline spaces on
  unstructured quadrilateral meshes with particular emphasis on extraordinary
  points: {G}eometric design and isogeometric analysis considerations, Computer
  Methods in Applied Mechanics and Engineering 327 (2017) 411--458.

\bibitem{nagy2015numerical}
A.~P. Nagy, D.~J. Benson, On the numerical integration of trimmed isogeometric
  elements, Computer Methods in Applied Mechanics and Engineering 284 (2015)
  165--185.

\bibitem{buffa2020minimal}
A.~Buffa, R.~Puppi, R.~V{\'a}zquez, A minimal stabilization procedure for
  isogeometric methods on trimmed geometries, SIAM Journal on Numerical
  Analysis 58~(5) (2020) 2711--2735.

\bibitem{leidinger2019explicit}
L.~Leidinger, M.~Breitenberger, A.~Bauer, S.~Hartmann, R.~W{\"u}chner, K.-U.
  Bletzinger, F.~Duddeck, L.~Song, Explicit dynamic isogeometric {B}-{R}ep
  analysis of penalty-coupled trimmed {NURBS} shells, Computer Methods in
  Applied Mechanics and Engineering 351 (2019) 891--927.

\bibitem{pasch2021priori}
T.~Pasch, L.~Leidinger, A.~Apostolatos, R.~W{\"u}chner, K.-U. Bletzinger,
  F.~Duddeck, A priori penalty factor determination for (trimmed) {NURBS}-based
  shells with {D}irichlet and coupling constraints in isogeometric analysis,
  Computer Methods in Applied Mechanics and Engineering 377 (2021) 113688.

\bibitem{Kiendl2009}
J.~Kiendl, K.-U. Bletzinger, J.~Linhard, R.~Wuchner, Isogeometric shell
  analysis with {K}irchhoff-{L}ove elements, Computer Methods in Applied
  Mechanics and Engineering 198 (2009) 3902--3914.

\bibitem{Kiendl2015}
J.~Kiendl, M.-C. Hsu, M.~C. Wu, A.~Reali, Isogeometric {K}irchhoff--{L}ove
  shell formulations for general hyperelastic materials, Computer Methods in
  Applied Mechanics and Engineering 291 (2015) 280--303.

\bibitem{casquero2017arbitrary}
H.~Casquero, L.~Liu, Y.~Zhang, A.~Reali, J.~Kiendl, H.~Gomez, Arbitrary-degree
  {T}-splines for isogeometric analysis of fully nonlinear {K}irchhoff-{L}ove
  shells, Computer-Aided Design 82 (2017) 140--153.

\bibitem{duong2017new}
T.~X. Duong, F.~Roohbakhshan, R.~A. Sauer, A new rotation-free isogeometric
  thin shell formulation and a corresponding continuity constraint for patch
  boundaries, Computer Methods in Applied Mechanics and Engineering 316 (2017)
  43--83.

\bibitem{roohbakhshan2017efficient}
F.~Roohbakhshan, R.~A. Sauer, Efficient isogeometric thin shell formulations
  for soft biological materials, Biomechanics and modeling in mechanobiology
  16~(5) (2017) 1569--1597.

\bibitem{leonetti2023mixed}
L.~Leonetti, J.~Kiendl, A mixed integration point ({MIP}) formulation for
  hyperelastic {K}irchhoff--{L}ove shells for nonlinear static and dynamic
  analysis, Computer Methods in Applied Mechanics and Engineering 416 (2023)
  116325.

\bibitem{bischoff2004models}
M.~Bischoff, K.-U. Bletzinger, W.~Wall, E.~Ramm, Models and finite elements for
  thin-walled structures, Encyclopedia of computational mechanics (2004).

\bibitem{bieber2018variational}
S.~Bieber, B.~Oesterle, E.~Ramm, M.~Bischoff, A variational method to avoid
  locking--independent of the discretization scheme, International Journal for
  Numerical Methods in Engineering 114~(8) (2018) 801--827.

\bibitem{greco2018reconstructed}
L.~Greco, M.~Cuomo, L.~Contrafatto, A reconstructed local $\overline{B}$
  formulation for isogeometric {K}irchhoff--{L}ove shells, Computer Methods in
  Applied Mechanics and Engineering 332 (2018) 462--487.

\bibitem{zou2021galerkin}
Z.~Zou, T.~J.~R. Hughes, M.~A. Scott, R.~A. Sauer, E.~J. Savitha, Galerkin
  formulations of isogeometric shell analysis: {A}lleviating locking with
  {G}reville quadratures and higher-order elements, Computer Methods in Applied
  Mechanics and Engineering 380 (2021) 113757.

\bibitem{stolarski1983shear}
H.~Stolarski, T.~Belytschko, Shear and membrane locking in curved {C}0
  elements, Computer methods in applied mechanics and engineering 41~(3) (1983)
  279--296.

\bibitem{echter2013hierarchic}
R.~Echter, B.~Oesterle, M.~Bischoff, A hierarchic family of isogeometric shell
  finite elements, Computer Methods in Applied Mechanics and Engineering 254
  (2013) 170--180.

\bibitem{oesterle2016shear}
B.~Oesterle, E.~Ramm, M.~Bischoff, A shear deformable, rotation-free
  isogeometric shell formulation, Computer Methods in Applied Mechanics and
  Engineering 307 (2016) 235--255.

\bibitem{oesterle2017hierarchic}
B.~Oesterle, R.~Sachse, E.~Ramm, M.~Bischoff, Hierarchic isogeometric large
  rotation shell elements including linearized transverse shear
  parametrization, Computer Methods in Applied Mechanics and Engineering 321
  (2017) 383--405.

\bibitem{zou2022efficient}
Z.~Zou, T.~J.~R. Hughes, M.~A. Scott, D.~Miao, R.~A. Sauer, Efficient and
  robust quadratures for isogeometric analysis: {R}educed gauss and
  {G}auss--{G}reville rules, Computer Methods in Applied Mechanics and
  Engineering 392 (2022) 114722.

\bibitem{hauptmann1998systematic}
R.~Hauptmann, K.~Schweizerhof, A systematic development of
  ‘solid-shell’element formulations for linear and non-linear analyses
  employing only displacement degrees of freedom, International Journal for
  Numerical Methods in Engineering 42~(1) (1998) 49--69.

\bibitem{liu1998multiple}
W.~K. Liu, Y.~Guo, S.~Tang, T.~Belytschko, A multiple-quadrature eight-node
  hexahedral finite element for large deformation elastoplastic analysis,
  Computer Methods in Applied Mechanics and Engineering 154~(1-2) (1998)
  69--132.

\bibitem{cardoso2008enhanced}
R.~P. Cardoso, J.~W. Yoon, M.~Mahardika, S.~Choudhry, R.~Alves~de Sousa,
  R.~Fontes~Valente, Enhanced assumed strain ({EAS}) and assumed natural strain
  ({ANS}) methods for one-point quadrature solid-shell elements, International
  Journal for Numerical Methods in Engineering 75~(2) (2008) 156--187.

\bibitem{bouclier2013development}
R.~Bouclier, T.~Elguedj, A.~Combescure, On the development of {NURBS}-based
  isogeometric solid shell elements: 2{D} problems and preliminary extension to
  3{D}, Computational Mechanics 52~(5) (2013) 1085--1112.

\bibitem{bouclier2013efficient}
R.~Bouclier, T.~Elguedj, A.~Combescure, Efficient isogeometric {NURBS}-based
  solid-shell elements: mixed formulation and {B}-method, Computer Methods in
  Applied Mechanics and Engineering 267 (2013) 86--110.

\bibitem{bouclier2015isogeometric}
R.~Bouclier, T.~Elguedj, A.~Combescure, An isogeometric locking-free
  {NURBS}-based solid-shell element for geometrically nonlinear analysis,
  International Journal for Numerical Methods in Engineering 101~(10) (2015)
  774--808.

\bibitem{macneal1978simple}
R.~H. MacNeal, A simple quadrilateral shell element, Computers \& Structures
  8~(2) (1978) 175--183.

\bibitem{hughes1981finite}
T.~J.~R. Hughes, T.~Tezduyar, Finite elements based upon {M}indlin plate theory
  with particular reference to the four-node bilinear isoparametric element
  (1981).

\bibitem{macneal1982derivation}
R.~H. Macneal, Derivation of element stiffness matrices by assumed strain
  distributions, Nuclear Engineering and Design 70~(1) (1982) 3--12.

\bibitem{dvorkin1984continuum}
E.~N. Dvorkin, K.-J. Bathe, A continuum mechanics based four-node shell element
  for general non-linear analysis, Engineering computations 1 (1984) 77--88.

\bibitem{zienkiewicz1971reduced}
O.~Zienkiewicz, R.~Taylor, J.~Too, Reduced integration technique in general
  analysis of plates and shells, International Journal for Numerical Methods in
  Engineering 3~(2) (1971) 275--290.

\bibitem{flanagan1981uniform}
D.~Flanagan, T.~Belytschko, A uniform strain hexahedron and quadrilateral with
  orthogonal hourglass control, International journal for numerical methods in
  engineering 17~(5) (1981) 679--706.

\bibitem{belytschko1984hourglass}
T.~Belytschko, J.~S.-J. Ong, W.~K. Liu, J.~M. Kennedy, Hourglass control in
  linear and nonlinear problems, Computer Methods in Applied Mechanics and
  Engineering 43~(3) (1984) 251--276.

\bibitem{belytschko1984explicit}
T.~Belytschko, J.~I. Lin, T.~Chen-Shyh, Explicit algorithms for the nonlinear
  dynamics of shells, Computer methods in applied mechanics and engineering
  42~(2) (1984) 225--251.

\bibitem{lsdyna}
{LS-DYNA}, {A}nsys, {I}nc., {L}ivermore, {CA}, {USA},
  \url{https://lsdyna.ansys.com}.

\bibitem{ansysmechanical}
{A}nsys {M}echanical, {A}nsys, {I}nc., {C}anonsburg, {PA}, {USA},
  \url{https://www.ansys.com/products/structures/ansys-mechanical}.

\bibitem{abaqus}
{A}baqus, {D}assault {S}ystemes, {J}ohnston, {RI}, {USA},
  \url{https://www.3ds.com/products-services/simulia/products/abaqus/}.

\bibitem{malkus1978mixed}
D.~S. Malkus, T.~J.~R. Hughes, Mixed finite element methods—reduced and
  selective integration techniques: a unification of concepts, Computer Methods
  in Applied Mechanics and Engineering 15~(1) (1978) 63--81.

\bibitem{simo1986variational}
J.~C. Simo, T.~J.~R. Hughes, {On the variational foundations of assumed strain
  methods}, Journal of Applied Mechanics 53~(1) (1986) 51--54.

\bibitem{adam2014improved}
C.~Adam, S.~Bouabdallah, M.~Zarroug, H.~Maitournam, Improved numerical
  integration for locking treatment in isogeometric structural elements. {P}art
  {I}: {B}eams, Computer Methods in Applied Mechanics and Engineering 279
  (2014) 1--28.

\bibitem{adam2015improved}
C.~Adam, S.~Bouabdallah, M.~Zarroug, H.~Maitournam, Improved numerical
  integration for locking treatment in isogeometric structural elements. {P}art
  {II}: {P}lates and shells, Computer Methods in Applied Mechanics and
  Engineering 284 (2015) 106--137.

\bibitem{casquero2022removing}
H.~Casquero, M.~Golestanian, Removing membrane locking in quadratic
  {NURBS}-based discretizations of linear plane {K}irchhoff rods: {CAS}
  elements, Computer Methods in Applied Mechanics and Engineering 399 (2022)
  115354.

\bibitem{casquero2023trods}
M.~Golestanian, H.~Casquero, Extending {CAS} elements to remove shear and
  membrane locking from quadratic {NURBS}-based discretizations of linear plane
  {T}imoshenko rods, International Journal for Numerical Methods in Engineering
  124~(18) (2023) 3997--4021.

\bibitem{adam2015selective}
C.~Adam, T.~J.~R. Hughes, S.~Bouabdallah, M.~Zarroug, H.~Maitournam, Selective
  and reduced numerical integrations for {NURBS}-based isogeometric analysis,
  Computer Methods in Applied Mechanics and Engineering 284 (2015) 732--761.

\bibitem{leonetti2018efficient}
L.~Leonetti, F.~Liguori, D.~Magisano, G.~Garcea, An efficient isogeometric
  solid-shell formulation for geometrically nonlinear analysis of elastic
  shells, Computer Methods in Applied Mechanics and Engineering 331 (2018)
  159--183.

\bibitem{leonetti2019simplified}
L.~Leonetti, D.~Magisano, A.~Madeo, G.~Garcea, J.~Kiendl, A.~Reali, A
  simplified {K}irchhoff--{L}ove large deformation model for elastic shells and
  its effective isogeometric formulation, Computer Methods in Applied Mechanics
  and Engineering 354 (2019) 369--396.

\bibitem{hokkanen2020quadrature}
J.~Hokkanen, D.~M. Pedroso, Quadrature rules for isogeometric shell
  formulations: Study using a real-world application about metal forming,
  Computer Methods in Applied Mechanics and Engineering 363 (2020) 112904.

\bibitem{hughes1980generalization}
T.~J.~R. Hughes, Generalization of selective integration procedures to
  anisotropic and nonlinear media, International Journal for Numerical Methods
  in Engineering 15~(9) (1980) 1413--1418.

\bibitem{thomas}
T.~Elguedj, Y.~Bazilevs, V.~M. Calo, T.~J.~R. Hughes, $\overline{B}$ and
  $\overline{F}$ projection methods for nearly incompressible linear and
  non-linear elasticity and plasticity using higher-order {NURBS} elements,
  Computer methods in applied mechanics and engineering 197 (2008) 2732--2762.

\bibitem{bouclier2012locking}
R.~Bouclier, T.~Elguedj, A.~Combescure, Locking free isogeometric formulations
  of curved thick beams, Computer Methods in Applied Mechanics and Engineering
  245 (2012) 144--162.

\bibitem{zhang2018locking}
G.~Zhang, R.~Alberdi, K.~Khandelwal, On the locking free isogeometric
  formulations for 3-{D} curved {T}imoshenko beams, Finite Elements in Analysis
  and Design 143 (2018) 46--65.

\bibitem{miao2018bezier}
D.~Miao, M.~J. Borden, M.~A. Scott, D.~C. Thomas, B{\'e}zier $\overline{B}$
  projection, Computer Methods in Applied Mechanics and Engineering 335 (2018)
  273--297.

\bibitem{greco2017efficient}
L.~Greco, M.~Cuomo, L.~Contrafatto, S.~Gazzo, An efficient blended mixed
  {B}-spline formulation for removing membrane locking in plane curved
  {K}irchhoff rods, Computer Methods in Applied Mechanics and Engineering 324
  (2017) 476--511.

\bibitem{zou2020isogeometric}
Z.~Zou, M.~A. Scott, D.~Miao, M.~Bischoff, B.~Oesterle, W.~Dornisch, An
  isogeometric {R}eissner--{M}indlin shell element based on {B}{\'e}zier dual
  basis functions: {O}vercoming locking and improved coarse mesh accuracy,
  Computer Methods in Applied Mechanics and Engineering 370 (2020) 113283.

\bibitem{bucalem1993higher}
M.~L. Bucalem, K.-J. Bathe, Higher-order {MITC} general shell elements,
  International Journal for Numerical Methods in Engineering 36~(21) (1993)
  3729--3754.

\bibitem{kim2022isogeometric}
M.-G. Kim, G.-H. Lee, H.~Lee, B.~Koo, Isogeometric analysis for geometrically
  exact shell elements using {B}{\'e}zier extraction of {NURBS} with assumed
  natural strain method, Thin-Walled Structures 172 (2022) 108846.

\bibitem{casquero2023le}
H.~Casquero, M.~Golestanian, Vanquishing volumetric locking in quadratic
  {NURBS}-based discretizations of nearly-incompressible linear elasticity:
  {CAS} elements, Computational Mechanics (2023) accepted for publication.

\bibitem{macneal1985proposed}
R.~H. Macneal, R.~L. Harder, A proposed standard set of problems to test finite
  element accuracy, Finite elements in analysis and design 1~(1) (1985) 3--20.

\bibitem{Belytschko1985}
T.~Belytschko, H.~Stolarski, W.~K. Liu, N.~Carpenter, J.~S. Ong, Stress
  projection for membrane and shear locking in shell finite elements, Computer
  Methods in Applied Mechanics and Engineering 51 (1985) 221 -- 258.

\bibitem{chapelle1998fundamental}
D.~Chapelle, K.-J. Bathe, Fundamental considerations for the finite element
  analysis of shell structures, Computers \& Structures 66~(1) (1998) 19--36.

\bibitem{bathe2000evaluation}
K.-J. Bathe, A.~Iosilevich, D.~Chapelle, An evaluation of the {MITC} shell
  elements, Computers \& Structures 75~(1) (2000) 1--30.

\bibitem{lovemathematical}
A.~Love, The mathematical theory of elasticity, 4nd (1927).

\bibitem{koiter1970mathematical}
W.~Koiter, On the mathematical foundation of shell theory, in: Proc. Int.
  Congr. of Mathematics, Nice, Vol.~3, 1970, pp. 123--130.

\bibitem{Hughes2012}
T.~J.~R. Hughes, The finite element method: {L}inear static and dynamic finite
  element analysis, Courier Corporation, 2012.

\bibitem{dalcin2016petiga}
L.~Dalcin, N.~Collier, P.~Vignal, A.~C{\^o}rtes, V.~M. Calo, Pet{IGA}: {A}
  framework for high-performance isogeometric analysis, Computer Methods in
  Applied Mechanics and Engineering 308 (2016) 151--181.

\bibitem{petsc-web-page}
S.~Balay, M.~F. Adams, J.~Brown, P.~Brune, K.~Buschelman, V.~Eijkhout, W.~D.
  Gropp, D.~Kaushik, M.~G. Knepley, L.~C. McInnes, K.~Rupp, B.~F. Smith,
  H.~Zhang, {PETS}c {W}eb page, \url{http://www.mcs.anl.gov/petsc} (2014).

\bibitem{greco2013b}
L.~Greco, M.~Cuomo, B-spline interpolation of {K}irchhoff-{L}ove space rods,
  Computer Methods in Applied Mechanics and Engineering 256 (2013) 251--269.

\bibitem{belytschko1981explicit}
T.~Belytschko, C.~Tsay, Explicit algorithms for nonlinear dynamics of shells,
  Nonlinear finite element analysis of plates and shells (1981) 209--231.

\bibitem{engelmann1989simple}
B.~E. Engelmann, R.~Whirley, G.~Goudreau, A simple shell element formulation
  for large-scale elastoplastic analysis, Tech. rep., Lawrence Livermore
  National Lab.(LLNL), Livermore, CA (United States) (1989).

\bibitem{pian1984rational}
T.~Pian, K.~Sumihara, Rational approach for assumed stress elements,
  International Journal of Numerical Methods in Engineering 20 (1984)
  1685--1695.

\bibitem{kam1985use}
W.~Kam~Liu, T.~Belytschko, J.~Shau-Jen~Ong, S.~E. Law, Use of stabilization
  matrices in non-linear analysis, Engineering computations 2~(1) (1985)
  47--55.

\bibitem{liu1994multiple}
W.~K. Liu, Y.-K. Hu, T.~Belytschko, Multiple quadrature underintegrated finite
  elements, International Journal for Numerical Methods in Engineering 37~(19)
  (1994) 3263--3289.

\bibitem{chapelle2010finite}
D.~Chapelle, K.-J. Bathe, The finite element analysis of shells-fundamentals,
  Springer Science \& Business Media, 2010.

\bibitem{bathe1986formulation}
K.-J. Bathe, E.~N. Dvorkin, A formulation of general shell elements—the use
  of mixed interpolation of tensorial components, International journal for
  numerical methods in engineering 22~(3) (1986) 697--722.

\end{thebibliography}






\end{document}